\documentclass[10pt,prd,onecolumn,
nofootinbib]{revtex4-2}
\usepackage[dvips]{graphics,graphicx}
\usepackage[usenames,dvipsnames]{xcolor}
\definecolor{darkblue}{RGB}{0,0,196}
\definecolor{darkgreen}{RGB}{0,120,0}

\usepackage[colorlinks=true,linktocpage=true,linkcolor=darkblue,citecolor=red,urlcolor=darkblue]{hyperref}
\usepackage{framed}
\usepackage{mathtools}
\usepackage{float}
\usepackage{color}
\usepackage{xcolor}
\usepackage{slashed}
\usepackage{wrapfig}
\usepackage{tikz-feynman}
\usepackage{subcaption}
\usepackage{amsmath}
\usepackage{amssymb}
\allowdisplaybreaks
\usepackage{listings}
\usepackage{tikz-feynhand}
\usepackage{bm}
\usepackage{fixmath} 
\usepackage{braket}
\usepackage{multirow}
\usepackage{lipsum} 
\usepackage[toc,page]{appendix}
\usepackage{setspace}
\usepackage{simpler-wick}
\usepackage{stackengine}
\usetikzlibrary{decorations.markings}
\usepackage{bbm}
\usepackage{booktabs}

\newcommand{\no}{\nonumber}
\definecolor {darkgreen}{rgb}{0.2,0.7,0.2}
\definecolor{lightblue}{RGB}{102,178,255}
\tikzset{
	dfermion/.style={
		decoration={
			markings,
			mark=at position 0.5 with {
				\draw[line width=0.4pt] (-1pt,-5pt) -- (-1pt,5pt);
				\draw[line width=0.4pt] (1pt,-5pt) -- (1pt,5pt);
			},
		},
		postaction={decorate},
		line width=0.8pt,
	}
}
\tikzset{
	dphoton/.style={
		/tikz/photon,
		postaction={
			decorate,
			decoration={
				markings,
				mark=at position 0.5 with {
					\draw[line width=0.3pt] (-1.2pt,-5pt) -- (-1.2pt,5pt);
					\draw[line width=0.3pt] (0.8pt,-5pt) -- (0.8pt,5pt);
				}
			}
		}
	}
}
\tikzset{
	mybullet/.style={
		shape=circle,
		fill=black,
		inner sep=1.5pt,
		minimum size=0pt
	}
}
\tikzfeynmanset{
	charged photon/.style={
		photon,
		line width=0.5pt,
		/tikz/decoration={markings,
			mark=at position 0.5 with {\arrow[line width=1.2pt, scale=0.8]{>}}},
		postaction=decorate
	},
	anti charged photon/.style={
		photon,
		line width=0.5pt,
		/tikz/decoration={markings,
			mark=at position 0.5 with {\arrow[line width=1.2pt, scale=0.8]{<}}},
		postaction=decorate
	}
}
\tikzfeynmanset{
	ct insertion/.style={
		/tikz/postaction={
			decorate,
			decoration={
				markings,
				mark=at position 0.5 with {
					\node[draw, circle, fill=white, inner sep=0.15pt, line width=0.35pt]
					{\scriptsize\(\boldsymbol{\times}\)};
				}
			}
		}
	}
}
\newcommand\hstar[1]{\ThisStyle{\ensurestackMath{%
			\setbox0=\hbox{$\SavedStyle#1$}%
			\stackengine{0pt}{\copy0}{\kern.2\ht0\smash{\SavedStyle\star}}{O}{c}{F}{T}{S}}}}
		\newcommand{\Qslash}{Q\!\!\!/}
		\newcommand{\uslash}{u\!\!\!/}

		\newcommand{\rev}{\mathrm{rev}}
		
		\newcommand{\Tr}{\operatorname{Tr}}
\begin{document}
\title{Hard-Region Fermion Self-Energy and Fermion--Photon Vertex in Thermal QED through Two Loops}
\author{Ali Hataei}
\email{a.hataei@alumni.sbu.ac.ir}
\email{Hataeiali76@gmail.com}
\affiliation{Department of Physics, Shahid Beheshti University, Tehran, Iran}
	\begin{abstract}
We compute hard-region contributions to the fermion self-energy and the
off-shell fermion--photon vertex in massless thermal QED in a covariant
\(R_\xi\) gauge. The calculation includes the one-loop
next-to-leading-power and two-loop leading-power terms in the expansion in
soft external momenta, together with the zero-temperature counterterms
required to remove ultraviolet subdivergences.
In the real-time formalism, we first derive symmetry constraints on the
two-fermion--\(N\)-photon amplitudes, which lead to distinct selection rules
in the soft-momentum expansion. The fermion--photon vertex is decomposed into
Dirac-vector and axial-vector coefficients. The former satisfies the
Ward--Takahashi identity, while the latter is transverse to the photon
momentum. These relations are verified at the integrand level.
The hard self-energy corrections computed here produce no finite
contribution to the fermion damping rate. Our results provide previously
unavailable two- and three-point hard-region data for higher-order analyses
of fermion mass shifts and inclusive rates, as well as for the construction
of a next-to-leading-order fermionic effective action.
\end{abstract}
\maketitle
\section{Introduction}\label{Introduction}
\subsection{General Framework}
Thermal field theory describes relativistic matter, from the quark--gluon plasma to the early Universe, and provides a framework for studying thermal-medium dynamics~\cite{Klimov:1982bv,Weldon,PhysRevLett.64.1338,Arnold_2002,PeterArnold_2001,PhysRevD.61.056003,Blaizot_1999,Mustafa_2023,Haque:2024gva,PhysRevD.83.075011}.
At weak coupling, temperature $T$ is usually treated as the dominant scale, leading to a separation between hard momenta of order \(T\) and soft momenta of order \(eT\), where \(e\) is the coupling \cite{Bellac:2011kqa}. More generally, hard momenta, semihard momenta (\(eT\ll Q\ll T\)) \cite{Wang_2000,PhysRevLett.134.011905,mzw1-bsvy}, and soft momenta probe microscopic, intermediate, and collective dynamics, respectively. The collective sector governs screening and linear response, quasiparticle dispersion and damping, relaxation toward equilibrium, infrared-sensitive thermodynamic contributions, transport coefficients, and thermal spectral and particle-production rates~\cite{BRAATEN1990569,BraatenPisarski_Ward,Blaizot_2002,Mustafa_2023,Bellac:2011kqa,Kapusta:2006pm}.

For the collective scale, the dominant one-loop contributions to gauge-theory self-energies arise from hard internal momenta and define the leading-order (LO) hard-thermal-loop (HTL) amplitudes. These amplitudes are gauge independent, satisfy tree-level-like Ward identities, and are generated by gauge-invariant nonlocal effective actions~\cite{BRAATEN1990569,FRENKEL1990199,TAYLOR1990115,BraatenPisarski_Ward,BraatenPisarski_HTLAction,Blaizot_2002,Mustafa_2023}. In particular, the fermionic sector is described by~\cite{Bellac:2011kqa}
\begin{align}\label{LO-effective-action}
	\Gamma_{\rm HTL}^{\psi}
	&=
	m_f^2\int_x\int\frac{d\Omega_{\bm v}}{4\pi}\,
	\bar\psi(x)\,
	\frac{\slashed v}{i\,v\!\cdot\!D}\,
	\psi(x),
\end{align}
where \(m_f^2\) is the fermionic HTL mass parameter, \(v^\mu=(1,\bm v)\) with \(|\bm v|=1\), and \(D_\mu\) is the covariant derivative. At vanishing background gauge field, corresponding to \(N=0\), Eq.~\eqref{LO-effective-action} reduces to the fermion self-energy. The replacement \(v\!\cdot\!\partial\to v\!\cdot\!D\), which leads to the complete effective action, has been established through direct hard-loop calculations and Ward identities~\cite{FRENKEL1990199,TAYLOR1990115,BraatenPisarski_Ward}, background-field methods~\cite{ElmforsHanssonZahed1999}, and Dyson--Schwinger and kinetic theory~\cite{BlaizotIancuQED1993,BlaizotIancuSoft1994,Blaizot_2002}. Expanding \((v\!\cdot\!D)^{-1}\) in the gauge field generates the complete leading-order two-fermion--\(N\)-gauge-boson amplitudes.

Beyond LO, thermal corrections can have a large numerical impact and may
occur in successive powers of the coupling rather than powers of
\(e^2\)~\cite{Braaten_1995,BlaizotIancuRebhan2003}. 
At next-to-leading order (NLO), the thermal-photon rate increases by approximately \(20\%\) for
\(\alpha_s=0.3\), while low-mass dilepton rates can increase by up to
\(30\)--\(40\%\)~\cite{Ghiglieri2013,GhiglieriMoore2014};
heavy-quark diffusion shows poor convergence
~\cite{CaronHuotMoorePRL,CaronHuotMooreJHEP}; and the ``almost NLO''
shear-viscosity result lowers \(\eta/s\) by more than a factor of three
at physically relevant couplings~\cite{GhiglieriMooreTeaney2018}.
Fermionic HTL structures also appear in the full leading-order
production rate of ultrarelativistic right-handed neutrinos
~\cite{BesakBodeker2012}, while relative-\(O(g)\) corrections reduce
active-neutrino interaction rates by about \(15\)--\(40\%\) in the
high-temperature regime below the electroweak crossover
~\cite{JacksonLaine2019}. These examples show that corrections
beyond LO can be numerically significant and motivate the systematic
determination of higher-order contributions, including the fermion-sector amplitudes
computed here and their future extension to non-Abelian gauge
theories. 

When soft internal lines are present, the bare perturbative expansion contains an infinite series of contributions at the same parametric order. HTL resummation introduces a single resummed amplitude that contains this infinite series. On the one hand, consider the one-loop fermion self-energy diagram in which all propagators and vertices are replaced by their HTL counterparts. The resummed soft one-loop self-energy can then be expanded schematically as
\begin{equation}
	\Sigma_{\rm soft}^{*(1)}=\Sigma_S+\Sigma_{HS}+\Sigma_{H^2S}+\cdots,
	\label{eq:resummed_soft_expansion_intro}
\end{equation}
where \(S\) denotes a soft loop, \(H\) a hard HTL subgraph, and the star the fully resummed contribution. On the other hand, at two loops, the momentum regions are hard--hard, hard--soft, and soft--soft. In Feynman gauge and in the limit \(\mathbf q\to0\), Ref.~\cite{Mottola_2010} showed that the bare hard--soft two-loop fermion self-energy, which scales as \(e^2T\), coincides with the \(\Sigma_{HS}\) term in Eq.~\eqref{eq:resummed_soft_expansion_intro}; it must therefore not be added independently. More generally, the resummed one-loop self-energy above contains the infinite series in which the one-loop contribution is soft and successive subgraphs are hard. Thus, HTL resummation is mandatory to obtain the complete set of contributions at the same parametric order. Most importantly, this reorganization leads to gauge-independent on-shell results
~\cite{BRAATEN1990569,FRENKEL1990199,TAYLOR1990115,
	BraatenPisarski_Ward,BraatenPisarski_HTLAction,
	PhysRevLett.64.1338,PhysRevD.45.4632,PhysRevD.46.1829,
	Blaizot_2002}.
 
The soft--soft region likewise belongs to the resummed soft sector.  
By contrast, in the hard--hard region, which is the subject of this study,
HTL insertions are suppressed relative to bare inverse propagators.
Therefore, strict perturbation theory applies here.
\begin{table}[t]
	\centering
	\caption{\label{tab:orders}Order and scale terminology.}
	\begin{tabular}{lp{0.67\columnwidth}}
		\hline\hline
		Expansion in \(Q/T\)
		& LP, NLP, \(\ldots\)
		\\
		Pole/observable hierarchy
		& \(\mathrm{LO}_{\rm pole}\),
		\(\mathrm{NLO}_{\rm pole}\),
		\(\mathrm{NNLO}_{\rm pole}\), \(\ldots\)
		\\
		Hard-matching hierarchy
		& \(\mathrm{LO}_{\rm hard}\): one-loop LP;
		\(\mathrm{NLO}_{\rm hard}\) at \(Q\sim eT\):
		one-loop NLP and two-loop LP
		\\
		External momenta
		& Generic \(Q\ll T\), including \(Q\sim eT\) and
		\(eT\ll Q\ll T\)
		\\
		\hline\hline
	\end{tabular}
\end{table}
\subsection{Power Counting and Existing Gaps}
Unless noted otherwise, LO, NLO, and NNLO refer to the perturbative order of the corresponding observable. For the HTL effective action, LO and NLO instead label successive hard matching contributions. Leading power (LP) and next-to-leading power (NLP) refer to the expansion in external momentum at a fixed loop order. Table~\ref{tab:orders} summarizes these conventions.  For the fermion self-energy, the main hard-region amplitudes scale as follows:
\begin{equation}
	\Sigma^{(1;\mathrm{LP})}(Q)
	\sim \frac{e^2T^2}{Q},
	\qquad
	\Sigma^{(1;\mathrm{NLP})}(Q)
	\sim e^2Q,
	\qquad
	\Sigma^{(2;\mathrm{LP})}(Q)
	\sim \frac{e^4T^2}{Q},
	\label{eq:semisoft_hierarchy_intro}
\end{equation}
where the superscripts denote loop and power orders. At \(Q\sim eT\), these terms scale as \(eT\), \(e^3T\), and \(e^3T\), respectively. The one-loop LP term generates the LO fermionic HTL action and determines the leading collective mass. The NLO fermion mass shift and damping rate instead arise from the fully HTL-resummed soft one-loop self-energy of order \(e^2T\)~\cite{Mirza_2013,Carrington_2007,PhysRevD.45.4632,PhysRevD.46.1829,PhysRevD.78.045018,Sumit_2023}. The hard one-loop NLP and hard--hard two-loop LP terms, studied here, first contribute to fermion pole quantities at NNLO, while they are also necessary building blocks to construct the NLO fermionic effective action. The gluonic sector follows the same hierarchy~\cite{PhysRevLett.64.1338,PhysRevD.42.2156,SCHULZ1994353}, whereas the photon sector obeys a different hard/soft ordering~\cite{Mirza_2013}. For the fermion sector, schematically,
\begin{equation}
	\underbrace{\Sigma^{(1;\mathrm{LP})}}_{\sim eT}
	\longrightarrow
	\{M_{\rm LO},\gamma_{\rm LO}=0\},
	\quad
	\underbrace{\Sigma^{*(1)}_{\rm soft}}_{\sim e^2T}
	\longrightarrow
	\{\delta M_{\rm NLO},\gamma_{\rm NLO}\},
	\quad
	\underbrace{\Sigma^{(1;\mathrm{NLP})}+\Sigma^{(2;\mathrm{LP})}+\Sigma_{\rm rem}^{(e^3T)}}_{\sim e^3T}
	\longrightarrow
	\{\delta M_{\rm NNLO},\gamma_{\rm NNLO}\},
	\label{eq:fermion_observable_hierarchy_intro}
\end{equation}
where \(\Sigma_{\rm rem}^{(e^3T)}\) denotes the additional same-order ingredients required for a complete pole quantity at the collective scale.
The self-energies $\Sigma^{(1;\mathrm{NLP})}$ and $\Sigma^{(2;\mathrm{LP})}$ are the one-loop NLP and two-loop LP, respectively.
This assignment is specific to the collective scale. At \(Q\sim\sqrt e\,T\), for example, the one-loop NLP term is a relative-\(O(e)\) correction to the LP kernel, whereas the two-loop LP term is relative \(O(e^2)\). Because the amplitudes derived here are valid for generic \(Q\ll T\), they apply both at \(Q\sim eT\) and throughout the semihard window \(eT\ll Q\ll T\).

Higher-order hard contributions to gauge-boson self-energies have been studied through two loops~\cite{Bellac:2011kqa,Carignano_2018,PhysRevD.94.025017,Gorda_2023,Carrington,Kaapo}. The corresponding photon corrections are gauge independent~\cite{Kaapo,Carrington}, whereas the one-loop NLP and two-loop LP hard gluon self-energies are separately gauge dependent~\cite{Gorda_2023}. This does not signal an inconsistency, because physical quantities require the combination of all hard and soft contributions at the same order. Related Feynman-gauge results have also been obtained using kinetic theory~\cite{Ekstedt_2023}.
The fermionic hard sector is less developed. The fermion self-energy was previously known at one-loop LP and NLP~\cite{Weldon,Carignano_2018}; the NLP term is gauge dependent and contains a local ultraviolet divergence removed by the zero-temperature counterterm~\cite{Carignano_2018}. The hard two-loop self-energy, however, had not been calculated. Similarly, the off-shell fermion--photon vertex has mainly been analyzed at one loop in LP~\cite{Ayala_2001,Bellac:2011kqa}; to our knowledge, its one-loop NLP and hard two-loop corrections were not previously available.

These vertex amplitudes generally cannot be reconstructed from the self-energy. The Ward--Takahashi identity (WTI) determines only the projection longitudinal to the photon momentum, leaving transverse structures unconstrained~\cite{PhysRevD.22.2542,Ayala_2001,Kizilers__1995,K_z_lers__2009}. This becomes important beyond LP, where axial-vector coefficients may appear and are themselves transverse. The NLO photon-sector effective action has been constructed in Refs.~\cite{Carrington,Carignano_2018}, but the analogous fermionic construction requires systematic matching of the two-fermion--\(N\)-photon amplitudes. Schematically, the fermion-bilinear action may be expanded in the background gauge field as
\begin{equation}
	\Gamma_{\bar\psi\psi}[A]
	=
	\sum_{N=0}^{\infty}
	\frac{(-1)^N}{N!}\,
	\Gamma^{(N)\mu_1\cdots\mu_N}
	A_{\mu_1}\cdots A_{\mu_N}.
\end{equation}
where the self-energy determines \(\Gamma^{(0)}\), the complete off-shell fermion--photon vertex determines \(\Gamma^{(1)\mu}\), and the two-fermion--two-photon kernel determines \(\Gamma^{(2)\mu\nu}\). The vertex therefore supplies independent transverse matching data that are absent from the self-energy. Closure through quadratic order in the background field additionally requires the two-fermion--two-photon kernel, followed by a determination of whether the higher-\(N\) amplitudes are generated by a common gauge-covariant operator, as at LO, or require additional independent structures.

This work first derives nonperturbative reality, momentum-reversal, and fermion-leg-exchange constraints from the symmetries of massless QED. Together with chiral invariance, these properties restrict the allowed vector and axial-vector structures and yield distinct selection rules in the soft-external-momentum expansion. Second, the one-loop NLP and two-loop LP hard-region corrections to the fermion self-energy and the off-shell fermion--photon vertex are computed in a general covariant \(R_\xi\) gauge, including the required zero-temperature counterterms.
The analysis is restricted to retarded amplitudes with a single causal
	\(a\)-index; real-time components with multiple \(a\)-indices are outside
	the scope of the present work.
At the integrand level, the vector sector satisfies the WTI and the axial sector satisfies photon-momentum transversality, independently in each statistical and gauge sector. We also show that hard self-energy corrections yield no finite contribution to the damping rate. These amplitudes provide hard-region matching data for higher-order fermionic effective descriptions and serve as ingredients for complete mass-shift and inclusive-rate analyses. Following the hard/soft matching logic of Refs.~\cite{Braaten_1995,Jantzen_2011,SCHULZ1994353}, the analysis also
summarizes how all same-order contributions must be combined to obtain
pole quantities at the collective scale.

The paper is organized at two levels. The first level consists of the
main text, the primary result appendices, and the Supplemental Material (SM)~\cite{SupplementalMaterial}.
App.~\ref{DR} contains the dimensional-regularization (DR) integrals,
while Apps.~\ref{self-energy computations two-loop}--\ref{final results}
contain the final and lengthy integrand-level results. These results,
while being among the main achievements of this work, are too lengthy
to include in the main text. Some of the integrand-level results are
even too lengthy to be incorporated in the appendices. Therefore, the complete integrand-level results together
with the WTI and axial-transversality checks are provided in the SM.
The second level contains material that can be skipped on a first
reading. Apps.~\ref{ra-graph-rules}--\ref{APP-ct} provide the full
technical derivations, while the final results needed for the physical
argument are given in the main text.

Sec.~\ref{Real-time formalism} introduces the real-time formalism and conventions.
Sec.~\ref{fermion self-energy} presents the renormalized hard self-energy corrections.
Sec.~\ref{N-Fermion--Two-Photon vertex} derives the symmetry constraints and selection rules for the two-fermion--\(N\)-photon amplitudes.
Sec.~\ref{vertex} presents the fermion--photon vertex results.
Sec.~\ref{Matching analysis} discusses the matching analysis for the mass shift and damping rate.
Sec.~\ref{sec:conclusion} summarizes the main findings.
\section{Real-time Formalism and Conventions}
\label{Real-time formalism}
We work in massless QED in a homogeneous, isotropic, equilibrium medium at zero chemical potential $\mu=0$.
We adopt the Schwinger--Keldysh real-time formalism
\cite{Keldysh:1964ud,LANDSMAN1987141,van_Eijck_1994,Carrington_1999,
	Carrington_2007_Real,Ghiglieri_2020}, in which the retarded, advanced, and symmetric components of self-energies and propagators are related to their
\(r/a\) counterparts as \cite{Carrington_2007_Real,CaronHuot_2009}
\begin{align}
	\label{3}
	\begin{pmatrix}
		\Delta_{rr} & \Delta_{ra}\\
		\Delta_{ar} & \Delta_{aa}
	\end{pmatrix}
	=
	\begin{pmatrix}
		\Delta_s & \Delta_R\\
		\Delta_A & 0
	\end{pmatrix},
	\quad
	\begin{pmatrix}
		\Pi_{rr} & \Pi_{ra}\\
		\Pi_{ar} & \Pi_{aa}
	\end{pmatrix}
	=
	\begin{pmatrix}
		0 & \Pi_A\\
		\Pi_R & \Pi_s
	\end{pmatrix}.
\end{align}
Here \(R\), \(A\), and \(s\) denote retarded, advanced, and symmetric
components, respectively. The free fermion and gauge-boson propagators are
\begin{align}
	\label{5}
	S_j(P)
	&=
	\slashed P\,\Delta_j(P),
	\qquad
	D_j^{\mu\nu}(P)
	=
	-g^{\mu\nu}\widetilde\Delta_j(P)
	-i(1-\xi)P^\mu P^\nu\widetilde\Delta_j^{(2)}(P),
	\qquad j=R, A,s,
\end{align}
where \(\xi\) is the covariant-gauge parameter. 
In the loop calculations, the contributions associated with the \(g^{\mu\nu}\) and \((1-\xi)\)-dependent parts are referred to as the metric and longitudinal sectors, respectively.
Here, the scalar functions are defined by
\begin{align}\label{scalar-functions}
	\Delta_R(P)
	&=
	\frac{i}{P^2+ip^0\eta},
	\qquad
	\Delta_A(P)
	=
	\frac{i}{P^2-ip^0\eta},
	\qquad
	\Delta_{R/A}^{(n)}(P)
	\equiv
	\left[\Delta_{R/A}(P)\right]^n,\no
	\\
	\Delta_d^{(n)}(P)
	&=
	\Delta_R^{(n)}(P)-\Delta_A^{(n)}(P),
	\qquad
	\Delta_s^{(n)}(P)
	=
	N_F(P)\Delta_d^{(n)}(P),
	\qquad
	\widetilde\Delta_s^{(n)}(P)
	=
	N_B(P)\Delta_d^{(n)}(P),
\end{align}
with \(\eta\to0^+\). The retarded, advanced, and difference scalar functions
are the same for fermions and bosons; the tilde distinguishes the bosonic
symmetric propagator from the fermionic one. The statistical factors are
\begin{align}
	N_F(P)
	=
	\frac12-n_F(p^0),
	\qquad
	N_B(P)
	=
	\frac12+n_B(p^0),
	\qquad
	N_{F/B}'(P)
	\equiv
	\frac{dN_{F/B}(P)}{dp^0}.
\end{align}
For \(n=1\), the difference propagator is the spectral function, and we write
\(\Delta_d^{(1)}\equiv\Delta_d\).
Table~\ref{tab:all_conventions} summarizes the remaining conventions. The
\(\overline{\rm MS}\) scheme and DR are used
throughout. 
We therefore assume $D=4-2\epsilon$ where $\epsilon\rightarrow 0$ is the DR regulator and $d=D-1$ is the spatial dimension.
For the integrated self-energy, rotational symmetry allows the spatial
integration measure to be reduced by choosing the external spatial
momentum as the polar axis.  We therefore write \cite{Kaapo}
\begin{align}
	\int_P\equiv \left(\frac{e^{\gamma_E}\bar{\Lambda}^2}{4\pi}\right)^{\frac{4-D}{2}}
	\int\frac{d^{D}P}{(2\pi)^{D}}
	\equiv \mathcal{N}\int_{-\infty}^{+\infty}\frac{d{p^0}}{2\pi}\int_p\int_z,\quad 
	z\equiv\cos\theta,
\end{align}
with 
\begin{align}\label{2}
	\int_p\equiv\int_0^\infty dp\, p^{d-1},\qquad
	\int_z\equiv \int_{-1}^{+1}(1-z^2)^{\frac{d-3}{2}} dz,\qquad
	\mathcal{N}=\frac{4}{(4\pi)^{\frac{d+1}{2}}\Gamma\!\left(\frac{d-1}{2}\right)}
	\left(\frac{e^{\gamma_E}{\bar{\Lambda}}^{2}}{4\pi}\right)^{\frac{3-d}{2}} .
\end{align}
The azimuthal angular factor has already been absorbed into
$\mathcal N$, and $z$ denotes the cosine of the relevant relative angle.
Here, \(\gamma_E\) and \(\bar\Lambda\) are the Euler--Mascheroni
constant and the \(\overline{\mathrm{MS}}\) scale, respectively. For $d\to 3-2\epsilon$,
\begin{align}\label{Eq:N-expansion}
	\mathcal{N}=\frac{1}{4\pi^2}+\frac{1}{4\pi^2}\ln({\bar{\Lambda}}^2)\,\epsilon+O(\epsilon^2).
\end{align}
The corresponding radial and angular integrals needed below are collected in
App.~\ref{DR}.
For the fermion--photon vertex, we retain the full \(D\)-dimensional
notation (\(\int_P\)), since no loop integration is performed.  Indeed, reducing the angular integral to a single variable integration ($\int_z$) requires checking that no azimuthal dependence remains.

Adopting the Breitenlohner--Maison--'t Hooft--Veltman (BMHV) scheme~\cite{tHooft:1972tcz,Breitenlohner:1977hr}, we used \textsc{FeynCalc}~\cite{Shtabovenko:2023idz,Shtabovenko:2020gxv,Shtabovenko:2016sxi,Mertig:1990an} for the Dirac algebra and scalar products. 

\begin{table}[t]
	\caption{Summary of conventions.}
	\label{tab:all_conventions}
	\centering
	\begin{tabular}{p{0.28\textwidth}p{0.72\textwidth}}
		\toprule
		\textbf{Convention/Definition} & \textbf{Expression} \\
		\midrule
		Metric and four-vector
		&
		\(g^{\mu\nu}=\mathrm{diag}(1,-1,-1,-1)\), \quad
		\(P^\mu\equiv(p^0,\mathbf p)\); with \(p\equiv|\mathbf p|\);\\
		& $u^\mu=(1,\mathbf{0})$ is the plasma four-velocity in its rest frame.
		\\
		Tree-level vertex components \cite{Carrington_2007_Real,Kaapo,Gorda_2023}
		&
		\(\gamma^\mu_{rrr}=\gamma^\mu_{raa}
		=\gamma^\mu_{ara}=\gamma^\mu_{aar}=0\), \quad
		\(\gamma^\mu_{rra}=\gamma^\mu_{rar}
		=\gamma^\mu_{arr}=\gamma^\mu\), \quad
		\(\gamma^\mu_{aaa}=\frac14\gamma^\mu\).
		\\
		Shorthand notation
		&
		\(\widetilde\Delta_s^{(1)}\equiv\widetilde\Delta_s\), \quad
		\(\Delta_s^{(1)}\equiv\Delta_s\), \quad
		\(\widetilde\Delta_d^{(1)}=\Delta_d^{(1)}\equiv\Delta_d\), \quad
		\(\mathcal D\equiv D-2\).
		\\
		Spectral function
		&
		\(\Delta_d(P)=2\pi\,\mathrm{sgn}(p^0)\delta(P^2)
		=
		\dfrac{\pi}{p}\left[\delta(p-p^0)-\delta(p+p^0)\right]\).
		\\
		Distribution functions
		&
		\(n_F(p^0)=\left(e^{p^0/T}+1\right)^{-1}\), \quad
		\(n_B(p^0)=\left(e^{p^0/T}-1\right)^{-1}\), with $T$ being temperature.
		\\
		Momentum-reversal identities 
		&
		\(\Delta_A(P)=\Delta_R(-P)\), \quad
		\(\Delta_d(P)=-\Delta_d(-P)\), \quad
		\(N_B(P)=-N_B(-P)\), \quad
		\(N_F(P)=-N_F(-P)\).\\
		\bottomrule
	\end{tabular}
\end{table}
\subsection{QED Conventions}
\begin{figure}[t]
	\centering
	\includegraphics[width=\textwidth]{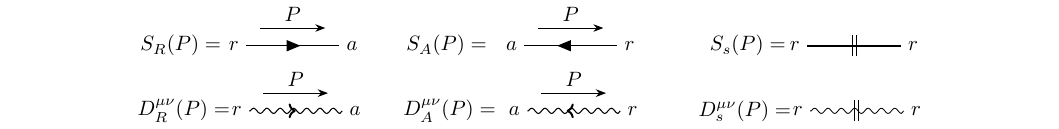}
	\caption{Free propagators in the \(r/a\) representation. Lorentz indices on
		gauge-boson lines are suppressed.}
	\label{Free propagators}
	\hrule
\end{figure}

We use the following QED renormalization convention
\begin{align}
	\psi_0=Z_2^{1/2}\psi,\qquad
	A_0^\mu=Z_3^{1/2}A^\mu,\qquad
	\xi_0=Z_3\xi,\qquad
	e_0=\bar\Lambda^\epsilon Z_1 Z_2^{-1} Z_3^{-1/2}e .
\end{align}
Using \(Z_i=1+\delta_i\), one writes the bare Lagrangian, expressed in terms of
renormalized fields and parameters, as
\begin{align}
	\mathcal L
	&=
	-\frac{1+\delta_3}{4}F_{\mu\nu}F^{\mu\nu}
	-\frac{1}{2\xi}(\partial_\mu A^\mu)^2
	+(1+\delta_2)\bar\psi i\slashed{\partial}\psi
	-e\,\bar\Lambda^\epsilon(1+\delta_1)
	\bar\psi\gamma^\mu\psi A_\mu,
\end{align}
with the local counterterm insertions
\begin{align}\label{Local-counterterms-insertions}
	-i\mathcal C_\psi(P)
	= i\delta_2\slashed P,\quad
	-i\mathcal C_A^{\mu\nu}(P)
	=- i\delta_3\left(P^2g^{\mu\nu}-P^\mu P^\nu\right),\quad
	-ie\mathcal C_{\bar\psi A\psi}^{\mu}
	= -ie\,\bar\Lambda^\epsilon\delta_1\gamma^\mu.
\end{align}
Here, the WTI gives \(Z_1=Z_2\) or $\delta_1=\delta_2$.
For one massless fermion flavor, the one-loop zero-temperature
counterterms in DR are
\begin{align}\label{Local-counterterms}
	\delta^{(1)}_2
	= -\frac{e^2}{16\pi^2}\frac{\xi}{\epsilon},
	\quad
	\delta^{(1)}_3
	= -\frac{e^2}{12\pi^2}\frac{1}{\epsilon},
\end{align}
where the superscript \((1)\) denotes the one-loop correction and $\epsilon\rightarrow 0$. Additionally, the dressed counterterm propagators are
\begin{align}\label{ra-ct}
	S_{j,{\rm CT}}(P)
	= i\delta_2 P^2\slashed P\,\Delta_j^{(2)}(P),\qquad
	D_{j,{\rm CT}}^{\mu\nu}(P)
	= -i\delta_3
	\left(P^2g^{\mu\nu}-P^\mu P^\nu\right)
	\widetilde\Delta_j^{(2)}(P),
	\qquad j=R, A,s.
\end{align} 
Figure~\ref{Free propagators} shows the free \(r/a\) propagators, where Lorentz
indices on gauge-boson lines are suppressed.
Throughout the appendices, a cross on an internal fermion or photon line denotes the insertion of these propagators. 
The ordinary QED Feynman rules \cite{Peskin:1995ev} are used for each $r/a$ assignment. Each Feynman diagram is replaced by the sum over its allowed $r/a$ contributions.

We denote virtual corrections by
\(\Sigma^{(l;r)}\) and \(\Gamma^{\mu\nu\cdots(l;r)}\). Lorentz, temporal, and
spatial indices are written before the parentheses, while \((l;r)\) labels the
loop order and the order in the soft-momentum expansion, with
\(l\in\{1,2\}\) and \(r\in\{\mathrm{LP},\mathrm{NLP},\ldots\}\). Before the soft expansion is performed, the power
label is omitted. Subscripts distinguish gauge sectors, statistical
sectors, and topologies when needed.
We use the following gauge sector conventions: the purely metric sector, corresponding to the \(g^{\mu\nu}\) part of the gauge-boson propagators, is denoted by the subscript \(m\); sectors proportional to a single factor of \((1-\xi)\) are denoted by the subscript \(\xi\); and the sector proportional to \((1-\xi)^2\) is denoted by the subscript \(\xi^2\).

\subsection{r/a constructions}
	In the \(r/a\) representation, retarded and advanced propagators encode causal
	propagation, while symmetric propagators carry the statistical factors
	\cite{Carrington_2007_Real,Carrington,Gorda_2023,Kaapo}.
For fixed external \(r/a\) indices, the amplitude is obtained by summing
over all allowed \(r/a\) assignments. Here, we first present the general
counting and then consider amplitudes with a single \(a\)-index. A
diagrammatic construction is then given only for convenience.
	
	The corresponding $r/a$-index counting and graph properties are derived in
	App.~\ref{ra-graph-rules}, while we summarize the main points here.
	For a connected amputated graph with
	\(N_v\) vertices and \(L\) loops, direct \(r/a\)-index counting gives
	\begin{equation}
		N_{rr}
		=
		L-1+E_a
		-\sum_{i=1}^{N_v}\bigl(k_{v_i}-1\bigr),\quad  N_{\mathrm C}=
		N_v-E_a+
		\sum_{i=1}^{N_v}\bigl(k_{v_i}-1\bigr).
		\label{eq:main-ra-count}
	\end{equation}
	where \(E_a\) is the number of external \(a\)-indices, \(k_{v_i}\)
	is the number of \(a\)-indices carried by vertex \(v_i\), $N_{rr}$ is the number of symmetric propagators, and $N_C$ is the number of causal propagators.

For cubic QED, where three fields interact with each other,
	Eq.~\eqref{eq:main-ra-count} becomes
	\begin{equation}
		N_{rr}=L-1+E_a-2N_{aaa},\quad N_C=N_v-E_a+2N_{aaa},
		\label{eq:main-qed-count}
	\end{equation}
	where $N_{aaa}$ is the number of three $a$-indexed vertices.
	Equation~\eqref{eq:main-qed-count} therefore provides a direct map
	between the external \(r/a\) index structure and the internal
	statistical--causal decomposition of any connected amputated cubic-QED
	diagram.
	For amplitudes with one external $a$-index, $E_a=1$, this becomes
	\begin{equation}
		N_{rr}
		=
		L-2N_{aaa}.
		\label{eq:single-a-count}
	\end{equation}
	At one loop, non-negativity of \(N_{rr}\) yields \(N_{aaa}=0\), so
	one symmetric propagator contributes. At two loops, the counting
	permits \(N_{aaa}=0\) or \(1\), corresponding to two or zero symmetric
	propagators, respectively. The latter assignment contains no statistical propagator and
	therefore is the temperature-independent sector, excluded from this study. At three
	loops, the counting permits \(N_{aaa}=0\) or \(1\), corresponding to
	three or one symmetric propagators, consistent with
	the Keldysh counting of Ref.~\cite{CaronHuot:2010zt}.

We now provide a diagrammatic construction for the sector in which every vertex carries exactly one $a$-index, $i.e.,$ $E_a=1$ and $N_{aaa}=0$, yielding
	\begin{equation}
		N_{rr}=L,
		\qquad
		N_{\mathrm C}=N_v-1.
		\label{eq:main-tree-count}
	\end{equation}
	For an \(L\)-loop diagram, we first choose \(L\) internal lines to carry
	symmetric propagators. These lines are chosen so that, once separated from the original diagram,
	the remaining graph has no loops and is tree-like. The remaining internal lines are then assigned retarded/advanced propagators, with a unique causal orientation.
	The \(r/a\) indices are assigned according to the \(N\)-point rules
	\cite{Carrington_2007,Ghiglieri_2020}. In our case, the required rules are
	given in Eqs.~\eqref{3} and \eqref{5} and Tab.~\ref{tab:all_conventions}.
	The complete thermal
	contribution is obtained by summing over all such choices of symmetric
	propagators and the retarded/advanced tree completions. We refer to this as the cutting procedure.
	This sum over allowed assignments is required by the \(r/a\) real-time
	formalism.

At arbitrary loop order, the construction above applies only
to the \(E_a=1\), \(N_{aaa}=0\) sector of \(N\)-point functions.  A representative assignment is shown in Sec.~\ref{Intermediate self-energy}.
	In imaginary time, Gaudin's method
	\cite{Blaizot_2006,Espinosa_2005,Mottola_2010} is the analog of the analysis above.
\section{Fermion Self-Energy}\label{fermion self-energy}
We begin this section with the symmetry constraints on the nonperturbative fermion spectral function and self-energy, with derivations given in App.~\ref{momentum reversal properties}. Chiral and parity invariance restrict the spectral function to
\begin{equation}
	\rho_F(Q)=a(Q)\,\slashed{Q}+b(Q)\,\slashed{u},
\end{equation}
where $a(Q)$ and $b(Q)$ are real scalar functions of the Lorentz invariants $Q^2$ and $u\cdot Q$. Furthermore,  $u^\mu=(1,\mathbf{0})$ denotes the four-velocity of the plasma in its rest frame.
Their reality follows from the pseudo-Hermiticity relation of the spectral function.
Charge conjugation further enforces the evenness of the spectral function under momentum reversal: $\rho_F(-Q)=\rho_F(Q)$.
Together with the spectral representation of the fermion propagator,
\begin{align}
	S_{R/A}(Q)= i\int_{-\infty}^\infty\frac{d\omega}{2\pi}\,
\frac{\rho_F(\omega,\mathbf{q})}{Q_0-\omega\pm i\eta},
\end{align}
this implies $S_A(-Q)=-S_R(Q)$ and hence
$\Sigma_A(-Q)=-\Sigma_R(Q)$.
The difference between the advanced and retarded cases lies in the explicit $\pm i\eta$ prescriptions, while their common principal-value part is the same.
At the integrand level, we suppress these prescriptions before integration and compare the algebraic part.
In this algebraic sense, the relation above becomes
\begin{align}\label{algebraic}
	\Sigma_{\rm alg}(-Q)=-\Sigma_{\rm alg}(Q),
\end{align}
which gives the oddness of the self-energy up to the retarded/advanced prescriptions.

In the hard sector, we seek to capture contributions from internal loop momenta
of order \(P\sim T\) (hard momenta), where the distribution functions peak.
For soft external momentum, \(Q\ll T\), the hard-region integrand is expanded in \(Q/T\).
Throughout, we refer to this expansion as the
	``soft expansion''. It is implemented by rescaling \(Q\to aQ\), performing
a Laurent expansion in the bookkeeping parameter \(a\), and then taking the
limit \(a\to1\)~\cite{Kaapo,Gorda_2023}.
Following Refs.~\cite{Kaapo,Carrington}, we first perform the soft expansion at the algebraic level and restore the retarded prescription ($q^0\to q^0+i\eta$) for the external momentum in the subsequent integrations, which yields the retarded fermion self-energy.
Expanded in the soft external momentum, Eq.~\eqref{algebraic} becomes
\begin{align}
	\Sigma_{\rm alg}(aQ)=\sum_r a^r \Sigma^{(r)}(Q),
\end{align}
which implies
\begin{align}
	\big[(-1)^r+1\big]\Sigma^{(r)}(Q)=0.
\end{align}
Thus, this relation gives a selection rule that forbids even powers in the soft expansion of the self-energy. 
For the self-energy, LP and NLP correspond to the $a^{-1}$ and $a^{1}$ terms in the soft expansion, respectively. 

In the plasma rest frame, the retarded self-energy takes the form
\begin{equation}\label{self-energy rest frame}
	\Sigma(Q)=	\Sigma^{0}(Q)\,\gamma^0+\Sigma^{i}(Q)\,\gamma^i,
\end{equation}
where we have dropped the retarded subscript $R$. The temporal and spatial components are then
\begin{equation}\label{projected self-energy}
	\Sigma^{0}(Q)\equiv\frac{1}{4}\,\mathrm{Tr}\!\left[\gamma^0\Sigma(Q)\right],
	\qquad
	\Sigma^{i}(Q)\equiv-\frac{1}{4}\,\mathrm{Tr}\!\left[\gamma^i\Sigma(Q)\right].
\end{equation}
\subsection{One-Loop Results}
The one-loop results, at both LP and NLP, are collected in App.~\ref{one-loop calculations}. Here, we present the final results. At one loop, the self-energy contains both metric and longitudinal contributions and takes the form
\begin{align}\label{Eq: LO self-energy general}
	\Sigma^{(1)}(Q)=\Sigma_{m}^{(1)}+\Sigma_{\xi}^{(1)},
\end{align}
with $m$ and $\xi$ denoting the metric and longitudinal sectors.
The retarded self-energy at LP is~\cite{Weldon, Bellac:2011kqa}
\begin{equation}\label{14}
	\Sigma^{(1;\rm LP)}(Q)\equiv	\mathcal{H}(T)=m^2(T)\left[\gamma^0
	L(Q)+\frac{\hat{\mathbf q}.\boldsymbol{\gamma}}{q}\bigl[1-q^0 L(Q)\bigr]
	\right],
\end{equation}
where
\begin{equation}
	L(Q)\equiv \frac{1}{2q}\ln\!\left(\frac{q^0+q+i\eta}{q^0-q+i\eta}\right)
	=\frac{1}{q}\mathcal{Q}_0\!\left(\frac{q^0}{q}\right).
\end{equation}
Here, the gauge-dependent sector does not contribute, in agreement with earlier results in the literature \cite{Carignano_2018,Mottola_2010}.
 Additionally, $\mathcal{Q}_0(q^0/q)$ denotes the Legendre function of the second kind, and $m^2(T)=\frac{e^2T^2}{8}$ is the (QED) fermion thermal mass \cite{Bellac:2011kqa}.
This LP expression is consistent with the momentum-reversal property:
under $Q\to -Q$, one has the integrated relations:
$\Sigma^{(1;LP)}_R(-Q)=-\Sigma^{(1;LP)}_A(Q)$.

There are two integration formulas frequently used in this study \cite{Gorda_2023,Kaapo}:
\begin{subequations}\label{a3}
	\begin{align}
		&\int_{-\infty}^{+\infty}\frac{d{p^0}}{2\pi} \Delta^{(n)}_d(P)f({p^0},p)=\frac{(-i)^{n+1}}{(n-1)!}\int_{-\infty}^{+\infty}\frac{d{p^0}}{2\pi}\Delta_d(P)\,\Bigl\{2{p^0} \Bigl[\frac{-1}{2{p^0}}\frac{\partial}{\partial {p^0}}\Bigr]^{n-1}\Bigl[\frac{-f({p^0,p})}{2{p^0}}\Bigr]\Bigr\},\label{a3-difference}\\
		&\int_P \Delta_d(P)\, f({p^0},\vec{p})=\int_{\vec{p} } \frac{f(p,\vec{p})-f(-p,\vec{p})}{2p}.\label{a3-spectral}
	\end{align}
\end{subequations}
The general higher-order difference propagator can equivalently be treated
through the residue representation given in Ref.~\cite{Gorda_2023}.
Evaluating those higher-order residues explicitly gives the differential
form in Eq.~\eqref{a3-difference}. For \(n=2\), this reduces to the form
used in Ref.~\cite{Kaapo}. Thus, Eq.~\eqref{a3-difference} is simply an
explicit representation of the same residue structure.
Eq.~\eqref{a3-spectral} follows from the spectral form of
\(\Delta_d(P)\) and vanishes when \(f(p^0,\vec p)\) is even in \(p^0\).
Moreover, the spectral function \(\Delta_d(P)\) puts the loop momentum on shell,
\(p^0=\pm p\), so \(P^2=0\). Higher difference propagators are first
reduced to \(\Delta_d(P)\) using Eq.~\eqref{a3-difference}, after which the
on-shell condition is imposed.

After the HTL expansion, the relevant one-loop integrands have the schematic form
\begin{align}
	\int_P \Delta_d(P) N_{B/F}(P)\,F(Q,P).
\end{align}
Based on Eq.~\eqref{a3-spectral}, the contribution above survives only if
	\(N_{B/F}(P)F(Q,P)\) is odd under \(P\to -P\). Since \(N_{B/F}(P)\) is odd,  the factor $F(Q,P)$ must be even under $P\to -P$ to yield a nonzero contribution. This integrand symmetry is compatible with the oddness selection rule: it retains only the odd-in-$Q$ terms. 
The same argument applies to higher difference propagators after using Eq.~\eqref{a3-difference}, since this reduction preserves the
	\(P\to-P\) parity of the factor multiplying the spectral function.

Since the spectral function sets $P^2=0$, we first use the reduction of the difference propagator and then set this on-shell condition. Subsequently,  we perform the soft expansion.
The NLP corrections are obtained by expanding to order $a^1$. The NLP corrections take the form
 \begin{subequations}\label{Eq: NLP final results}
  \begin{align}
 	&\Sigma^{(1;\rm NLP)}_{m}
 	=\frac{e^2\, \slashed{Q}}{16\pi^2}\Bigl\{-\frac{1}{\epsilon}-2\ln\left(\frac{e^{\gamma_E}\bar{\Lambda}}{2\pi T}\right)+1-2q^0L(Q)\Bigr\},
  	\label{Eq:NLP Physical}
  	\\
 	&\Sigma^{(1;\rm NLP)}_{\xi}=(1-\xi)\frac{e^2}{16\pi^2}\Bigl\{\slashed{Q}\Bigl[\frac{1}{\epsilon}+2\ln\left(\frac{e^{\gamma_E}\bar{\Lambda}}{2\pi T}\right)-1+2q^0 L(Q)\Bigr]+\frac{Q^2}{q^2}\mathbf{q}.\boldsymbol{\gamma}\Bigl[1-q^0 L(Q)\Bigr]\Bigr\}	\label{Eq:NLP xi}.
 \end{align}
 \end{subequations}
The divergences appearing in the equations above are ultraviolet (UV) in origin. To see this explicitly, we include the one-loop local counterterm, Eqs.~\eqref{Local-counterterms-insertions}--\eqref{Local-counterterms}, which yields
\begin{align}\label{renormalized-NLP}
	\Sigma^{(1;\rm NLP)}_{\rm Ren}(Q)&=\Sigma^{(1;\rm NLP)}_{m}+\Sigma^{(1;\rm NLP)}_{\xi}-\delta_2^{(1)}\slashed{Q}\no\\
	&=\frac{e^2\, }{16\pi^2}\Bigl\{-\xi\, \slashed{Q}\Bigl[2\ln\left(\frac{e^{\gamma_E}\bar{\Lambda}}{2\pi T}\right)-1+2q^0L(Q)\Bigr]+(1-\xi)\frac{Q^2}{q^2}\mathbf{q}.\boldsymbol{\gamma}\Bigl[1-q^0 L(Q)\Bigr]\Bigr\}.
\end{align}
The singularities cancel out, yielding a finite result.
The second square bracket cancels in Feynman gauge ($\xi=1$), whereas the first square bracket cancels in Landau gauge ($\xi=0$).
Our result agrees with Ref.~\cite{Carignano_2018}.
For \(q^0>q\), the functions above are real, so the one-loop NLP gives no damping contribution.
\subsection{Two-loop organization before the HTL expansion}\label{Intermediate self-energy}
\begin{figure}
	\centering
	\includegraphics[width=\textwidth]{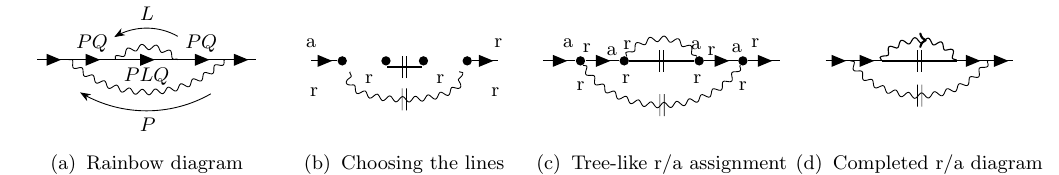}
	\caption{The cutting-open procedure for one contribution to the rainbow self-energy.}
	\label{Rainbow-diagram}
	\phantomsubcaption\label{Rainbow-original}
	\phantomsubcaption\label{choosing-symmetric lines}
	\phantomsubcaption\label{tree-like graph}
	\phantomsubcaption\label{Completed r/a diagram}
	\hrule
\end{figure}
At two loops, three distinct topologies contribute to the fermion self-energy: the rainbow, cross-photon, and bubble diagrams, as shown in Figs.~\ref{Rainbow-diagram} and \ref{topologies of fermion self-energy}. At this stage, we focus on the rainbow diagram to illustrate the $r/a$ assignment using the method explained in Sec.~\ref {Real-time formalism}. We choose two lines to be symmetric propagators such that, after separating them from the main diagram, the remaining graph is connected and tree-like (Fig.~\ref{choosing-symmetric lines}). We then assign the remaining $r/a$ labels to determine the causality flow (Figs.~\ref{tree-like graph}--\ref{Completed r/a diagram}). Momenta are omitted in the $r/a$ assignment figures since the corresponding Feynman diagram already fixes them. The full set of $r/a$ diagrams is obtained by summing over all possible choices of symmetric propagators and completing the resulting tree-like graphs with $r/a$ labels, as represented in Fig.~\ref{rainbow self-energy assignments}. The remaining Feynman diagrams are treated in the same way.

Using Eq.~\eqref{5} and the propagator conventions of Fig.~\ref{Free propagators}, this $r/a$ assignment gives
\begin{align}\label{represetive}
	-i\Sigma^{{\rm Rep}}_{{R}}=-ie^4 \int_P\int_L 
	&\Delta_s(PLQ)\Delta_R^{(2)}(PQ)
	\Bigl\{
	\mathcal{S}_{R_{11}}\,
	\tilde{\Delta}_s(P)\Delta_A(L)
	-(1-\xi)^2
\mathcal{S}_{R_{22}}\,
	\tilde{\Delta}_s^{(2)}(P)\Delta_A^{(2)}(L)
	\nonumber\\
	&\hspace{2.8cm}
	+i(1-\xi)
	\Bigl[
\mathcal{S}_{R_{12}}\tilde{\Delta}_s(P)\Delta_A^{(2)}(L)
	+\mathcal{S}_{R_{21}}\tilde{\Delta}_s^{(2)}(P)\Delta_A(L)
	\Bigr]
	\Bigr\},
\end{align}
where
\begin{align}\label{Dirac rainbow}
	&\mathcal{S}_{R_{11}}\equiv 
	\gamma^\alpha(\slashed{P}+\slashed{Q})
	\gamma^\nu(\slashed{P}+\slashed{L}+\slashed{Q})
	\gamma_\nu(\slashed{P}+\slashed{Q})
	\gamma_\alpha ,\quad
\mathcal{S}_{R_{22}}\equiv 
	\slashed{P}(\slashed{P}+\slashed{Q})
	\slashed{L}(\slashed{P}+\slashed{L}+\slashed{Q})
	\slashed{L}(\slashed{P}+\slashed{Q})
	\slashed{P},
	\nonumber\\
	&\mathcal{S}_{R_{12}}\equiv 
	\gamma^\alpha(\slashed{P}+\slashed{Q})
	\slashed{L}(\slashed{P}+\slashed{L}+\slashed{Q})
	\slashed{L}(\slashed{P}+\slashed{Q})
	\gamma_\alpha ,\quad
\mathcal{S}_{R_{21}}\equiv 
	\slashed{P}(\slashed{P}+\slashed{Q})
	\gamma^\nu(\slashed{P}+\slashed{L}+\slashed{Q})
	\gamma_\nu(\slashed{P}+\slashed{Q})
	\slashed{P}.
\end{align}
Here, $R$ denotes the rainbow topology, and the compact momentum notation introduced in Tab.~\ref{tab:all_conventions} is used.
Three gauge sectors contribute in Eq.~\eqref{represetive}: the purely metric sector, the mixed metric-longitudinal sector proportional to $(1-\xi)$, and the purely longitudinal sector proportional to $(1-\xi)^2$.
The metric sector originates from the $g^{\mu\nu}$ part of the gauge-boson propagators;
the mixed metric--longitudinal sector originates from the product of
$g^{\mu\nu}$ and the $(1-\xi)$-dependent longitudinal structure.
In the mixed sector, there are two inequivalent metric-longitudinal contractions, leading to the two Dirac structures $\mathcal{S}_{R_{12}}$ and $\mathcal{S}_{R_{21}}$.
The two-loop contributions can then be decomposed in gauge sectors, as 
\begin{align}
\Sigma^{(2)}\equiv	\Sigma^{(2)}_{m}+\Sigma_\xi^{(2)}+\Sigma_{\xi^2}^{(2)}.
\end{align}
The same applies to the fermion--photon vertex. After the soft expansions through LP, the corresponding label ``LP'' is added.

The symmetric propagators in the curly brackets of Eq.~\eqref{represetive} contain spectral and difference functions, namely $\Delta_d(P)$ and $\Delta_d^{(2)}(P)$; the latter can be reduced to a spectral function  using  Eq.~\eqref{a3-difference}.
 The spectral function then sets the corresponding momentum on shell, $P^2=0$, as in Eq.~\eqref{a3-spectral}. For this representative assignment, the other internal momentum can be put on shell, $L^2=0$, by shifting $L\to L-P-Q$, so that the other symmetric propagator depends only on $L$.
At two loops, however, the computation is not that simple. Products such as $\Delta_R(PQ)\Delta_A(PQ)$ give rise to pinch structures in some topologies. These pinch contributions are removed by using the Kubo--Martin--Schwinger (KMS) relation
\begin{align}\label{KMS relation}
	N_F(P_1)N_F(P_2) + N_F(P_2)N_B(P_3) + N_B(P_3)N_F(P_1) + \frac{1}{4} = 0,
\end{align}
with the momenta satisfying $P_1+P_2+P_3=0$. This relation is sufficient for the purpose of this study, while its generalization is provided in Ref.~\cite{Carrington_2007_Real}.
 Thus, before the HTL expansion, we perform two reorganizations. First, we rewrite the full set of $r/a$ contributions in a pinch-free form. Second, we shift the loop momenta so that each spectral or difference propagator depends only on one internal momentum, either $P$ or $L$. 
This sequence allows us to impose the on-shell condition before the soft expansion. The detailed implementation of these steps is deferred to App.~\ref{two-loop self-energy computations}; see also Ref.~\cite{Kaapo} for a more pedagogical discussion. We refer to these pinch-free expressions, in which the spectral and difference propagators depend solely on a single internal momentum, as the pre-HTL expressions.

From Fig.~\ref{Rainbow-original}, the different possible choices of symmetric propagators correspond to choosing two photon lines, one photon line and one fermion line, or two fermion lines. At two loops, the corresponding statistical factors therefore fall into the three independent products
\begin{align}\label{main statistical sectors}
	N_F(L)N_B(P),\quad
	N_B(L)N_B(P),\quad
	N_F(L)N_F(P).
\end{align}
Combining this statistical decomposition with the gauge-sector
decomposition $g\in\{m,\xi,\xi^2\}$, the two-loop self-energy is organized as
\begin{equation}
	\Sigma^{(2)}(Q)
	=
	\sum_{g\in\{m,\xi,\xi^2\}}
	\sum_{s\in\{FB,BB,FF\}}
	\Sigma^{(2)}_{g,s}(Q).
	\label{eq:gauge-stat-decomposition}
\end{equation}
The same gauge--statistical organization is used for the
fermion--photon vertex. After the soft expansion, the corresponding power label, such as LP, is added. Derivatives of the statistical distributions generated by difference propagators are not counted as separate sectors; for example, $N_F' N_F$ is included in the $N_F N_F$ sector.

Readers interested only in the physical results and the sector-by-sector
consistency relations can safely skip the detailed pre-HTL construction
of the self-energy and fermion--photon vertex given in
Apps.~\ref{two-loop self-energy computations} and
\ref{intermediate vertex}.
\subsection{Two-Loop Results}
We apply the soft expansion to the pinch-free expressions collected in
Eqs.~\eqref{rainbow compact} and \eqref{self-energy compact}. The LP metric
and mixed metric--longitudinal contributions are given in
Eqs.~\eqref{self-energy results} and \eqref{self-energy results xi},
respectively. These integrand-level results provide the self-energy side of
the WTI and allow the identity to be checked separately in each gauge sector
and for each independent statistical sector.
The metric self-energy, in addition to the sector decomposition given in Eq.~\eqref{eq:gauge-stat-decomposition}, separates into terms proportional to
	\(\mathcal D^{2}\), \(\mathcal D\), and \(\mathcal D^{0}=1\), schematically,
	\begin{equation}
		\Sigma_{m}^{(2;\mathrm{LP})}
		=
		\mathcal D^{2}(\cdots)
		+
		\mathcal D(\cdots)
		+
		\mathcal D^{0}(\cdots).
\end{equation}
Each of these three sectors satisfies the WTI independently.
At this LP order, the fully longitudinal $(1-\xi)^2$ sector vanishes by
integrand symmetry, in analogy with the corresponding LP cancellation at one
loop. The relevant terms are odd under $P\to -P$ and/or $L\to -L$ and
therefore integrate to zero. Consequently, the three fully longitudinal
statistical sectors vanish at LP, while the six non-vanishing sectors are
verified explicitly: three metric sectors and three mixed metric--longitudinal
sectors.
At NLP, however, the fully longitudinal sector no longer vanishes and
contributes to the self-energy. A representative example, together with its
Ward--Takahashi verification, is provided in the SM.

The self-energy integrands and their reductions to the DR integrals
defined in App.~\ref{DR} are given in
App.~\ref{self-energy computations two-loop}.
The metric sector contains collinear singularities. Up to overall phases,
the relevant terms have the form
\begin{equation}
	\begin{aligned}
		\sim\int_P\int_L \Delta_d(L)\Delta_d(P)\,
		\Bigl\{&
		- N_B(P)\bigl[N_B(L)-N_F(L)\bigr]
		\frac{\mathcal{D}^2\slashed{P}}
		{(L\!\cdot P)(L\!\cdot Q)}
		\\
		&+
		N_F(L)\bigl[N_B(P)-N_F(P)\bigr]
		\Bigl[
		\frac{8\slashed{P}\,(L\!\cdot Q)}
		{(L\!\cdot P)(P\!\cdot Q)^2}
		+
		\frac{(\mathcal{D}^2+8)\slashed{L}}
		{(L\!\cdot P)(P\!\cdot Q)}
		\Bigr]
		\Bigr\}.
	\end{aligned}
\end{equation}
Here, the collinear singularity comes from the relative angle between the
two hard momenta,
\begin{align*}
	\frac{1}{L\!\cdot P}
	=
	\frac{1}{pl}\frac{1}{1-z},
	\qquad
	z\equiv \hat{\mathbf p}\!\cdot\!\hat{\mathbf l},
\end{align*}
where the collinear region corresponds to $z\to1$.
Since this divergence comes from the angular integration, an overlap with
radial divergences can in principle occur. However, the relevant
statistical combinations contain differences such as
$N_B(P)-N_F(P)$ and $N_B(L)-N_F(L)$. 
Although the bosonic and fermionic
parts separately contain radial divergences, their divergent parts are
the same and therefore cancel in these differences, leading to a finite result. 
Schematically,
	\begin{align}
		\mathcal{R}_4=\int_p \frac{1}{p^3}N_F(p)
		= \mathcal R_3+\ln 2\Big[1+\epsilon\big(2\ln(e^{\gamma_E}/2\pi T)+\ln 2\big)\Big],\quad  \mathcal R_3 =\int_p \frac{1}{p^3}N_B(p),
	\end{align}
	where the relation between $\mathcal{R}_3$ and $\mathcal{R}_4$ holds only through $O(\epsilon)$ truncation, sufficient for our purpose. The radial differences are finite and proportional to $\ln(2)$.
	Thus, the collinear pole is proportional to $\ln(2)/\epsilon$.
	As a result, no radial--collinear double divergence remains.

After performing these integrations, one obtains
\begin{subequations}\label{self-energy-integrated}
	\begin{align}
		\Sigma_m^{(2;\rm LP)}(Q)
		&=
		\frac{e^2 \mathcal{H}(T)}{2\pi^2}
		\Bigl[
		\mathcal{K}-\frac{7}{3}
		+\frac{1}{2}\ln\Bigl(\!-\frac{q^2}{Q^2}\Bigr)\bigl[1-\frac{7}{3}\ln(2)\bigr]
		-\frac{7\mathcal{K}}{3}\ln(2)
		+\frac{17}{6}\ln^2(2)\Bigr]\no
		\\
		&\quad
		+\frac{e^4T^2}{16\pi^2}
		\Bigl\{
		\gamma^0
		\Bigl[
		q_0L(Q)^2
		+
		\frac{q_0}{Q^2}\bigl[
		-\frac{1}{6}+\frac{2}{3}\ln(2)\bigr]
		+
		\frac{\Delta_2(Q)}{4q}
		\bigl[1-\frac{7}{3}\ln(2)\bigr]
		+
		L(Q)
		\bigl[\frac{5}{6}+\frac{7}{3}\ln(2)\bigr]
		\Bigr]\no
		\\
		&\hspace{1.1cm}
		+
		\frac{\boldsymbol{\gamma}.\hat{\mathbf q}}{q}
		\Bigl[
		2
		-\frac{1}{2}\ln\Bigl(\!-\frac{q^2}{Q^2}\Bigr)
		+
		\ln(2)\bigl[
		-\frac{10}{3}
		+\frac{7}{6}\ln\Bigl(\!-\frac{q^2}{Q^2}\Bigr)\bigr]
		+
		\frac{7}{3}\ln^2(2)+
		\frac{q^2}{Q^2}\bigl[\frac{1}{6}-\frac{2}{3}\ln(2)\bigr]\no
		\\
		&\hspace{2.1cm}-\frac{q_0\Delta_2(Q)}{4q}\bigl[1-\frac{7}{3}\ln(2)\bigr]-q_0^2L(Q)^2\Bigr]\Bigr\},
	\end{align}
	\begin{align}
		\Sigma_\xi^{(2;\rm LP)}(Q)
		&=
		-\frac{e^2\mathcal{H}(T)}{8\pi^2}(1-\xi)\Bigl[
			\mathcal{K}-\frac{7}{2}
		+\frac{2}{3}\ln(2)
		+\frac{1}{2}\ln\Bigl(\!-\frac{q^2}{Q^2}\Bigr)\Bigr]\no
		\\
		&\quad
		-\frac{e^4T^2}{64\pi^2}(1-\xi)
		\Bigl\{
		\gamma^0
		\Bigl[
		\frac{q_0}{q^2}
		+
		\frac{\Delta_2(Q)}{4q}
		+
		q_0
		\bigl[1+\frac{q_0^2}{q^2}\bigr]
		L(Q)^2
		-
		\frac{2q_0\ln(2)}{Q^2}
		-
		\frac{2Q^2}{q^2}L(Q)
		\Bigr]\no
		\\
		&\hspace{2.1cm}
		+
		\frac{\boldsymbol{\gamma}.\hat{\mathbf q}}{q}
		\Bigl[
		3
		-\ln(2)
		-\frac{1}{2}\ln\Bigl(\!-\frac{q^2}{Q^2}\Bigr)
		+
		\frac{2q^2\ln(2)}{Q^2}
		-
		\frac{q_0\Delta_2(Q)}{4q}
		-
		2q_0^2L(Q)^2
		\Bigr]
		\Bigr\},
	\end{align}
\end{subequations}
where
\begin{align}
	\Delta_2(Q)\equiv\text{Li}_2\bigl(\frac{q^0+q}{{q^0}-q}\bigr)-\text{Li}_2\bigl(\frac{q^0-q}{{q^0}+q}\bigr),\quad \mathcal{K}\equiv -\gamma_E+\frac{1}{2\epsilon}+12\ln(A)+2\ln\!\Bigl(\frac{e^{\gamma_E}\bar{\Lambda}}{2\pi T}\Bigr),
\end{align} 
where \(A\) is the Glaisher--Kinkelin constant, and
\(\mathcal H(T)\) denotes the one-loop LP self-energy in
Eq.~\eqref{14}. All singular terms in the contributions above are
proportional to \(\mathcal H(T)\) and are collected in \(\mathcal K\).
The $\frac{1}{\epsilon}\ln(2)$ term arises from the collinear divergence multiplying the finite radial difference $\mathcal{R}_4-\mathcal{R}_3\sim \ln(2)$.

Let us now evaluate the imaginary part of the corrections above on the
timelike branch: $q^0>q$. With the retarded prescription
\(q^0\to q^0+i0^+\), the relevant functions have
\begin{align}
\Im\Bigl[ L(Q)\Bigr]=0,\quad
\Im\Bigl[\Delta_2(Q)\Bigr]=-2\pi q\,L(Q),\quad
	\Im\Bigl[\ln\!\left(-\frac{q^2}{Q^2}\right)\Bigr]=\pi .
\end{align}
Using these relations, the imaginary parts of the metric and mixed
metric--longitudinal sectors cancel separately:
\begin{align}
\Im\Sigma^{(2;{\rm LP})}_{m}(Q)
	=
	\Im\Sigma^{(2;{\rm LP})}_{\xi}(Q)
	=0,
	\qquad q^0>q .
\end{align}
Thus, on the timelike branch, the hard two-loop LP corrections computed here
do not generate a finite damping contribution.  This behavior is consistent with the corresponding timelike hard corrections to the photon polarization tensor \cite{Kaapo}.

At the same parametric order, zero-temperature counterterm insertions must
also be included to remove the UV subdivergences.
These insertions with the one-loop self-energies are detailed in Eq.~\eqref{mainCT-Self}. The final results therefore become
\begin{align}\label{self-summing-cts}
	\Sigma^{(1;LP)}_{{\rm CT}}(Q)
	=
	\left(\delta_2^{(1)}-\delta_3^{(1)}\right)\mathcal H(T)=\Bigl(-\frac{e^2}{16\pi^2}\frac{\xi}{\epsilon}+ \frac{e^2}{12\pi^2}\frac{1}{\epsilon}\Bigr)\mathcal{H}(T).
\end{align}
The renormalized hard-region contribution is therefore
\begin{align}\label{renormalized-two-loop-LP}
	\Sigma_{{\rm ren}}^{(2;{\rm LP})}(Q)
	=
	\Sigma^{(2;{\rm LP})}(Q)
	+
	\Sigma_{{\rm CT}}^{(1;{\rm LP})}(Q).
\end{align}
	The divergent part of the ordinary two-loop self-energy in
	Eq.~\eqref{self-energy-integrated} is
	\begin{align}
		\Sigma_{\mathrm{Div}}^{(2;\mathrm{LP})}(Q)
		=
		\frac{e^2}{48\pi^2\epsilon}
		\bigl(9+3\xi-28\ln 2\bigr)\mathcal{H}(T).
	\end{align}
	Adding the counterterm contribution in Eq.~\eqref{self-summing-cts}
	then gives
\begin{align}\label{divergences in hard sector}
	\Sigma_{\rm Div}(Q)=\frac{e^2}{48\pi^2\epsilon}\Bigl(13-28\ln(2)\Bigr)\mathcal{H}(T).
\end{align}
The \(\xi\)-dependent divergence is removed by the zero-temperature counterterm
insertion. After this subtraction, the remaining singularities are
gauge-independent collinear and hard-region infrared (IR) divergences. The hard-region IR divergence
refers to the radial endpoint singularity of the hard expansion, where the
expanded hard integrals overlap with softer momentum regions.
 In the following sections, we explain how these singular structures enter the matching at full parametric order
\(e^3T\), assuming that the external momentum is of order $Q\sim eT$. The final physical requirement is that, after all contributions at
the same order are included, observables contain no uncancelled singularities
or gauge-dependent structures. 
\section{Two-Fermion–N-Photon Vertex}\label{N-Fermion--Two-Photon vertex}
This section summarizes the symmetry constraints on the
amputated two-fermion–$N$-photon vertex kernel. Detailed
derivations are provided in Appendix~\ref{momentum reversal properties}.
Chiral invariance restricts the most general Dirac structure of the
amputated two-fermion--$N$-photon kernel, with fermion momenta
$Q_{1,2}$, to
\begin{align}\label{general results of n kernel}
	&\Gamma_{c}^{(N)}(Q_2,Q_1;\{K_i\})
	=
	V_{c}^{(N)}(Q_2,Q_1;\{K_i\})\cdot\gamma
	+
	A_{c}^{(N)}(Q_2,Q_1;\{K_i\})\cdot\gamma\,\gamma_5,\quad
	c=(c_0,c_1,\ldots,c_N,c_{N+1}),\quad c_i\in \{r,a\},
\end{align}
where all photon momenta are taken incoming. 
Here, the subscript c denotes the external $r/a$ index string, and the superscript $(N)$ serves as shorthand for the Lorentz indices.
The $N-1$ independent
photon momenta are denoted by
$\{K_i\}\equiv\{K_1,\ldots,K_{N-1}\}$, while $K_N$ is fixed by
momentum conservation, $
Q_2=Q_1+\sum_{i=1}^{N}K_i$.
For the $c$-number tensors $V_c^{(N)}$ and $A_c^{(N)}$, ``$\cdot\gamma$'' indicates contraction with the corresponding Dirac matrices. The explicit index form is
given in App.~\ref{momentum reversal properties}.

The combination of parity $P$ and time reversal $\mathcal{T}$ invariance yields
$P\mathcal{T}$
\footnote{We use $\mathcal{T}$ for time reversal to avoid confusion
	with temperature and transpose.}
constraints
\begin{align}\label{reality-constraints}
	V_{c}^{(N)}(Q_2,Q_1;\{K_i\})
	&=
	\Big[
	V_{c}^{{\rm rev},(N)}(Q_2,Q_1;\{K_i\})
	\Big]^*,
	\nonumber\\
	A_{c}^{(N)}(Q_2,Q_1;\{K_i\})
	&=
	-
	\Big[
	A_{c}^{{\rm rev},(N)}(Q_2,Q_1;\{K_i\})
	\Big]^*,
\end{align}
Here, ``rev'' denotes the reversed real-time process and is
distinct from complex conjugation. The \({\rm rev}\) reverses the causal prescriptions while
leaving the external \(r/a\) string fixed.  Consequently,
\(\Delta_{ra}=\Delta_R\) and \(\Delta_{ar}=\Delta_A\) are interchanged,
whereas the symmetric \(rr\) propagators carry no causal orientation
and remain unchanged:
\begin{align}
	{\rm rev}: \qquad
	\Delta_R \leftrightarrow \Delta_A,\quad
	\Delta_s,\widetilde{\Delta}_s\to
	\Delta_s,\widetilde{\Delta}_s,
	\quad
	c \to c .
\end{align}
For instance, suppose 
\begin{align}
	\Gamma_{arra\cdots}\sim \bigl[\Delta_A \Delta_A \Delta_R\cdots\bigr] \bigl[\Delta_s \tilde{\Delta}_s\cdots\bigr].
\end{align}
Its $\rm rev$ construction then becomes
\begin{align}
	\Gamma^{\rm rev}_{arra\cdots}\sim \bigl[\Delta_R \Delta_R \Delta_A\cdots\bigr] \bigl[\Delta_s \tilde{\Delta}_s\cdots\bigr].
\end{align}

When the retarded/advanced causal prescriptions are suppressed,
Eq.~\eqref{reality-constraints} reduces at the algebraic level to
\begin{align}
	V_{c,\mathrm{alg}}^{(N)}
=
	\Big[V_{c,\mathrm{alg}}^{(N)}\Big]^*,\quad
	A_{c,\mathrm{alg}}^{(N)}=
	-\Big[A_{c,\mathrm{alg}}^{(N)}\Big]^* .
\end{align}
This indicates that the vector coefficients and axial coefficients become real and purely imaginary, respectively.

Hermiticity constrains the amplitudes as
\begin{align}\label{eq:HermSummary}
	V_{c}^{(N)}(Q_2,Q_1;\{K_i\})
	&=
	(-1)^{E_a(c)+1}
	\Big[
	V_{\tilde c}^{(N)}
	(Q_1,Q_2;\{-K_i\})
	\Big]^*,\quad  
	\nonumber\\
	A_{c}^{(N)}(Q_2,Q_1;\{K_i\})
	&=
	(-1)^{E_a(c)+1}
	\Big[
	A_{\tilde c}^{(N)}
	(Q_1,Q_2;\{-K_i\})
	\Big]^*,\quad E_a(c)\equiv \#\{i:c_i=a\},
\end{align}
where $E_a(c)$ is the number of external $a$ indices of the amplitude. Here, $\tilde{c}$ denotes the exchange of the two fermion endpoint indices: 
\begin{align}
	\tilde{c}\equiv
	(c_{N+1},c_1,\ldots,c_N,c_0).
\end{align}
The combination of $P\mathcal{T}$ and Hermiticity yields the fermion-leg exchange constraints
\begin{align}\label{eq:exchangeSummaryGeneral}
	V_{c}^{(N)}(Q_2,Q_1;\{K_i\})
	&=
	(-1)^{E_a(c)+1}\,
	V_{\tilde c}^{{\rm rev},(N)}
	(Q_1,Q_2;\{-K_i\}),
	\nonumber\\
	A_{c}^{(N)}(Q_2,Q_1;\{K_i\})
	&=
	-(-1)^{E_a(c)+1}\,
	A_{\tilde c}^{{\rm rev},(N)}
	(Q_1,Q_2;\{-K_i\}).
\end{align}

Charge conjugation $C$ further gives
\begin{align}\label{eq:CsummaryGeneral}
	V_{c}^{(N)}(Q_2,Q_1;\{K_i\})
	&=
	(-1)^{N+1}
	V_{\tilde c}^{(N)}
	(-Q_1,-Q_2;\{K_i\}),
	\nonumber\\
	A_{c}^{(N)}(Q_2,Q_1;\{K_i\})
	&=
	(-1)^N
	A_{\tilde c}^{(N)}
	(-Q_1,-Q_2;\{K_i\}).
\end{align}
Since charge conjugation does not reverse the real-time evolution,
the same relations hold separately within the reversed process.
Combining Eqs.~\eqref{eq:exchangeSummaryGeneral} and
\eqref{eq:CsummaryGeneral}, and using
$\tilde{\tilde c}=c$, one obtains
\begin{align}\label{combined-exC-general}
	X_{c}^{(N)}(Q_2,Q_1;\{K_i\})
	=
	(-1)^{N+E_a(c)}
	X_{c}^{{\rm rev},(N)}
	(-Q_2,-Q_1;\{-K_i\}),
	\qquad
	X\in\{V,A\}.
\end{align}
\subsection{Selection Rules In the Soft Expansion}
Since ${\rm rev}$ changes only the retarded/advanced causal
prescriptions, suppressing these prescriptions gives
\[
X_{c,\mathrm{alg}}^{{\rm rev},(N)}
=
X_{c,\mathrm{alg}}^{(N)} .
\]
Thus, Eq.~\eqref{combined-exC-general} reduces at the algebraic level to
\begin{align}\label{algebraic-results}
	X_{c,\mathrm{alg}}^{(N)}
(Q_2,Q_1;\{K_i\})
=
(-1)^{N+E_a(c)}
X_{c,\mathrm{alg}}^{(N)}
(-Q_2,-Q_1;\{-K_i\}) .
\end{align}
Under the  soft rescaling $Q_{1,2}\to aQ_{1,2}$ and $\{K_i\}\to\{aK_i\}$,
one writes
\begin{align}
	&X_{c,{\rm alg}}^{(N)}
	(aQ_2,aQ_1;\{aK_i\})
	=
	\sum_{r\in\mathbb Z}
	a^r
	X_{c,r}^{(N)}(Q_2,Q_1;\{K_i\}),\no\\
	&X_{c,\mathrm{alg}}^{(N)}
	(-aQ_2,-aQ_1;\{-aK_i\})
	=
	\sum_{r\in\mathbb{Z}}
	(-1)^r a^r
	X_{c,r}^{(N)}(Q_2,Q_1;\{K_i\}).
\end{align}
Using Eq.~\eqref{algebraic-results} and matching equal powers of $a$
gives
\begin{align}
	\left[
	1-(-1)^{N+E_a(c)+r}
	\right]
	X_{c,r}^{(N)}(Q_2,Q_1;\{K_i\})
	=0.
\end{align}
Thus, for an arbitrary fixed $c$, a contribution at order $a^r$ can
be nonzero only if
\begin{align}\label{general-selection-rule}
	r+N+E_a(c)
	\quad\text{is even}.
\end{align}

For amplitudes with a single $a$ index, this reduces to 
\begin{align}\label{selection-rule-single-a}
	r+N+1
	\quad\text{is even}.
\end{align}
Thus, for odd $N$ and single $a$, only even soft powers can be nonzero, while for even
$N$, only odd soft powers can be nonzero. For example, the
fermion--photon vertex, $N=1$, can receive contributions from even
powers of the soft expansion, $\{a^{-2},a^0,\ldots\}$. For $N=0$,
the condition reduces to $r+1$ even, so only odd powers contribute,
reproducing the self-energy selection rule.

As an example, consider the retarded two-fermion--two-photon vertex, with a single external $a$ index: $\Gamma^{\mu\nu}(Q_2,Q_1;K_1)$, where $K_1$ is an independent incoming photon momentum. The one-loop LP HTL contribution is~\cite{Bellac:2011kqa} 
\begin{align}
	\Gamma^{\mu\nu}(Q_2,Q_1;K_1) &= - m^2(T) \int\frac{d\Omega_v}{4\pi} \frac{{v}^\mu\,{v}^\nu\, \slashed{v}}{\big[(Q_1+K_1)\!\cdot\! v\big]\big[(Q_2-K_1)\!\cdot\! v\big]} \Bigl[\frac{1}{Q_1\!\cdot\!v}+\frac{1}{Q_2\!\cdot\!v}\Bigr],
\end{align}
where $v^\mu=(1,\hat{\mathbf v})$, $d\Omega_{\mathbf v}$ denotes
	the solid-angle element associated with the direction
	$\hat{\mathbf v}$, and the angular measure is normalized to unity: $\int \frac{d\Omega_v}{4\pi}=1$.
	After suppressing the retarded/advanced prescriptions, this
	vector-like result satisfies the algebraic forms of reality,
	fermion-leg-exchange, and charge conjugation constraints, respectively,  in Eqs.~\eqref{reality-constraints}, \eqref{eq:exchangeSummaryGeneral}, and \eqref{eq:CsummaryGeneral}.

Note that Levi--Civita tensor structures are allowed in the axial coefficients.
For a fermion--photon vertex, this implies
\begin{equation}\label{Levi-Civita-structure}
	\epsilon^{\mu\nu\rho\sigma} Q_{1\rho} Q_{2\sigma}\,\gamma_\nu\gamma_5
	= -i\Big(
	\gamma^\mu\,\slashed Q_1\,\slashed Q_2
	- Q_1^\mu\,\slashed Q_2
	+ Q_2^\mu\,\slashed Q_1
	- (Q_1\!\cdot\!Q_2)\,\gamma^\mu
	\Big).
\end{equation}
In vacuum, such axial contributions in on-shell matrix elements $\bar u(Q_2)\Gamma_\mu(Q_2, Q_1)u(Q_1)$ can be reduced to vector structures by using the on-shell Dirac equations (see e.g.\ Ch.~10 of Ref.~\cite{Weinberg:1995mt}) \[ \slashed Q\,u(Q)=m\,u(Q),\qquad \bar u(Q)\slashed Q=m\,\bar u(Q), \] so that for $m=0$ the right-hand sides vanish. Equivalently, the renormalized on-shell external states satisfy the free Dirac equation (up to an overall wavefunction renormalization), which enforces this on-shell reduction. At finite temperature, however, quasiparticle dispersion relations and spinors are modified \cite{PhysRevD.28.340, PhysRevD.35.4020, PhysRevD.35.1861, Donoghue:1984zz}, so this reduction need not hold. We therefore keep the full off-shell decomposition to avoid missing contributions relevant for subsequent analysis.
\section{Fermion--Photon Vertex}\label{vertex}
We use the vertex component $\Gamma^\mu_{arr}(Q_2,Q_1)$, where the three $r/a$ indices are ordered, from left to right, as the incoming fermion with momentum $Q_1$, the photon with momentum $K=Q_2-Q_1$, and the outgoing fermion with momentum $Q_2$.
Within our conventions, the WTI reads
\cite{Carrington_1998_Ward}
\begin{align}\label{WT identity}
	K_\mu \Gamma_{arr}^\mu(Q_2,Q_1)=\Sigma_{ar}(Q_1)-\Sigma_{ar}(Q_2),
\end{align}
with the causal prescription imposed at the end,
\(q^0_{1,2}\to q^0_{1,2}+i\eta_{1,2}\).
Other fermion--photon vertices, with exactly one external \(a\) index, differ by the corresponding retarded/advanced assignments of the external legs
~\cite{Fueki_2002,Carrington_1998_Ward,Hou_1997,Hou_1999}.
Components with multiple external \(a\) indices, such as
	\(\Gamma^{aaa}\), have distinct real-time statistical and causal
	structures~\cite{CaronHuot_2009} and are outside the scope of this study.
Here, we drop the $r/a$ indices: \(\Gamma^\mu\equiv \Gamma^\mu_{arr}\).
Chiral invariance restricts the vertex correction to 
\begin{align}\label{vertex decomposition}
	\Gamma^\mu(Q_2,Q_1)=	V^{\mu\nu}\gamma_\nu+ A^{\mu\nu}\gamma_{\nu}\gamma_5,
\end{align}
with the coefficients given by
\begin{align}\label{Eq:vertex projections}
	V^{\mu \nu}=
	\frac{1}{4}\,
	\mathrm{Tr}\!\left[\gamma^\nu\, \Gamma^\mu(Q_2,Q_1)\right],\quad
	A^{\mu \nu}=
	-\frac{1}{4}\,
	\mathrm{Tr}\!\left[\gamma^\nu\gamma^5\, \Gamma^\mu(Q_2,Q_1)\right].
\end{align}
The temporal and spatial components are read directly from these covariant expressions. The symmetry constraints follow from the \(N=1\) case of the
general constraints discussed above.
Contracting Eq.~\eqref{Eq:vertex projections} with the photon momentum gives
\begin{align}
	K_\mu A^{\mu\nu}
	&\propto
	\mathrm{Tr}\!\left[\gamma^\nu\gamma^5\gamma^\rho\right]
	=0,
\end{align}
where Eqs.~\eqref{Eq:vertex projections}, \eqref{WT identity} and \eqref{self-energy rest frame} are used.
Thus, $A^{\mu\nu}$ is transverse to the photon momentum. The WTI fixes the longitudinal projection of the vector coefficient, $K_\mu V^{\mu\nu}$, via the self-energy difference. Below, we compute the one- and two-loop contributions to the fermion--photon vertex and explicitly verify these constraints.
\subsection{One-Loop LP and NLP Corrections}
 \begin{figure}
	\centering
	\includegraphics[width=\textwidth]{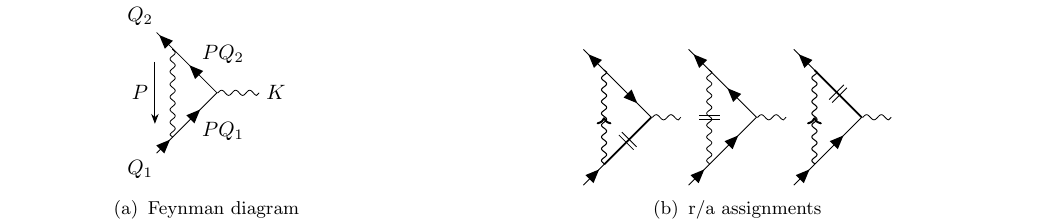}
	\caption{Feynman diagram and r/a assignments of one-loop fermion--photon vertex.}\label{one-loop fermion--photon vertex}
	\phantomsubcaption\label{one-loop three vertex}
	\phantomsubcaption\label{ra one-loop three vertex}
	\hrule
\end{figure}
Before proceeding, we mention that
Tab.~\ref{tab:ward_identity_verification} summarizes all 
one- and two-loop contributions computed in this study, for both the
self-energy and the fermion--photon vertex. As mentioned previously, the complete
results are also provided in machine-readable form in the SM \cite{SupplementalMaterial}.
	The one-loop fermion--photon diagram is shown in Fig.~\ref{one-loop fermion--photon vertex}. After applying loop-momentum shifts, the vertex contributions become
\begin{subequations}\label{Eq:one-loop vertex}
	\begin{align}
		&-ie\Gamma^{\mu(1)}_{m}(Q_2,Q_1)=-ie^3\int_P
		\Delta_d(P)\Bigl[N_F(P)\Delta_A(P-Q_1)\Delta_A(P+Q_2-Q_1)\mathcal{V}_{{{m}(a_v)}}+N_B(P)\Delta_{R}(PQ_1)\Delta_{R}(PQ_2)\mathcal{V}_{m}\no\\
		&\hspace{5.5cm}+N_F(P)\Delta_{R}(P+Q_1-Q_2)\Delta_A(P-Q_2)\mathcal{V}_{m{(m_v)}}\Bigr],\\
		&-ie\Gamma_\xi^{\mu(1)}(Q_2,Q_1)=e^3\int_P(1-\xi)\Bigl[\Delta_d(P)N_F(P)\Delta_{A}^{(2)}(P-Q_1)\Delta_A(P+Q_2-Q_1)\mathcal{V}_{{{\xi}}{(a_v)}}\no\\
		&\hspace{5.1cm}+\Delta_d^{(2)}(P)N_B(P)\Delta_{R}(PQ_1)\Delta_{R}(PQ_2)\mathcal{V}_{\xi}\no\\
		&\hspace{5.1cm}+\Delta_d(P)N_F(P)\Delta_{R}(P+Q_1-Q_2)\Delta_{A}^{(2)}(P-Q_2)\mathcal{V}_{{\xi}{(m_v)}}\Big],
	\end{align}
\end{subequations}
where
\begin{align}\label{Eq:one-loopDirac}
	\mathcal{V}_{{m}}\equiv\gamma^\nu (\slashed{P}+\slashed{Q}_1) \gamma
	^\mu (\slashed{P}+\slashed{Q}_2)\gamma_\nu,\qquad 	\mathcal{V}_{\xi}\equiv\slashed{P} (\slashed{P}+\slashed{Q}_1) \gamma
	^\mu (\slashed{P}+\slashed{Q}_2)\slashed{P}.
\end{align}
Here, the labels $a_v:P\to P-Q_1$ and $m_v:P\to P-Q_2$ indicate that the corresponding shift is applied to the Dirac numerator.
To perform the power expansion in the soft external momenta, we use
\begin{align}
	K =Q_2-Q_1 ,
	\qquad
	Q_1\to aQ_1,\quad Q_2\to aQ_2,\quad K\to aK ,
\end{align}
and rewrite all denominators containing \(Q_2-Q_1\) in terms of \(K\), together
with the corresponding numerator factors. For the metric
Fermi--Dirac sector, this amounts to the replacement
\begin{align}
	N_F(P)\Bigl\{\Delta_A(P-Q_1)\Delta_A(P+K)\left[\gamma^\nu \slashed{P} \gamma^\mu (\slashed{P}+\slashed{K})\gamma_\nu\right]+\Delta_{R}(P-K)\Delta_A(P-Q_2)\left[\gamma^\nu (\slashed{P}-\slashed K) \gamma^\mu \slashed{P}\gamma_\nu\right]\Bigr\}.
\end{align}
The leading non-vanishing contribution scales as \(a^{-2}\) and lies in the
vector coefficient, while the axial coefficient vanishes by integrand
symmetry. Hence, the LP result becomes
\begin{align}\label{LP-Vertex}
	-ie\,\Gamma^{\mu(1;\rm LP)}(Q_2,Q_1)
	&=
	-\frac{ie^3\mathcal{D}}{4}
	\int_P \Delta_d(P)\,
	\bigl[N_B(P)-N_F(P)\bigr]\,
	\frac{2P^\mu\slashed P}{(P\cdot Q_1)(P\cdot Q_2)},
\end{align}
which satisfies the WTI; see Tab.~\ref{tab:ward_identity_verification}.
We use the same prescription in the remaining cases: denominators involving a single external momentum are left unchanged, whereas those involving linear combinations of external momenta are rewritten in terms of the momentum transfer, with the corresponding numerator factors transformed in the same way; see App.~\ref{intermediate vertex} for a two-loop example\footnote{This reorganization rewrites soft combinations that may appear in different forms, such as $P\cdot Q_1-P\cdot Q_2$ or $P\cdot(Q_1-Q_2)$, in terms of $\pm P\cdot K$ consistently. This makes the power expansion transparent.}.

The NLP corrections arise at order $a^0$. At this order, both vector and axial coefficients receive contributions from both metric and longitudinal sectors, as given in Eqs.~\eqref{metric-NLP} and \eqref{longitudinal-NLP}. 
The different gauge and statistical sectors satisfy these constraints independently:
\begin{align}
	K_\mu \Gamma^{\mu(1;\rm NLP)}_{g,s}(Q_2,Q_1)
	&=
	\Sigma^{(1;\rm  NLP)}_{g,s}(Q_1)
	-
	\Sigma^{(1;\rm  NLP)}_{g,s}(Q_2),
	\quad
	g\in\{m,\xi\},\quad
	s\in\{N_B(P),N_F(P)\},
\end{align}
with
\begin{align}
	K_\mu A^{\mu\nu(1;\rm NLP)}_{g,s}(Q_2,Q_1)=0 .
\end{align}
The SM provides the Mathematica expressions together with the corresponding WTI and axial-transversality checks \cite{SupplementalMaterial}.
Notably, the axial coefficients carry an explicit imaginary factor, consistent with their symmetry properties. Using Eq.~\eqref{Levi-Civita-structure}, however, the corresponding vertex contributions can be rewritten in terms of Dirac-vector structures and products of Dirac slashes. In this representation, the explicit imaginary factor is absorbed, and the rewritten contributions contain no explicit overall imaginary factor. The renormalized vertex is also written as
\begin{align}
	\Gamma^{\mu(1;NLP)}_{Ren}=\sum_{g,s}\Gamma_{g,s}^{\mu(1;NLP)}+\bar{\Lambda}^\epsilon \delta_1 \gamma^\mu.
\end{align}
\begin{table}[h]
	\centering
	\begin{tabular}{p{0.32\textwidth}p{0.14\textwidth}p{0.12\textwidth}p{0.21\textwidth}}
		\toprule
			Order & Gauge Sector & Self-Energy &Fermion--Photon Vertex\\
		\midrule
		One-loop Pre-HTL&m, $\xi$&\eqref{basic-self-energy}&\eqref{Eq:one-loop vertex}\\
		Two-loop Pre-HTL&m, $\xi$, $\xi^2$&\eqref{rainbow compact}--\eqref{self-energy compact}&\eqref{vertex compact}\\
		One-loop Counterterms Pre-HTL&m, $\xi$&\eqref{mainCT-Self}&\eqref{Vertex-CT-one-loop}\\
		\midrule
	One-loop LP& m & \eqref{A2}--\eqref{A3} & \eqref{LP-Vertex}\\
	One-loop NLP & m, $\xi$ & \eqref{Eq:NLP self-energy} & \eqref{metric-NLP}--\eqref{longitudinal-NLP} \\
	Two-loop LP&m, $\xi$ &\eqref{HTL-self-energy-results}& \eqref{NBNB-metricsector-vector-vertex}--\eqref{NBNB-metricsector-axial-vertex}, SM \cite{SupplementalMaterial}\\	
	\midrule
	One-loop Counterterms LP& m, $\xi$& \eqref{LP-CT-Self}&\eqref{Vertex-CT-one-loop-LP}\\
		\bottomrule
	\end{tabular}
	\caption{Pre-HTL results are pinch-free, with spectral and higher-difference
		functions depending only on a single internal momentum, allowing the
		on-shell conditions to be imposed before the soft expansion.
		Integrand-level results are labeled by their power order, LP or NLP.}
	\label{tab:ward_identity_verification}
\end{table}
\subsection{Two-Loop Corrections}\label{vertex two-loop section}
At two-loop order, seven distinct topologies contribute to the fermion--photon vertex; see Fig.~\ref{vertex diagrams at two-loop order}. The pinch-free pre-HTL expressions are collected in Eq.~\eqref{vertex compact}, using the notation of Tab.~\ref{tab:conventions}. In these expressions, the spectral and higher-difference functions depend only on the internal momenta $P$ or $L$, allowing the on-shell conditions to be imposed before the soft expansion. We then apply the soft expansion introduced above.

The non-planar and self-energy-corrected topologies occur in two orientations, denoted type~I and type~II. At the level of the full amputated vertex at zero temperature, the type-II contribution is obtained from type~I by the exchange $Q_1\leftrightarrow Q_2$ \cite{Bonciani_2004(7),Bonciani_2004(5)}.
However, in real-time formalism and the $r/a$ basis, the fermion-leg exchange given in Eq.~\eqref{eq:exchangeSummaryGeneral} determines the behavior of amplitudes. 
	For $c=arr$, Eq.~\eqref{eq:exchangeSummaryGeneral} gives $\tilde c=rra$ and
	\begin{align}
		V^{\mu\nu}_{arr}(Q_2,Q_1)
		&=
		V^{{\rm rev},\mu\nu}_{rra}(Q_1,Q_2), \nonumber\\
		A^{\mu\nu}_{arr}(Q_2,Q_1)
		&=
		-
		A^{{\rm rev},\mu\nu}_{rra}(Q_1,Q_2),
	\end{align}
	with $K\to-K$ under $Q_1\leftrightarrow Q_2$.
	For the prescription-suppressed algebraic soft expansion used here,
	the retarded/advanced causal assignments are suppressed. For the
	type-I/type-II pairs considered below, the exchanged and directly
	constructed expressions have the same momentum routing and symmetric-propagator
	assignment, differing only in the retarded/advanced causal labels.
	Therefore, the type-I to type-II map is implemented algebraically by
	\begin{align}
		Q_1\leftrightarrow Q_2,\qquad K\to-K,
	\end{align}
	with the corresponding additional minus sign in the axial sector. This approach yields the same algebraic integrand as obtained by directly determining the $r/a$ construction. 

At two loops, in contrast to the one-loop LP vertex, the axial coefficient contributes already at LP. Moreover, the mixed metric--longitudinal sector, proportional to $(1-\xi)$, also contributes, in parallel with the self-energy. The fully longitudinal sector, proportional to $(1-\xi)^2$, vanishes at LP by integrand symmetry, but appears at NLP and obeys the  WTI and axial-transversality constraints.
In the metric sector for the $N_B(L)N_B(P)$ statistical sector,
the LP results are presented in Eqs.~\eqref{NBNB-metricsector-vector-vertex}--\eqref{NBNB-metricsector-axial-vertex}. 
Because of the antisymmetry of the Levi--Civita tensor, the photon-momentum contraction of the axial Dirac structure, $K_\mu A^{\mu\rho}\gamma_\rho\gamma_5$, vanishes identically for this configuration. In contrast, the vector Dirac structure reproduces the corresponding self-energy difference.
The remaining sectors are collected in the SM \cite{SupplementalMaterial}, and the identities are also checked sector by sector. In the bookkeeping used here, LP terms scale as $a^{-2}$, NLP terms scale as $a^0$, and all odd powers vanish, as required by the selection rule for the fermion--photon vertex. 
Tab.~\ref{tab:ward_identity_verification} summarizes the explicit self-energy and fermion--photon vertex relations verified in this study.
As an additional check, we extended the algorithmic expansion to $\mathcal{O}(a^{10})$ for the vector sector. At this order, the expansion initially generates $\sim 4.9\times10^7$ terms, which reduce to roughly $\mathcal{O}(10^5)$ terms after using integrand symmetries and contracting with $K_\mu$. The WTI is preserved throughout this reduction: after contraction, all terms separate into $Q_1$-only and $Q_2$-only structures, and these isolated structures reproduce the corresponding self-energy difference.

Each two-loop vertex sector satisfies
\begin{align}
	K_\mu\Gamma_{g,s}^{\mu,(2;r)}(Q_2,Q_1)
	&=
	\Sigma^{(2;r)}_{g,s}(Q_1)
	-
	\Sigma^{(2;r)}_{g,s}(Q_2),\quad	K_\mu A^{\mu\nu,(2;r)}_{g,s}(Q_2,Q_1)=0,
	\quad
	s\in\{N_BN_B,\,N_FN_F,\,N_FN_B\},
\end{align}
with $g\in\{m,\xi\}$ at LP and $g\in\{m,\xi,\xi^2\}$ at NLP and higher orders, since the fully longitudinal $(1-\xi)^2$ sector does not contribute at LP. 
This implies that each distinct sector satisfies the corresponding constraint independently. 

The SM \cite{SupplementalMaterial} associated with the two-loop vertex contains the complete LP vector and axial coefficients, and verifies the WTI and axial transversality sector by sector. It also demonstrates the preservation of the WTI through $\mathcal{O}(a^{10})$ for the $N_B(P)N_F(L)$ sector in the $(1-\xi)$ gauge sector.\footnote{To illustrate the WTI constraint at higher orders, we focus on the $N_B(P)N_F(L)$ sector, where no $L\leftrightarrow P$ symmetry is present, and the isolation into $Q_1$-only and $Q_2$-only structures is most transparent.} 
For the $N_B(P)N_F(L)$ sector, it explicitly shows that the fully longitudinal $(1-\xi)^2$ sector vanishes at LP but contributes at NLP, while maintaining the WTI and axial transversality.
The remaining statistical sectors follow the same algorithmic procedure. Unlike $N_B(P)N_F(L)$, the $N_F(L)N_F(P)$ and $N_B(L)N_B(P)$ sectors are symmetric under $L\leftrightarrow P$. 
For instance,
their intermediate Dirac vector expressions can contain mixed $Q_1$ and $Q_2$ terms after contraction with $K_\mu$. However, these terms vanish by integrand symmetries, leaving a final result depending only on $Q_1$ or only on $Q_2$, as shown for the LP cases in the SM. 

The one-loop counterterm insertions are given in
Eq.~\eqref{Vertex-CT-one-loop}, and their LP contributions are collected in
Eq.~\eqref{Vertex-CT-one-loop-LP}. The renormalized LP vertex at two-loop order
is therefore
\begin{align}
	\Gamma^{\mu(2;LP)}_{\rm Ren}(Q_2,Q_1)
	=
\sum_{g,s}\Gamma^{\mu(2;\mathrm{LP})}_{g,s}(Q_2,Q_1)
	+
	\Gamma^{\mu(1;LP)}_{\rm CT}(Q_2,Q_1),
\end{align}
where the sum runs over the two-loop gauge and statistical sectors listed in the SM, and
\begin{align}\label{CT-vertex-result}
	\Gamma^{\mu(1;LP)}_{\rm CT}(Q_2,Q_1)
	=
	\Bigl(\delta_2^{(1)}-\delta_3^{(1)}\Bigr)
	\Gamma^{\mu(1;LP)}(Q_2,Q_1).
\end{align}
Comparison with Eq.~\eqref{self-summing-cts} shows that the counterterm
insertions in the self-energy and in the vertex satisfy the WTI independently
of the ordinary two-loop contributions:
\begin{align}
		K_\mu\Gamma^{\mu(1;LP)}_{\rm CT}(Q_2,Q_1)=	\Sigma^{(1;LP)}_{{\rm CT}}(Q_1)-\Sigma^{(1;LP)}_{{\rm CT}}(Q_2).
\end{align}
\section{Self-Energy Region Matching}\label{Matching analysis}
We now summarize the region analysis for the fermion self-energy at the
collective scale, $Q\sim eT$\footnote{The same region analysis applies to the vertex
	corrections.}. Schematically,
\begin{equation}
	\Sigma(Q)
	=
	\Sigma^{(eT)}(Q)
	+
	\Sigma^{(e^2T)}(Q)
	+
	\Sigma^{(e^3T)}(Q)
	+\cdots .
	\label{Self-Energy-Hierarchy}
\end{equation}
The first term is the one-loop LP hard contribution,
Eq.~\eqref{14}, which enters the LO pole equation. The
second term is the soft one-loop self-energy evaluated with LO HTL-resummed
propagators and vertices \cite{Carrington_2007}. This term already contains the leading tower with one soft loop and \(N\) LP
hard insertions, with \(N=0,1,2,\ldots\)~. 
At order \(e^3T\), the matched self-energy takes the form
\begin{align}
	\Sigma^{(e^3T)}(Q)
	=
	\Sigma_{\rm Ren}^{(1;\mathrm{NLP})}(Q)
	+
	\Sigma_{\rm Ren}^{(2;\mathrm{LP})}(Q)
	+
	\Sigma_{N_BN_B}^{*(2)}(Q)
	-
	\Sigma_{\rm overlap}^{(e^3T)}(Q).
	\label{eq:e3T_self_energy_structure}
\end{align}
The first two terms are the hard one-loop NLP and hard-hard two-loop LP
contributions, given in Eqs.~\eqref{renormalized-NLP} and
\eqref{renormalized-two-loop-LP}. The third term denotes the two-loop soft
self-energy in the \(N_BN_B\) sector, with the internal lines and vertices
replaced by their LO HTL-resummed counterparts. The LO HTL
fermion--photon vertex is \(O(1)\) at the soft scale, as follows from
Eq.~\eqref{LP-Vertex}. 
We denote the HTL-resummed symmetric photon propagator by
	$\widetilde D_s^*$, its spectral density by $\rho_B^*$, and the
	HTL-resummed fermion propagator by $S^*$.
For the representative topology with two symmetric
photon propagators and three retarded/advanced fermion propagators, the power
counting is
\begin{equation}
	d^4P\,d^4L\sim (eT)^8,
	\qquad
	N_B(P)\sim n_B(P)\sim \frac{T}{e T}\sim \frac1e,
\end{equation}
and therefore
\begin{equation}
	\widetilde D_s^*(P)
	\sim
	N_B(P)\rho_B^*(P)
	\sim
	\frac1e\,\frac1{(eT)^2}
	\sim
	\frac1{e^3T^2},
	\qquad
	S^*(P)\sim \frac1{eT},
\end{equation}
so the power counting becomes
\begin{equation}
	\Sigma_{N_BN_B}^{*(2)}(Q)
	\sim
	e^4
	(eT)^8
	\left(\frac1{e^3T^2}\right)^2
	\left(\frac1{eT}\right)^3
	\sim
	e^3T.
\end{equation}
The other soft statistical sectors are suppressed at this order.
The overlap term is the asymptotic part of the hard and soft
descriptions,
\begin{equation}
	\Sigma_{\rm overlap}^{(e^3T)}
	=
	\left(\Sigma_{\rm hard}\right)_{\rm soft\;limit}
	=
	\left(\Sigma_{\rm soft}\right)_{\rm hard\;limit}.
	\label{eq:overlap_definition}
\end{equation}
It is obtained by matching the IR limit of the hard-region result against the
UV limit of the soft-sector calculation. In the present work, the two-loop soft
self-energy and the corresponding overlap are not evaluated.
 Once these pieces are computed, the self-energy at order \(e^3T\) is complete within this matched expansion.
Since the soft sector and overlap subtraction are not evaluated here, the precise
cancellation channel of the hard-region singularities is not assigned at this stage.
Pole observables are obtained from the zeros of the retarded inverse
propagator. For a massless fermion self-energy, we write
\begin{equation}
	\Sigma_R(Q)
	=
	\Sigma_0(Q)\gamma^0
	+
	\Sigma_v(Q)\,\hat{\mathbf q}\cdot\boldsymbol{\gamma},
\end{equation}
where
\begin{equation}
	\Sigma_h(Q)=\Sigma_0(Q)+h \Sigma_v(Q),
	\qquad
	D_h(Q)=q^0-h q-\Sigma_h(Q),
	\qquad h=\pm .
\end{equation}
The leading pole is obtained by
\begin{equation}
	D_h^{\rm LO}(\omega_h^{\rm LO},q)=0,
	\qquad
	D_h^{\rm LO}(Q)
	=
	q^0-h q-\Sigma_h^{(eT)}(Q),
\end{equation}
where \(\Sigma_h^{(eT)}(Q)\) is the helicity projection of the one-loop LP
self-energy \cite{Bellac:2011kqa}. Expanding the self-energy and pole gives
\begin{equation}
	\Sigma_h
	=
	\Sigma_h^{(eT)}
	+
	\Sigma_h^{(e^2T)}
	+
	\Sigma_h^{(e^3T)}
	+\cdots ,
	\qquad
	\omega_h
	=
	\omega_h^{\rm LO}
	+
	\delta\omega_h^{(e^2T)}
	+
	\delta\omega_h^{(e^3T)}
	+\cdots ,
\end{equation}
and therefore the NLO and NNLO pole shifts are
\begin{subequations}
	\begin{align}
		&\delta\omega_h^{(e^2T)}
		=\frac{\Sigma_h^{(e^2T)}(\omega_h^{\rm LO},q)}{Z_h^{\rm LO}},
		\quad
	Z_h^{\rm LO}=1-\left.\partial_{q^0}\Sigma_h^{(eT)}(q^0,q)\right|_{q^0=\omega_h^{\rm LO}},
		\label{eq:first_pole_shift}\\
		&\delta\omega_h^{(e^3T)}
		=\frac{\Sigma_h^{(e^3T)}+\delta\omega_h^{(e^2T)}\partial_{q^0}\Sigma_h^{(e^2T)}
			+\frac12\left(\delta\omega_h^{(e^2T)}\right)^2\partial_{q^0}^2\Sigma_h^{(eT)}}{Z_h^{\rm LO}}\bigg|_{q^0=\omega_h^{\rm LO}}.
		\label{eq:second_pole_shift}
	\end{align}
\end{subequations}
Thus, the nonlinear expansion of the pole equation also generates iteration
terms at order \(e^3T\).
Similar collinear structures have been reported in the leading hard--soft two-loop self-energy~\cite{Mottola_2010}, which is already contained in the HTL-resummed one-loop soft self-energy. These structures may provide a natural source for the cancellation of the hard-region collinear divergence.
Until the full matching calculation is completed, the precise cancellation channel of the divergence cannot be identified.
 Moreover, gauge-dependent hard-region self-energy contributions must cancel in the complete same-order pole equation, as required by gauge consistency.

For a complex pole,
\begin{equation}
	\omega_h(q)=\Omega_h(q)-i\gamma_h(q),
\end{equation}
the damping rate is
\begin{equation}
	\gamma_h(q)
	=
	-\Im\,\delta\omega_h(q).
\end{equation}
The imaginary part of the one-loop LP hard self-energy vanishes on the timelike
branch, Eq.~\eqref{14}, so the first nonzero damping contribution comes from
the soft HTL-resummed sector \cite{Carrington_2007}. The hard virtual corrections computed here do not generate a finite damping contribution on this branch.
 Therefore, the two-loop soft \(N_BN_B\) sector, the corresponding overlap
 subtraction, and the pole-iteration terms are among the same-order
 ingredients required for the complete damping rate at NNLO.
\section{Conclusion}
\label{sec:conclusion}
We have derived the symmetry constraints of massless QED. Chiral invariance restricts the off-shell two-fermion--\(N\)-photon vertex to vector and axial-vector Dirac structures, each of which has definite behavior under momentum reversal and fermion-leg exchange. With the retarded/advanced prescriptions omitted, the vector structures
are tensors with real amplitudes, while the axial structures are
pseudotensors with purely imaginary amplitudes. The WTI constrains the vector structure of the fermion--photon vertex by relating it to the self-energies. On the other hand, the axial structure is transverse to the photon momentum. The self-energy is odd under momentum reversal up to retarded/advanced prescriptions, while the fermion spectral function is even. These symmetry constraints further lead to the selection rules for the soft expansions: at fixed power \(r\), a contribution can be nonzero only when \(r+N+E_a(c)\) is even, where $N=0$ corresponds to the fermion self-energy and $E_a(c)$ denotes the number of external ``$a$'' indices.

Using DR throughout, we have computed the renormalized one-loop NLP and two-loop LP contributions to the fermion self-energy and the fermion--photon vertex in a general covariant \(R_\xi\) gauge. The insertions of the zero-temperature counterterms remove the UV subdivergences.
Axial-vector structures are absent in the one-loop LP vertex, but appear beyond that order, including in the two-loop LP vertex. The results are organized by gauge sector, statistical sector, and soft power. Each sector at fixed soft power is checked to preserve the WTI and axial transversality, as required by gauge consistency. The counterterm insertions also obey the WTI independently. Notably, the fully longitudinal sectors vanish at two-loop LP, consistent with the vanishing of the one-loop LP longitudinal sector. As an additional algorithmic test, the vector sector was extended through \(O(a^{10})\), for which the same selection rule and WTI persist. The SM contains the machine-readable one-loop NLP and two-loop LP coefficients, along with the corresponding WTI and axial-transversality checks \cite{SupplementalMaterial}. In addition, for the fully longitudinal contribution in the \(N_B(P)N_F(L)\) statistical sector, the SM shows that all LP contributions vanish. In contrast, they appear at  NLP, where the vector and axial coefficients satisfy the WTI and transversality constraints, respectively.

For the self-energy, the hard contribution at one-loop NLP is gauge dependent and contains a local UV singularity, which is removed by the zero-temperature fermion counterterm. At two-loop LP, the metric and mixed metric--longitudinal sectors contribute. After the corresponding counterterm subtraction, the remaining divergence is gauge independent, \(\Sigma_{\mathrm{Div}}(Q)=(e^2/48\pi^2\epsilon)(13-28\ln(2))\mathcal{H}(T)\), where $\mathcal{H}(T)$ is the LO self-energy contribution (one-loop LP). The singularity corresponding to \(\ln(2)\) is collinear, while the remaining singularity is the hard-region IR  singularity. 
Since the soft sector and overlap subtraction are not evaluated here, the precise cancellation channel of these singularities is not assigned at this stage. The cancellation of the singularities can be checked in the pole equation, after the iteration terms in Eq.~\eqref{eq:second_pole_shift} are also included. 
Furthermore, the gauge-dependent contributions that appear in the hard-region self-energy must cancel out in the complete same-order pole equation.
On the timelike branch \(q^0>q\), although individual functions at two-loop LP contain imaginary parts, these contributions cancel separately in the metric and mixed metric--longitudinal sectors. Thus, the two-loop LP terms do not generate a finite damping contribution.
The one-loop NLP hard contribution is real on the timelike branch and therefore gives no damping contribution.

At the same order, a complete calculation of quasiparticle observables must combine the hard contributions with the HTL-resummed soft sector, the appropriate overlap subtraction, and the iteration terms. Inclusive rates also require the corresponding real-emission and absorption contributions according to the Kinoshita--Lee--Nauenberg (KLN) theorem~\cite{Kinoshita:1962ur, PhysRev.133.B1549}. The present work supplies the hard-region part of this program, along with its symmetry constraints, WTI, and axial-transversality checks. These hard-region amplitudes can also be used for the future construction of the fermionic NLO effective action.
\section*{Acknowledgments}
I am grateful to Kaapo Seppänen for helpful correspondence on hard-thermal-loop calculations, especially on higher-order difference propagators and hard-region expansions.
I am especially grateful to S. S. Gousheh for helpful discussions, steady encouragement, and support at difficult stages of this work.
\begin{appendices}
\begin{appendix}
\section{Dimensional Regularization}\label{DR}
			We define the on-shell unit vectors $v^\mu=(1,\hat{\mathbf p})$ and ${v'}^\mu=(1,\hat{\mathbf l})$ as well as \cite{Kaapo}
		\begin{equation}
			L(Q) \equiv \frac{1}{2q}\ln\!\left[\frac{q^0 + q + i\eta}{q^0 - q + i\eta}\right].
		\end{equation}
		The relevant radial integrals used in this work are
		\begin{align}\label{Rdial integral}
			\mathcal R_1 &\equiv \int_p \frac{1}{p}N_B(p)
			= \frac{\pi^2T^2}{6}\Big[1+\epsilon\big(24\ln A-2-2\gamma_E+2\ell_T\big)\Big]
			+\mathcal O(\epsilon^2),\no\\
			\mathcal R_2 &\equiv -\!\int_p \frac{1}{p}N_F(p)
			= \frac{1}{2}\mathcal R_1-\frac{\pi^2T^2}{6}\ln(2)\,\epsilon +\mathcal O(\epsilon^2),\no\\
			\mathcal R_3 &\equiv \int_p \frac{1}{p^3}N_B(p)
			= \frac{1}{4\epsilon}+\frac{1}{2}\ell_T
			+\frac{\epsilon}{8}\Big(\pi^2-4\gamma_E^{\,2}+4\ell_T^{\,2}-8\gamma_1\Big)
			+\mathcal O(\epsilon^2),\no\\
			\mathcal R_4 &\equiv \int_p \frac{1}{p^3}N_F(p)
			= \mathcal R_3+\ln 2\Big[1+\epsilon\big(2\ell_T+\ln 2\big)\Big]
			+\mathcal O(\epsilon^2),\no\\
			\mathcal R_5 &\equiv \int_p \frac{1}{p^2}\frac{dN_B(p)}{dp}
			=2\epsilon\,\mathcal R_3,\qquad
			\mathcal R_6 \equiv \int_p \frac{1}{p^2}\frac{dN_F(p)}{dp}
			=2\epsilon\,\mathcal R_4.
		\end{align}
Here \(\ell_T\equiv \ln(e^{\gamma_E}/2\pi T)\), while \(\gamma_1\) and \(A\) denote the first Stieltjes constant and the Glaisher--Kinkelin constant, respectively.
The relations for \(\mathcal R_2\) and \(\mathcal R_4\)  are valid only up to \(\mathcal O(\epsilon^2)\).
In addition, the last line follows from an integration-by-parts (IBP) identity\footnote{
	The IBP step for
	\(\mathcal{R}_5\equiv\int_p \frac{1}{p^2}\frac{dN_B(p)}{dp}\)
	must be applied with care because \(N_B(p)=\tfrac12+n_B(p)\sim T/p\) as \(p\to0\), so the boundary term \(\big[p^{d-3}N_B(p)\big]_{0}^{\infty}\) is ill-defined.
	To implement IBP within DR, we decompose
	\begin{equation*}
		N_B(p)=\frac{T}{p}+\widetilde N_B(p),
	\end{equation*}
	where \(\widetilde N_B(p)=\mathcal{O}(p)\) as \(p\to0\). This yields
	\begin{equation*}
		\int_0^\infty dp\,p^{d-3}\frac{d}{dp}\Big(\frac{T}{p}\Big)
		+\int_0^\infty dp\,p^{d-3}\widetilde N_B'(p).
	\end{equation*}
	The first term is a scaleless power integral and therefore vanishes, while the second term is well-defined and its boundary contribution also vanishes. Hence, IBP together with the scaleless-integral prescription implies \(\mathcal R_5=2\epsilon\,\mathcal R_3\).
}. 
The radial integrands also contain vacuum (\(T=0\)) contributions. Indeed, after performing the soft-momentum expansion, the vacuum part reduces to a scaleless power integral of the form 
\begin{equation}\label{Eq:scaleless}
	\Bigl(\text{$p$-independent factor}\Bigr)\times \int_{0}^{\infty} dp\, p^{\alpha},
\end{equation}
which is set to zero in DR. Therefore, all non-vanishing contributions arise from the finite-temperature part, which is not scaleless due to the presence of the distribution functions \(n_{B/F}\) (Tab.~\ref{tab:all_conventions}).

The angular integrals involving temporal components of $v^\mu$ are
\begin{equation}\label{Angular Integrals}
		\begin{split}
			\mathcal{A}_0 &\equiv \!\int_z 1
			= 2+4\epsilon [1 - \ln 2] + O(\epsilon^2),\\
			\mathcal{A}_1 &\equiv \!\int_z \frac{1}{v\!\cdot\! Q}
			= 2L(Q) +2\epsilon\!\left\{
			L(Q)\ln\!\left(-\frac{q^2}{Q^2}\right)
			- \frac{1}{2q}\!\left[
			\operatorname{Li}_2\!\left(\frac{q^0 - q}{q^0 + q}\right)
			- \operatorname{Li}_2\!\left(\frac{q^0 + q}{q^0 - q}\right)
			\right]\!\right\}+O(\epsilon^2),\\
			\mathcal{A}_2 &\equiv \!\int_z \frac{1}{(v\!\cdot\! Q)^2}
			= \frac{2}{Q^2}
			- \frac{4\epsilon}{Q^2}[\ln(2) - q^0 L(Q)] + O(\epsilon^2),\\
			\mathcal{A}_3 &\equiv \!\int_z \frac{1}{(v\!\cdot\! Q)^3}
			= \frac{2q^0}{(Q^2)^2}
			+ \frac{2\epsilon}{(Q^2)^2}\!\left[L(Q)\!\left(q^2+(q^0)^2\right)
			+ q^0(1-2\ln(2))\right] + O(\epsilon^2),
		\end{split}
	\end{equation}
	with $z=\hat{q}.\hat{v}=\cos\theta$.
Here \(\operatorname{Li}_2(x)\) denotes the dilogarithm,
\[
\operatorname{Li}_2(x)=-\int_0^x dt\,\frac{\ln(1-t)}{t}.
\]
		For spatial components, rotational symmetry implies
		\begin{equation}\label{Angular integrals-spatial}
			\mathcal{A}^i_\alpha \equiv \!\int_z \frac{v^i}{(v\!\cdot\! Q)^\alpha}
			= \hat{q}^i\!\int_z \frac{z}{(v\!\cdot\! Q)^\alpha}
			= \frac{q^i}{q^2}\!\left(q^0 \mathcal{A}_\alpha - \mathcal{A}_{\alpha-1}\right).
		\end{equation}
		The collinear denominators arise from
		\[
		\frac{1}{v\!\cdot v'}=\frac{1}{1-z},\qquad z\equiv \hat v\!\cdot\hat v',
		\]
		with the collinear limit $z\to1$. Holding $v$ fixed and integrating over $v'$, DR gives
		\begin{align}\label{collinear integrals1}
			\mathcal{C}_1 \equiv \int_{z} \frac{1}{1-z}
			&= -\frac{1}{\epsilon}+2\ln2+\frac{\epsilon}{6}\bigl(\pi^2-12\ln^2 2\bigr)+O(\epsilon^2),\no\\
			\mathcal{C}^{\,i}_1 \equiv \int_{z} \frac{v'^i}{1-z}
			&= \hat v^{\,i}\int_{z}  \frac{z}{1-z}
			= \hat v^{\,i}\,(\mathcal{C}_1-\mathcal{A}_0),
		\end{align}
		where the vector relation follows from axial symmetry about $\hat v$. The remaining integration is 
		\begin{align}\label{collinear integrals3}
			&\int_{z}\frac{v'\!\cdot Q}{v\!\cdot v'}=q^0\,\mathcal{A}_0+(\mathcal{C}_1-\mathcal{A}_0)\,v\!\cdot Q.
		\end{align}
		The slashed notation is also introduced to represent more compact relations 
		\begin{align}\label{slashed angular notations}
			\slashed{\mathcal A}_\alpha \equiv \mathcal A_\alpha\,\gamma^0-\mathcal A_\alpha^{\,i}\gamma^i.
		\end{align}
		\section{r/a Graph Properties}\label{ra-graph-rules}
We consider connected amputated diagrams in the real-time \(r/a\)
basis. Let \(N_v\) be the number of vertices, \(I\) the number of
internal propagators,
\begin{equation}
	L=I-N_v+1,
\end{equation}
the number of loops \cite{Peskin:1995ev}, and \(E_a\) the number of
external \(a\)-indices. We label the vertices by \(v_i\) with
\(i=1,\ldots,N_v\), and let \(k_{v_i}\) be the number of \(a\)-indices
carried by vertex \(v_i\). The total number of \(a\)-indices is
\begin{equation}
	N_a^{\mathrm{vert}}
	\equiv
	\sum_{i=1}^{N_v} k_{v_i},
	\label{eq:vertex-a-count}
\end{equation}
where \(E_a\) lie on external legs.
Let \(N_{\mathrm C}\) be the number of causal internal propagators and
\(N_{rr}\) the number of symmetric propagators. An \(rr\) propagator contains no \(a\)-labelled end,
every nonzero \(ra/ar\) propagator contains exactly one, and an \(aa\)
propagator vanishes. Hence,
\begin{equation}
	N_{\mathrm C}=
	N_a^{\mathrm{vert}}-E_a,
	\quad
	I=N_{rr}+N_{\mathrm C}.
	\label{eq:causal-count}
\end{equation}
It follows that
\begin{align}
	N_{rr}
	&=
	I-\sum_{i=1}^{N_v}k_{v_i}+E_a
	\notag\\
	&=
	L+N_v-1-\sum_{i=1}^{N_v}k_{v_i}+E_a
	\notag\\
	&=
	L-1+E_a
	-\sum_{i=1}^{N_v}\bigl(k_{v_i}-1\bigr),
	\label{eq:general-rr-count}
\end{align}
where $N_v=\sum_{i=1}^{N_v}1$ is used. 
Eq.~\eqref{eq:causal-count} can also be written as
\begin{align}
	N_{\mathrm C}
	&=
	\sum_{i=1}^{N_v}k_{v_i}-E_a=
	N_v-E_a+
	\sum_{i=1}^{N_v}\bigl(k_{v_i}-1\bigr).
	\label{eq:general-causal-count}
\end{align}

For cubic QED,  since only vertices with one and three $a$-indices are present, we can write
\begin{align}
	\sum^{N_v}_{i=1}(k_{v_i}-1)=2N_{aaa}.
\end{align}
Single a-indexed vertex contributions to the relation above vanish since $k_{v_i}=1$, while three $a$-indexed vertices with $k_{v_i}=3$  yield the relation above, with $N_{aaa}$ being the number of three $a$-indexed vertices.
Thus, Eqs.~\eqref{eq:general-rr-count} and \eqref{eq:general-causal-count} reduce to
\begin{equation}
	N_{rr}
	=
	L-1+E_a-2N_{aaa},\quad   N_{\mathrm C}=N_v-E_a+2N_{aaa}.
	\label{eq:qed-rr-count}
\end{equation}
\subsection{Special case}
We now restrict the sector with one external
\(a\)-index and one \(a\)-index at every vertex:
\begin{equation}
	E_a=1,
	\quad
	N_{aaa}=0.
\end{equation}
Eq.~\eqref{eq:qed-rr-count} then gives
\begin{equation}\label{special-reduction}
	N_{rr}=L,
	\qquad
	N_{\mathrm C}=N_v-1.
\end{equation}
Thus, when no \(\Gamma^{aaa}\) vertex occurs in the assignment,
\(L\) symmetric propagators and \(N_v-1\) causal propagators
contribute.
We choose the \(L\) symmetric propagators and remove them temporarily
from the original graph. The remaining causal graph, denoted by
\(H\), contains all \(N_v\) vertices because no vertex has been
removed. It also contains the remaining \(N_v-1\) causal
propagators.

\subsection{Acyclicity of the Causal Subgraph \(H\)}
Let \(v_\star\) be the vertex attached to the external \(a\)-leg.
Orient each causal propagator from its \(a\)-labeled end toward its
\(r\)-labeled end. The vertex \(v_\star\) has no outgoing internal
causal propagator because its unique \(a\)-index is already used by
the external leg. Every non-\(v_\star\) vertex has exactly one
outgoing internal causal propagator because its \(a\)-index must lie
on an internal line.

Suppose that a cycle \(\mathcal C\) passes through \(m\) distinct
vertices. A closed path through \(m\) vertices contains exactly \(m\)
causal propagators, which require exactly \(m\) \(a\)-labeled ends.
If \(v_\star\in\mathcal C\), only \(m-1\) vertices can supply such
ends, while the \(m\) causal propagators require \(m\). Therefore,
no consistent \(r/a\) assignment exists. If
\(v_\star\notin\mathcal C\), all \(m\) vertices on the cycle are
non-\(v_\star\) vertices. Since each has exactly one outgoing causal
propagator, all \(m\) outgoing propagators must lie on the cycle.
Hence, the cycle can be written
\begin{equation}
	v_1\longrightarrow v_2\longrightarrow\cdots
	\longrightarrow v_m\longrightarrow v_1.
	\label{eq:directed-cycle}
\end{equation}
In coordinate space, one orientation contains
\begin{equation}
	\theta(t_2-t_1)
	\theta(t_3-t_2)
	\cdots
	\theta(t_1-t_m),
	\label{eq:theta-cycle}
\end{equation}
which would require
\begin{equation}
	t_1<t_2<\cdots<t_m<t_1.
\end{equation}
This condition is impossible and indicates that every
nonvanishing causal subgraph \(H\) is acyclic (tree-like), consistent with Refs.
\cite{CaronHuot_2009,CaronHuot:2010zt}.
\subsection{Connectivity and Unique Orientation of $H$}
Suppose that the acyclic graph \(H\) has \(C_H\) connected
components, and let \(n_j\) be the number of vertices in its
\(j\)-th component. Since the components contain all vertices of
\(H\),
\begin{equation}
	\sum_{j=1}^{C_H}n_j=N_v.
	\label{eq:component-vertices}
\end{equation}
Each component is connected and acyclic and is therefore a tree. The
\(j\)-th component contains \(n_j-1\) causal propagators, so
\begin{align}
	N_{\mathrm C}=
	(n_1-1)+(n_2-1)+\cdots+(n_{C_H}-1)=
	N_v-C_H.
\end{align}
Equating Eq.~\eqref{special-reduction} with the equation above yields
\begin{equation}
	N_v-C_H=N_v-1\longrightarrow C_H=1.
\end{equation}
Thus, \(H\) is connected. Since \(H\) is both connected and acyclic and contains all \(N_v\) vertices, it is a spanning tree
\cite{CaronHuot:2010zt,Runkel:2019yrs}.

Since $H$ is a spanning tree, starting from any
vertex $v_i\neq v_*$ and repeatedly following its unique outgoing causal
propagator cannot revisit a vertex. Since $H$ is finite, the walk must
terminate, and the only vertex with no outgoing causal propagator is
$v_\star$. Hence every vertex has a unique directed path toward $v_\star$.
Thus, $H$ is an in-branching rooted at $v_\star$; see Ch.~1 of Ref.~\cite{BangJensenGutin2009}. In conclusion, once the allowed set of
symmetric propagators and $v_\star$ are fixed, the causal orientation of
$H$ is unique.
\section{Symmetry Constraints on Building Blocks}\label{momentum reversal properties}
	In massless QED,  with no background source that breaks the discrete symmetries and chiral invariance, the Lagrangian and the generating functional are invariant under \(P\), \(\mathcal T\), \(C\). Invariance under $\mathcal{T}$ is the invariance under an antiunitary transformation. 
While the classical action is invariant under global chiral transformations, this axial symmetry is anomalous at the quantum level \cite{Peskin:1995ev}. Since the anomalous contribution depends only on the gauge field, it does not affect the two-fermion--\(N\)-photon sectors considered here. Here, ``chiral invariance'' refers to this perturbative conservation of chirality.
	We adopt the notation
	$X\,\Phi\,X^{-1}\equiv \Phi_X$ with $X\in \{P,\mathcal{T}, C\}$ and $\Phi$ a generic field. In addition,
	\begin{equation}
		Q^\mathcal{T}=Q^P=(q^0,-\mathbf{q}),\qquad
		x^P=(t,-\mathbf{x}),\qquad
		x^\mathcal{T}=(-t,\mathbf{x}),
	\end{equation}
	with Lorentz indices suppressed.
	A complete basis of Dirac matrices in four dimension
	 is~\cite{Peskin:1995ev}:
	\(
	\{\mathbbm{1},\gamma_5,\gamma_\mu,\gamma_\mu\gamma_5,\sigma_{\mu\nu}\}
	\).
	
In the BMHV scheme, writing
			$\gamma^\mu=\bar\gamma^\mu+\hat\gamma^\mu$ and
			$g^{\mu\nu}=\bar g^{\mu\nu}+\hat g^{\mu\nu}$, with
			$\bar g^\mu{}_\mu=4$ and $\hat g^\mu{}_\mu=D-4$, all external momenta
			and free Lorentz indices are taken to lie in the physical four-dimensional
			subspace, so $\hat Q_i^\mu=\hat u^\mu=0$. 
			 In the loop integration, odd-rank evanescent tensors
			vanish under simultaneous reflection of the evanescent components of all
			loop momenta; schematically,
			\[
			\int_{P,L}\hat P^\rho f=0,
			\]
			where the scalar invariants in $f$, including $L\!\cdot\!P$, are unchanged
			under this simultaneous reflection. Even-rank evanescent tensors reduce to the symmetric evanescent metric; for
			example,
			\[
			\int_{P,L}\hat P^\rho\hat L^\sigma f
			=
			\frac{\hat g^{\rho\sigma}}{D-4}
			\int_{P,L}(\hat P\!\cdot\!\hat L)f .
			\]
			If such tensors multiply independent antisymmetrized evanescent gamma
			structures, as can occur in an enlarged $D$-dimensional Dirac decomposition,
			their contribution vanishes, e.g.
			\[
			\hat g^{\rho\sigma}
			\hat\gamma_{[\rho}\hat\gamma_{\sigma]}=0.
			\]
			Thus no independent evanescent Dirac structure survives in the physical
			amplitude, and for four-dimensional external momenta and free Lorentz indices
			the physical Dirac decomposition closes on the four-dimensional structures.
			
		Evanescent indices contracted internally, however, do survive and reduce to
			ordinary four-dimensional structures with explicit $D$-dependent
			coefficients; for example,
			\[
			\hat\gamma^\mu\bar\gamma^\alpha\hat\gamma_\mu
			=-(D-4)\bar\gamma^\alpha .
			\]
			Accordingly, bars are omitted in the displayed physical decompositions and
			independent hatted structures are not shown, purely for notational
			convenience. The loop algebra itself remains fully $D$-dimensional, and all
			resulting $D$-dependent factors, including their $O(\epsilon)$ parts, are
			retained throughout.
	
	The chiral anticommutation relations are
	\begin{align}\label{chiral}
		\{\gamma_5,\mathbbm{1}\}=2\gamma_5,\quad
		\{\gamma_5,\gamma_5\}=2,\quad
		\{\gamma_5,\gamma_\mu\}=0,\quad
		\{\gamma_5,\gamma_\mu\gamma_5\}=0,\quad
		\{\gamma_5,\sigma_{\mu\nu}\}=2\,\sigma_{\mu\nu}\gamma_5\neq 0.
	\end{align}
	The Dirac matrices satisfy
	\begin{align}\label{B2}
		\gamma_0\gamma_\mu\gamma_0=\Lambda^{\nu}{}_\mu\gamma_\nu,\quad
		\gamma_0(\gamma_\mu\gamma_5)\gamma_0=-\Lambda^{\nu}{}_\mu\gamma_\nu\gamma_5,
	\end{align}
	with $\Lambda^{\nu}{}_\mu=\mathrm{diag}(1,-1,-1,-1)$.
	The fermion and vector fields transform under parity as~\cite{Peskin:1995ev}
	\begin{align}\label{parity properties}
		\psi_P(x)= \gamma_0\,\psi(x^P),\qquad
		\bar\psi_P(x)= \bar\psi(x^P)\gamma_0,\qquad
		A^\mu_P(x)=\Lambda^{\mu}{}_\nu A^\nu(x^P).
	\end{align}
	The overall fermionic phase is omitted since it cancels between $\psi$ and $\bar\psi$ in bilinears.
	Similarly, charge conjugation acts as~\cite{Peskin:1995ev}
	\begin{align}\label{charge conjugation}
		&\psi_C(x)= -i\big[\bar{\psi}(x) M\big]^{T},\qquad
		\bar{\psi}_C(x)=-i\big[M\psi(x)\big]^{T},\qquad
		A^\mu_C(x)=-A^\mu(x),\nonumber\\
		&\psi_C(Q)=-i\big[\bar\psi(-Q)M\big]^T,\qquad
		\bar\psi_C(Q)=-i\,\big[M\psi(-Q)\big]^T,
	\end{align}
	with \(M\equiv\gamma_0\gamma_2\) and \(T\) denoting transpose in Dirac space. The matrix \(M\) satisfies
	\begin{align}\label{M properties}
		M^{T}(\gamma_\mu)^T M^T=-\gamma_\mu,\quad
		M^{T}(\gamma_\mu\gamma_5)^T M^T=\gamma_\mu\gamma_5,\quad
		M^T=-M,\quad
		M^{-1}=M.
	\end{align}
	Time reversal acts on fields as~\cite{Peskin:1995ev}
	\begin{equation}\label{B6}
		\psi_{\mathcal{T}}(Q)=B\,\psi(Q^{\mathcal{T}}),\qquad
		\bar\psi_{\mathcal{T}}(Q)=-\bar\psi(Q^{\mathcal{T}})B,\qquad 
		A^\mu_{\mathcal T}(Q)
		=
		\Lambda^\mu{}_\nu A^\nu(Q^{\mathcal T}),
	\end{equation}
	with $B=\gamma_{1}\gamma_{3}$ (up to a phase). The required gamma-matrix identities  are
	\begin{equation}\label{B77}
		B\gamma_{\mu}^{*}B=-\Lambda^{\nu}{}_{\mu}\gamma_{\nu},
		\qquad
		B(\gamma_{\mu}\gamma_{5})^{*}B=-\Lambda^{\nu}{}_{\mu}\gamma_{\nu}\gamma_{5}.
	\end{equation}
	Since \(\mathcal T\) is antiunitary, it complex conjugates the \(c\)-number coefficients and the matrix entries of the Dirac structures.
	\subsection{Spectral Function and Fermion Self-Energy}
	We start with the fermionic spectral function defined in position space by
	\(\rho_F(x,y)=\big\langle \{\psi(x),\bar\psi(y)\}\big\rangle_T\),
	with Fourier transform \(\rho_F(Q)\). The Dirac decomposition gives
	\begin{equation}\label{eq:rho-decomp}
		\rho_F(Q)
		= \mathcal{S}\,\mathbbm{1} + \mathcal{P}\,\gamma_5
		+ V^\mu\,\gamma_\mu
		+ A^\mu\,\gamma_\mu\gamma_5
		+ \tfrac{1}{2}\,T^{\mu\nu}\,\sigma_{\mu\nu}.
	\end{equation}
	Chiral invariance implies $\mathcal{S}=\mathcal{P}=T^{\mu\nu}=0$,  yielding
	\begin{equation}\label{eq:rho-VA}
		\rho_F(Q)= V^\mu(Q)\,\gamma_\mu + A^\mu(Q)\,\gamma_\mu\gamma_5.
	\end{equation}
	We now impose discrete symmetries. Under parity invariance, the spectral matrix transforms as
	\begin{equation}\label{eq:parity-rho}
		\rho_F(q^0,\mathbf{q})
		= \gamma_0\,\rho_F(q^0,-\mathbf{q})\,\gamma_0,
	\end{equation}
	where we used Eq.~\eqref{parity properties}.
	Using Eqs.~\eqref{B2}, \eqref{eq:rho-VA} and \eqref{eq:parity-rho}, one finds that $V^\mu$ transforms as a polar vector, while $A^\mu$ transforms as a
	pseudovector. 
	For the latter, no nonzero pseudovector can be constructed from $Q^\mu$ and $u^\mu$ alone. Equivalently, any Levi--Civita construction built only from $Q^\mu$ and $u^\mu$ vanishes without an additional
	pseudovector/pseudoscalar background. Hence, setting $A^\mu(Q)=0$, we obtain
	\begin{equation}\label{eq:rho-final}
		\rho_F(Q)=a(Q)\,\Qslash + b(Q)\,\uslash,
	\end{equation}
	with real scalar coefficients $a(Q)$ and $b(Q)$ depending on the invariants $Q^2$ and $u\!\cdot\!Q$. The reality is ensured by Dirac pseudo-Hermiticity
	\begin{align}\label{spectral-pseudo-Hermiticity}
		\rho_F^\dagger(Q)=\gamma_0\rho_F(Q)\gamma_0.
	\end{align}
	Next, using the charge-conjugation properties in Eq.~\eqref{charge conjugation} and translation invariance, \(\rho_F(x,y)=\rho_F(x-y)\), one obtains\footnote{  The corresponding position-space relation is
		\begin{equation*}\label{eq:C-rho}
			\rho_F(x,y)= -M^{T}\rho_F(y,x)^{T}M^{T},
	\end{equation*}}
	\begin{equation}
		\rho_F(Q) = -M^{T}\,\rho_F(-Q)^{T}\,M^{T}.
	\end{equation}
	Substituting Eq.~\eqref{eq:rho-final} and using Eq.~\eqref{M properties}, one obtains
	\begin{equation}
		a(-Q) = -a(Q),
		\qquad
		b(-Q) = b(Q),
	\end{equation}
	which yields
	\begin{equation}\label{spectral property}
		\rho_F(-Q)=\rho_F(Q).
	\end{equation}
	The evenness above, together with the spectral representation~\cite{Bellac:2011kqa}, yields
	\begin{align}\label{advanced property}
		S_A(-Q)
		&=i\int_{-\infty}^\infty\frac{d\omega}{2\pi}\,
		\frac{\rho_F(\omega,-\mathbf{q})}{-Q_0-\omega-i\eta}
		= -i\int_{-\infty}^\infty\frac{d\omega}{2\pi}\,
		\frac{\rho_F(-\omega,-\mathbf{q})}{Q_0-\omega+i\eta}
		= -S_R(Q),
	\end{align}
	where we used $\rho_F(-\omega,-\mathbf{q})=\rho_F(\omega,\mathbf{q})$.
	Using the Dyson relation $S^{-1}_{R/A}=S^{-1}_{0,R/A}-\Sigma_{R/A}$, one obtains
	\begin{equation}
		\Sigma_A(-Q)=-\Sigma_R(Q).
	\end{equation}
	\subsection{Two-Fermion--\(N\)-Photon Vertices}
We start with the closed-time-path (CTP) formalism \cite{Bellac:2011kqa}, where the generic fields in the $r/a$ basis are defined by
\begin{equation}
	\begin{pmatrix}
		\phi_r\\[1mm]\phi_a
	\end{pmatrix}
	=
	U
	\begin{pmatrix}
		\phi_+\\[1mm]\phi_-
	\end{pmatrix},
	\qquad
	U=
	\begin{pmatrix}
		\tfrac12&\tfrac12\\
		1&-1
	\end{pmatrix}.
\end{equation}
Here, $\phi_+$ and $\phi_-$ live on the forward and backward branches, respectively; $U$ is the operator that relates them to the $r/a$ space.
In the CTP, interchanges of the
forward and backward branches are introduced by
\begin{equation}
	\Phi\equiv\begin{pmatrix}
		\phi_+\\
		\phi_-
	\end{pmatrix}
	\xrightarrow{\rm Intch}
	X
	\begin{pmatrix}
		\phi_+\\
		\phi_-
	\end{pmatrix},
	\qquad
	X=
	\begin{pmatrix}
		0&1\\
		1&0
	\end{pmatrix},
\end{equation}
In the $r/a$ basis, the branch interchange reads as
\begin{equation}
	\begin{pmatrix}
		\phi_r\\
		\phi_a
	\end{pmatrix} \xrightarrow{\rm Intch} UXU^{-1} \begin{pmatrix}
		\phi_r\\
		\phi_a
	\end{pmatrix},
	\quad
	UXU^{-1}
	=
	\begin{pmatrix}
		1&0\\
		0&-1
	\end{pmatrix},
\end{equation}
Therefore, the $r/a$ fields behave under interchange of the two CTP branches as
\begin{equation}\label{exchange}
	\phi_r\xrightarrow{\rm Intch}+\phi_r,
	\qquad
	\phi_a\xrightarrow{\rm Intch}-\phi_a.
\end{equation}

We now determine how the discrete operations act on the effective action\footnote{One can see the details of the generating functional and $W$ in Refs.~\cite{Crossley:2015evo, Haehl:2016pec}}.
Let $\mathcal U_J$ denote the source-dependent evolution operator and $\rho$ the
density matrix, with $\rho^\dagger=\rho$. Define
\begin{equation}\label{eq:Zdef}
	Z[J_+,J_-]\equiv\Tr\big(\mathcal U_{J_+}\,\rho\,\mathcal U^\dagger_{J_-}\big),
	\qquad
	Z_\rev[J_+,J_-]\equiv\Tr\big(\mathcal U^\dagger_{J_-}\,\rho\,\mathcal U_{J_+}\big).
\end{equation}
The second is the reversed-process functional: the two evolution operators
appear in the opposite order. 
The generating functional is invariant under (unitary operators) $P$ and $C$:
\begin{align}
	Z[J_+,J_-]= Z[J^X_+,J^X_-],\quad  Z_\rev[J_+,J_-]=Z_\rev[J^X_+,J^X_-],\quad X\in{\{P,C\}}.
\end{align}
To derive the consequence of antiunitary time reversal for $Z$, we proceed as follows.
If $Y$ is an arbitrary operator, and $\Theta$ is an antiunitary operator, we can obtain the following relations
\begin{align}
	\Tr\big(\Theta\,Y\,\Theta^{-1}\big)=(\Tr Y)^* ,\quad \Theta\,Y^\dagger\,\Theta^{-1}=\big(\Theta\,Y\,\Theta^{-1}\big)^\dagger.
\end{align}
Then, the operation of the antiunitary operator on the evolution operator and density matrix becomes
\begin{equation}\label{eq:Tassump}
	\Theta\,\mathcal U_J\,\Theta^{-1}=\mathcal U^\dagger_{J^{\mathcal T}},
	\qquad
	\Theta\rho\,\Theta^{-1}=\rho .
\end{equation}
Using $(\Tr Y)^*=\Tr(Y^\dagger)$ and $\rho^\dagger=\rho$, one first obtains
\begin{equation}\label{eq:reflect}
	Z[J_+,J_-]^*=Z[J_-,J_+],
	\qquad
	Z_{\rm rev}[J_+,J_-]^*=Z_{\rm rev}[J_-,J_+] .
\end{equation}

Using Eq.~\eqref{eq:Tassump} and the antiunitary identities above, one finds
\begin{align}
	Z[J_+,J_-]^*
	&=
	Z_{\rm rev}
	[J_-^{\mathcal T},J_+^{\mathcal T}],
	\nonumber\\
	Z[J_+,J_-]
	&=
	Z_{\rm rev}
	[J_+^{\mathcal T},J_-^{\mathcal T}],
	\label{eq:Zrev}
\end{align}
where the second line follows from Eq.~\eqref{eq:reflect}.

Since $Z=\exp[i W]$, we use the Legendre transform
\begin{align}
	\mathbf{\Gamma}\bigl[\phi_+,\phi_-\bigr] =W[J_+,J_-]-\int\bigl(J_+\phi_+-J_-\phi_-\bigr),
\end{align}
The source-field pairing is invariant under the time-reversal map.
Using the unitary $C/P$ invariance above, together with Eqs.~\eqref{eq:reflect} and \eqref{eq:Zrev}, we obtain
\begin{align}
	&\mathbf{\Gamma}[\phi_+,\phi_-]
	=
	\mathbf{\Gamma}[\phi_+^{X},\phi_-^{X}],
	\qquad X\in\{C,P\},
	\label{effective action invariance}\\
	&\mathbf{\Gamma}[\phi_+,\phi_-]
	=
	\mathbf{\Gamma}_{\rm rev}
	[\phi_+^{\mathcal T},\phi_-^{\mathcal T}],
	\label{eq:T-Gamma-fixed}\\
	&
	\mathbf{\Gamma}[\phi_+,\phi_-]^*
	=
	-\mathbf{\Gamma}[\phi_-,\phi_+],\label{eq:Gamma-reflectivity}
\end{align}
where $\phi=(\bar\psi,\psi,A^\mu)$.
Equivalently, using the branch-interchange rule in Eq.~\eqref{exchange}, the reflectivity condition in the $r/a$ basis is
\begin{equation}\label{Gamma real}
		-\mathbf{\Gamma}[\phi_r,\phi_a]^*
		=
		\mathbf{\Gamma}[\phi_r,-\phi_a].
\end{equation}
Notably, the action of the discrete symmetries does not change the branch ordering, whereas Eq.~\eqref{eq:Gamma-reflectivity} is the CTP reflectivity (reality) condition and explicitly interchanges the two branches.

In the following, each symmetry constraint is derived in three steps. First,
the transformed current is constructed explicitly from the transformed
effective action appearing on the right-hand side of the corresponding
effective-action relation. Second, since the right- and left-hand sides of
that relation are equal, the same transformed current is expressed using the
effective action on the left-hand side, with the appropriate transformation
of its external Lorentz indices and momenta. Finally, the two expressions for
the same current are equated, yielding the corresponding constraints on the
vertex coefficients. The current can be determined by the part of the one-particle-irreducible (1PI) effective action containing two fermion fields and \(N\)
external photon insertions,
\begin{equation}
	\mathbf{\Gamma}[\psi,\bar\psi,A]\supset
	\sum_{N\geq 1}\frac{(-1)^N}{N!}
	\int d^4y\,d^4z\,d^4x_1\cdots d^4x_N\;
	\bar\psi_{c_{N+1}}(y)\,
	\Gamma_{c}^{\mu_1\cdots\mu_N}(y,z;x_1,\ldots,x_N)\,
	\psi_{c_0}(z)\,
	A^{c_1}_{\mu_1}(x_1)\cdots A^{c_N}_{\mu_N}(x_N),
\end{equation}
where \(\Gamma_c^{\mu_1\cdots\mu_N}(y,z;x_1,\ldots,x_N)\) is the two-fermion--\(N\)-photon vertex function in position space and is a \(4\times4\) matrix in Dirac space. Subscript $c$ denotes the $r/a$ labels as
\begin{equation}\label{eq:string}
	c=(c_0,c_1,\ldots,c_N,c_{N+1}),
	\qquad c_i\in\{r,a\},
	\qquad
	\tilde{c}\equiv
	(c_{N+1},c_1,\ldots,c_N,c_0),
	\qquad
	E_a(c)\equiv\#\{i:c_i=a\},
	\quad i=0,\ldots,N+1,
\end{equation}
so that $c_0$, $c_i$, and $c_{N+1}$ label incoming fermion, incoming photon $i$, and outgoing fermion, respectively. 
The last relation in Eq.~\eqref{eq:string}, $E_a(c)$, counts the number of (external) $a$ indices of the kernel, while
$\tilde{c}$ denotes the exchange of the two
fermion endpoint labels.  
Throughout the component relations below, $c$ denotes a fixed external
$r/a$ string; no sum over $c$ is implied.
The position of $c$ as a superscript or subscript carries no distinction and is chosen only for convenience.
\subsubsection{\bf Discrete Symmetry Operations}
The corresponding dressed \(N\)-current kernel is defined by
\begin{align}
	&j_{c}^{\mu_1\cdots\mu_N}(x_1,\ldots,x_N)
	\equiv
	(-1)^N
	\frac{\delta^N\mathbf{\Gamma}[\psi,\bar\psi,A]}
	{\delta A^{c_1}_{\mu_1}(x_1)\cdots\delta A^{c_N}_{\mu_N}(x_N)}
	=
	\int d^4y\,d^4z\;
	\bar\psi_{c_{N+1}}(y)\,
	\Gamma_{c}^{\mu_1\cdots\mu_N}(y,z;x_1,\ldots,x_N)\,
	\psi_{c_0}(z),
	\label{eq:Ncurrent_kernel}
\end{align}
In a translationally invariant state,
\(\Gamma_{c}^{\mu_1\cdots\mu_N}(y,z;x_1,\ldots,x_N)\) depends only on relative coordinates. For example, one may choose \(x_N\) as a reference point and write it as a function of
\(y-x_N\), \(z-x_N\), and \(x_i-x_N\) with \(i=1,\ldots,N-1\). Equivalently, its Fourier transform contains the overall momentum-conservation delta function. Therefore, the Fourier-space kernel can be written as
\begin{equation}
	j_{c}^{\mu_1\cdots\mu_N}(K_1,\ldots,K_N)
	=
	\int\!\frac{d^4Q_1}{(2\pi)^4}\;
	\bar\psi_{c_{N+1}}(Q_2)\,
	\Gamma_{c}^{\mu_1\cdots\mu_N}(Q_2,Q_1;\{K_i\})\,
	\psi_{c_0}(Q_1),
	\qquad
	Q_2=Q_1+\sum_{i=1}^N K_i ,
	\label{eq:Ncurrent_momentum}
\end{equation}
where all photon momenta are taken incoming. By momentum conservation, only \(N-1\) photon momenta are independent. We denote them by
\(\{K_i\}\equiv\{K_1,\ldots,K_{N-1}\}\), with \(K_N\) fixed by momentum conservation.

Chiral invariance implies that the vertex matrix anticommutes with \(\gamma_5\).
Therefore, scalar, pseudoscalar, and tensor Dirac structures are forbidden, and the most general decomposition of the current is
\begin{equation}\label{B45}
	j_{c}^{\mu_1\cdots\mu_N}(K_1,\dots,K_N)
	= \int\!\frac{d^4Q_1}{(2\pi)^4}\;
	\bar\psi_{c_{N+1}}(Q_2)\,
	\Big[
	V_{c}^{\mu_1\cdots\mu_N,\nu}(Q_2,Q_1;\{K_i\})\,\gamma_\nu
	+ A_{c}^{\mu_1\cdots\mu_N,\nu}(Q_2,Q_1;\{K_i\})\,\gamma_\nu\gamma_5
	\Big]\,
	\psi_{c_0}(Q_1),
\end{equation}
where \(V_{c}^{\mu_1\cdots\mu_N,\nu}\) and
\(A_{c}^{\mu_1\cdots\mu_N,\nu}\) are coefficient tensors constructed from the available
four-vectors \(Q_{1,2}^\mu\), \(K_i^\mu\), and \(u^\mu\), together with \(g^{\mu\nu}\) and \(\epsilon^{\mu\nu\rho\sigma}\) (see, e.g., the zero-temperature discussion~\cite{Weinberg:1995mt}, Ch.~10). In our case, \(u^\mu=(1,0,0,0)\).
\paragraph{\bf Parity.}
Using Eq.~\eqref{parity properties}, one obtains the parity transform of the current as
\begin{align}
	j_{P,c}^{\mu_1\cdots\mu_N}(K_1,\ldots,K_N)
	=&\int\!\frac{d^4Q_1}{(2\pi)^4}\no\\
	&\times\bar\psi_{c_{N+1}}(Q_2^P)\Big[
	V_{c}^{\mu_1\cdots\mu_N,\nu}(Q_2,Q_1;\{K_i\})
	\Lambda^{\beta}{}_\nu\,\gamma_\beta-
	A_{c}^{\mu_1\cdots\mu_N,\nu}(Q_2,Q_1;\{K_i\})
	\Lambda^{\beta}{}_\nu\,\gamma_\beta\gamma_5
	\Big]\psi_{c_0}(Q_1^P),
\end{align}
with \(Q^P=(q^0,-\mathbf q)\) and
\(\Lambda^\mu{}_\nu=\mathrm{diag}(1,-1,-1,-1)\).
Since each photon field is a polar vector, parity invariance of the effective action implies that the current transforms as a rank-\(N\) polar tensor:
\begin{align}\label{Polar-tensor}
	j_{P,c}^{\mu_1\cdots\mu_N}(K_1,\ldots,K_N)
	=\int\!\frac{d^4Q_1}{(2\pi)^4}\;
	&\bar\psi_{c_{N+1}}(Q_2^P)\,
	\Lambda^{\mu_1}{}_{\rho_1}\cdots
	\Lambda^{\mu_N}{}_{\rho_N}\nonumber\\
	&\times\Big[
	V_{c}^{\rho_1\cdots\rho_N,\beta}
	(Q_2^P,Q_1^P;\{K_i^P\})\gamma_\beta
	+
	A_{c}^{\rho_1\cdots\rho_N,\beta}
	(Q_2^P,Q_1^P;\{K_i^P\})\gamma_\beta\gamma_5
	\Big]\psi_{c_0}(Q_1^P).
\end{align}
Since these expressions agree for arbitrary spinor fields, one obtains
\begin{align}
	V_{c}^{\mu_1\cdots\mu_N,\nu}(Q_2,Q_1;\{K_i\})
	&=
	\Lambda^{\mu_1}{}_{\rho_1}\cdots
	\Lambda^{\mu_N}{}_{\rho_N}\Lambda^\nu{}_\beta\,
	V_{c}^{\rho_1\cdots\rho_N,\beta}
	(Q_2^P,Q_1^P;\{K_i^P\}),
	\nonumber\\
	A_{c}^{\mu_1\cdots\mu_N,\nu}(Q_2,Q_1;\{K_i\})
	&=
	-\Lambda^{\mu_1}{}_{\rho_1}\cdots
	\Lambda^{\mu_N}{}_{\rho_N}\Lambda^\nu{}_\beta\,
	A_{c}^{\rho_1\cdots\rho_N,\beta}
	(Q_2^P,Q_1^P;\{K_i^P\}).
	\label{eq:parity_Nvertex}
\end{align}
Thus, \(V_{c}^{\mu_1\cdots\mu_N,\nu}\) is a rank-\((N+1)\) tensor, while
\(A_{c}^{\mu_1\cdots\mu_N,\nu}\) is a rank-\((N+1)\) pseudotensor.
\paragraph{\bf Time reversal and \(P\mathcal T\).}
For the reversed process, define
\begin{equation}\label{eq:D-rev-def}
	\Gamma_{c}^{{\rm rev},\mu_1\cdots\mu_N}(Q_2,Q_1;\{K_i\})
	\equiv
	V_{c}^{{\rm rev},\mu_1\cdots\mu_N,\rho}(Q_2,Q_1;\{K_i\})\gamma_\rho
	+
	A_{c}^{{\rm rev},\mu_1\cdots\mu_N,\rho}(Q_2,Q_1;\{K_i\})
	\gamma_\rho\gamma_5 ,
\end{equation}
so that
\begin{equation}
	j_{c}^{{\rm rev},\mu_1\cdots\mu_N}
	(K_1,\ldots,K_N)
	=
	\int\!\frac{d^4Q_1}{(2\pi)^4}\,
	\bar\psi_{c_{N+1}}(Q_2)\,
\Gamma_{c}^{{\rm rev},\mu_1\cdots\mu_N}
	(Q_2,Q_1;\{K_i\})\,
	\psi_{c_0}(Q_1).
\end{equation}
Since $\mathcal T$ is antiunitary, its action is antilinear.  Thus, for
a c-number $z$ and an operator $\mathcal O$,
\begin{equation}\label{antiunitary}
	\mathcal T\,(z\mathcal O)\,\mathcal T^{-1}
	=
	z^*\,
	\mathcal T\mathcal O\mathcal T^{-1}.
\end{equation}
One obtains the time-reversal operation on the reversed process (right-hand side of Eq.~\eqref{eq:T-Gamma-fixed}) as
\begin{align}\label{eq:T-current-direct-N}
	j_{{\mathcal T},c}^{{\rm rev},\mu_1\cdots\mu_N}
	&(K_1,\ldots,K_N)
	=
	\int\!\frac{d^4Q_1}{(2\pi)^4}\,
	\bar\psi_{c_{N+1}}(Q_2^{\mathcal T})
	\nonumber\\
	&\times
	\Big[
	\bigl[
	V_{c}^{{\rm rev},\mu_1\cdots\mu_N,\rho}
	(Q_2,Q_1;\{K_i\})
	\bigr]^*
	\Lambda^\nu{}_{\rho}\gamma_\nu
	+
	\bigl[
	A_{c}^{{\rm rev},\mu_1\cdots\mu_N,\rho}
	(Q_2,Q_1;\{K_i\})\bigr]^*
	\Lambda^\nu{}_{\rho}\gamma_\nu\gamma_5
	\Big]
	\psi_{c_0}(Q_1^{\mathcal T}),
\end{align}
where Eqs.~\eqref{B6} and \eqref{B77} are taken into account. The star on $V$ and $A$ comes from Eq.~\eqref{antiunitary}.
Eq.~\eqref{eq:T-current-direct-N} is written in the fixed-branch representation of
Eq.~\eqref{eq:T-Gamma-fixed}. Hence no entry of $c$ is exchanged and no additional
factor $(-1)^{E_a(c)}$ appears. 

We can obtain a second expression for the same transformed current
directly from the effective-action identity.  Using the left-hand side of Eq.~\eqref{eq:T-Gamma-fixed}, one obtains
\begin{align}\label{eq:T-current-tensor-N}
	j_{{\mathcal T},c}^{{\rm rev},\mu_1\cdots\mu_N}
	(K_1,\ldots,K_N)
	=
	\Lambda^{\mu_1}{}_{\alpha_1}\cdots
	\Lambda^{\mu_N}{}_{\alpha_N}\,
	j_{c}^{\alpha_1\cdots\alpha_N}
	(K_1^{\mathcal T},\ldots,K_N^{\mathcal T}) .
\end{align}
Equating Eqs.~\eqref{eq:T-current-direct-N} and \eqref{eq:T-current-tensor-N} gives
\begin{align}\label{eq:T-Nvertex}
	V_{c}^{\mu_1\cdots\mu_N,\nu}
	(Q_2,Q_1;\{K_i\})
	&=
	\Lambda^{\mu_1}{}_{\alpha_1}\cdots
	\Lambda^{\mu_N}{}_{\alpha_N}
	\Lambda^\nu{}_{\beta}
	\Big[
	V_{c}^{{\rm rev},\alpha_1\cdots\alpha_N,\beta}
	(Q_2^{\mathcal T},Q_1^{\mathcal T};
	\{K_i^{\mathcal T}\})
	\Big]^*,
	\nonumber\\
	A_{c}^{\mu_1\cdots\mu_N,\nu}
	(Q_2,Q_1;\{K_i\})
	&=
	\Lambda^{\mu_1}{}_{\alpha_1}\cdots
	\Lambda^{\mu_N}{}_{\alpha_N}
	\Lambda^\nu{}_{\beta}
	\Big[
	A_{c}^{{\rm rev},\alpha_1\cdots\alpha_N,\beta}
	(Q_2^{\mathcal T},Q_1^{\mathcal T};
	\{K_i^{\mathcal T}\})
	\Big]^* .
\end{align}
The star in Eq.~\eqref{eq:T-Nvertex} acts on the complete real-time
coefficient and, in particular, complex conjugates its $i\eta$
prescriptions.  

We next combine Eq.~\eqref{eq:T-Nvertex} with parity.  With our
conventions,
\begin{equation}
	Q^{\mathcal T}=Q^P=(q^0,-\mathbf q),
	\qquad
	K_i^{\mathcal T}=K_i^P .
\end{equation}
Parity does not reverse the time evolution, so the parity relations
Eq.~\eqref{eq:parity_Nvertex} apply separately to the forward and
reversed processes.  In particular,
\begin{align}
	V_{c}^{{\rm rev},\alpha_1\cdots\alpha_N,\beta}
	(Q_2^P,Q_1^P;\{K_i^P\})
	&=
	\Lambda^{\alpha_1}{}_{\rho_1}\cdots
	\Lambda^{\alpha_N}{}_{\rho_N}
	\Lambda^\beta{}_{\sigma}\,
	V_{c}^{{\rm rev},\rho_1\cdots\rho_N,\sigma}
	(Q_2,Q_1;\{K_i\}),
	\no
	\\
	A_{c}^{{\rm rev},\alpha_1\cdots\alpha_N,\beta}
	(Q_2^P,Q_1^P;\{K_i^P\})
	&=
	-\Lambda^{\alpha_1}{}_{\rho_1}\cdots
	\Lambda^{\alpha_N}{}_{\rho_N}
	\Lambda^\beta{}_{\sigma}\,
	A_{c}^{{\rm rev},\rho_1\cdots\rho_N,\sigma}
	(Q_2,Q_1;\{K_i\}).
	\label{eq:P-rev-A}
\end{align}
Substituting the reversed-process parity relations given in Eq.~\eqref{eq:P-rev-A}  into
Eq.~\eqref{eq:T-Nvertex}, along with using
$\Lambda^\mu{}_{\alpha}\Lambda^\alpha{}_{\nu}
=\delta^\mu{}_{\nu}$, yields the $P\mathcal T$ constraints
\begin{align}\label{eq:PT-Nvertex}
	V_{c}^{\mu_1\cdots\mu_N,\nu}
	(Q_2,Q_1;\{K_i\})
	&=
	\Big[
	V_{c}^{{\rm rev},\mu_1\cdots\mu_N,\nu}
	(Q_2,Q_1;\{K_i\})
	\Big]^*,
	\nonumber\\
	A_{c}^{\mu_1\cdots\mu_N,\nu}
	(Q_2,Q_1;\{K_i\})
	&=
	-
	\Big[
	A_{c}^{{\rm rev},\mu_1\cdots\mu_N,\nu}
	(Q_2,Q_1;\{K_i\})
	\Big]^* .
\end{align}
\paragraph{\bf Charge conjugation.}
The transformed current under $C$ is written
\begin{align}
	j_{C,c}^{\mu_1\cdots\mu_N}
	&=
	(-i)^2
	\int\frac{d^4Q_1}{(2\pi)^4}\,
	\big[M\psi_{c_{N+1}}(-Q_2)\big]^T
	\Gamma_{c}^{\mu_1\cdots\mu_N}
	(Q_2,Q_1;\{K_i\})
	\big[\bar\psi_{c_0}(-Q_1)M\big]^T
	\nonumber\\
	&=
	-\int\frac{d^4Q_1}{(2\pi)^4}\,
	\psi_{c_{N+1}}^T(-Q_2)\,
	M^T
	\Gamma_{c}^{\mu_1\cdots\mu_N}
	(Q_2,Q_1;\{K_i\})
	M^T\,
	\bar\psi_{c_0}^T(-Q_1),
	\label{eq:C-before-reorder}
\end{align}
where Eq.~\eqref{charge conjugation} is used. 
Using the Grassmann identity
$\psi^T X\bar\psi^T=-\bar\psi X^T\psi$ and Eq.~\eqref{M properties}, one obtains
\begin{align}
	j_{C,c}^{\mu_1\cdots\mu_N}
	&=
	-\int\frac{d^4Q_1}{(2\pi)^4}\,
	\bar\psi_{c_0}(-Q_1)
	\Big[
	V_{c}^{\mu_1\cdots\mu_N,\nu}
	(Q_2,Q_1;\{K_i\})\gamma_\nu
	-
	A_{c}^{\mu_1\cdots\mu_N,\nu}
	(Q_2,Q_1;\{K_i\})\gamma_\nu\gamma_5
	\Big]
	\psi_{c_{N+1}}(-Q_2).
	\label{eq:C-transformed}
\end{align}
Now define
\begin{align}
	Q_1'\equiv-Q_2,
	\quad
	Q_2'\equiv-Q_1,\quad
	Q_2'-Q_1'
	=
	\sum_{i=1}^{N}K_i.
\end{align}
The photon momenta \(K_i\) are unchanged.  Eq.~\eqref{eq:C-transformed}
therefore becomes
\begin{align}
	j_{C,c}^{\mu_1\cdots\mu_N}(Q'_2)
	&=
	-\int\frac{d^4Q'_1}{(2\pi)^4}\,
	\bar\psi_{c_0}(Q'_2)
	\Big[
	V_{c}^{\mu_1\cdots\mu_N,\nu}
	(-Q'_1,-Q'_2;\{K_i\})\gamma_\nu
	-
	A_{c}^{\mu_1\cdots\mu_N,\nu}
	(-Q'_1,-Q'_2;\{K_i\})\gamma_\nu\gamma_5
	\Big]
	\psi_{c_{N+1}}(Q'_1).
\end{align}
One can drop the primes so that
\begin{align}
	j_{C,c}^{\mu_1\cdots\mu_N}(Q_2)
	&=
	-\int\frac{d^4Q_1}{(2\pi)^4}\,
	\bar\psi_{c_0}(Q_2)
	\Big[
	V_{c}^{\mu_1\cdots\mu_N,\nu}
	(-Q_1,-Q_2;\{K_i\})\gamma_\nu
	-
	A_{c}^{\mu_1\cdots\mu_N,\nu}
	(-Q_1,-Q_2;\{K_i\})\gamma_\nu\gamma_5
	\Big]
	\psi_{c_{N+1}}(Q_1).
	\label{eq:C-standard}
\end{align}
Comparing Eq.~\eqref{eq:C-standard} with the standard ordering
\(\bar\psi_{c'_{N+1}}(Q_2)\cdots\psi_{c'_0}(Q_1)\) identifies
\begin{align}\label{recovering}
		c_{N+1}\equiv \tilde{c}_0,\qquad
		c_i\equiv \tilde{c}_i, \quad (i=1,\ldots,N)\qquad
		c_0\equiv \tilde{c}_{N+1},
\end{align}
Hence, the string defined in Eq.~\eqref{eq:string} recovers:
$\tilde{c}=
(c_{N+1},c_1,\ldots,c_N,c_0)$.

Since each of the \(N\) photon fields is \(C\) odd, invariance of the
effective action gives
\begin{align}
	j_{C,c}^{\mu_1\cdots\mu_N}
	(K_1,\ldots,K_N)
	=
	(-1)^N
	j_{\tilde{c}}^{\mu_1\cdots\mu_N}
	(K_1,\ldots,K_N).
	\label{eq:C-current-relation}
\end{align}
Equating Eqs.~\eqref{eq:C-current-relation} and \eqref{eq:C-standard}, along with using Eq.~\eqref{recovering}, one obtains
\begin{align}
	V_{\tilde{c}}^{\mu_1\cdots\mu_N,\nu}
	(Q_2,Q_1;\{K_i\})
	&=
	(-1)^{N+1}
	V_{c}^{\mu_1\cdots\mu_N,\nu}
	(-Q_1,-Q_2;\{K_i\}),\no
	\\
	A_{\tilde{c}}^{\mu_1\cdots\mu_N,\nu}
	(Q_2,Q_1;\{K_i\})
	&=
	(-1)^N
	A_{c}^{\mu_1\cdots\mu_N,\nu}
	(-Q_1,-Q_2;\{K_i\}).
\end{align}
Finally, since
\(\tilde{\tilde{c}}=c\), one obtains
\begin{align}
	&V_{c}^{\mu_1\cdots\mu_N,\nu}
	(Q_2,Q_1;\{K_i\})
	=
	(-1)^{N+1}
	V_{\tilde{c}}^{\mu_1\cdots\mu_N,\nu}
	(-Q_1,-Q_2;\{K_i\}),\no
	\\
	&A_{c}^{\mu_1\cdots\mu_N,\nu}
	(Q_2,Q_1;\{K_i\})
	=
	(-1)^N
	A_{\tilde{c}}^{\mu_1\cdots\mu_N,\nu}
	(-Q_1,-Q_2;\{K_i\}).
\end{align}
\paragraph{\bf Hermiticity and fermion-leg exchange.}
We begin with the right-hand side of Eq.~\eqref{Gamma real}, from which  the 1PI action, with two fermion fields and $N$ external photon fields, becomes
\begin{equation}
	\mathbf{\Gamma}[\phi_r,-\phi_a]\supset
	\sum_{N\geq 1}\frac{(-1)^N}{N!}(-1)^{E_a(c)}
	\int d^4y\,d^4z\,d^4x_1\cdots d^4x_N\;
	\bar\psi_{c_{N+1}}(y)\,
	\Gamma_{c}^{\mu_1\cdots\mu_N}(y,z;x_1,\ldots,x_N)\,
	\psi_{c_0}(z)\,
	A^{c_1}_{\mu_1}(x_1)\cdots A^{c_N}_{\mu_N}(x_N),
\end{equation}
Once in the $r/a$ basis, the two CTP branches are interchanged, while the $r/a$ labels themselves are not relabeled. Each external $a$ field acquires a minus sign. Thus, $(-1)^{E_a(c)}$ counts the number of external $a$ indices as given in Eq.~\eqref{eq:string}. The corresponding current for this interchanged branch reads
\begin{align}\label{interchange-current}
		j^{\mu_1,\cdots,\mu_N}_{c,\rm Intch}(Q_2)
		=
		(-1)^{E_a(c)}
		j^{\mu_1,\cdots,\mu_N}_{c}(Q_2).
\end{align}
The label ``$\rm Intch$'' denotes the interchanged branch. Thus, the only difference between the interchanged-branch and fixed-branch currents is the factor $(-1)^{E_a(c)}$.

We now consider the left-hand side of Eq.~\eqref{Gamma real}. Since the dressed current has no free Dirac indices, it is a scalar in Dirac space and therefore $\bigl[j_c^{\mu_1\cdots\mu_N}\bigr]^*=\bigl[j_c^{\mu_1\cdots\mu_N}\bigr]^\dagger$. Hence
\begin{align}\label{dressed-current-star}
		- \bigl[j_{c}^{\mu_1\cdots\mu_N}\bigr]^\dagger
		=
		- \int\frac{d^4Q_1}{(2\pi)^4}
		\bar\psi_{c_0}(Q_1)\,
		\gamma_0\Gamma_{c}^\dagger(Q_2,Q_1;\{K_i\})\gamma_0\,
		\psi_{c_{N+1}}(Q_2).
\end{align}
Hermitian conjugation exchanges the two fermion endpoint positions. 
We therefore relabel the incoming and outgoing momenta as
$Q_1\leftrightarrow Q_2$. Since
\[
Q_1-Q_2=\sum_i(-K_i),
\]
the photon momenta simultaneously transform as $K_i\to-K_i$. 
Using $\tilde{c}=(c_{N+1},c_1,\ldots,c_N,c_0)$ and suppressing the
new momentum labels, Eq.~\eqref{dressed-current-star} becomes
\begin{align}\label{dressed-dagger}
	- \bigl[j_{c}^{\mu_1\cdots\mu_N}\bigr]^\dagger
	=
	- \int\frac{d^4Q_1}{(2\pi)^4}\,
	\bar\psi_{\tilde{c}_{N+1}}(Q_2)\,
	\gamma_0\Gamma_{c}^\dagger
	(Q_1,Q_2;\{-K_i\})\gamma_0\,
	\psi_{\tilde{c}_{0}}(Q_1).
\end{align}

Comparing Eq.~\eqref{dressed-dagger} with Eq.~\eqref{interchange-current} and using $E_a(\tilde{c})=E_a(c)$, 
one obtains
\begin{align}\label{eq:Herm-VA-Nvertex}
	V_{c}^{\mu_1\cdots\mu_N,\nu}
	(Q_2,Q_1;\{K_i\})
	&=
	(-1)^{E_a(c)+1}
	\Big[
	V_{\tilde{c}}^{\mu_1\cdots\mu_N,\nu}
	(Q_1,Q_2;\{-K_i\})
	\Big]^*,
	\nonumber\\
	A_{c}^{\mu_1\cdots\mu_N,\nu}
	(Q_2,Q_1;\{K_i\})
	&=
	(-1)^{E_a(c)+1}
	\Big[
	A_{\tilde{c}}^{\mu_1\cdots\mu_N,\nu}
	(Q_1,Q_2;\{-K_i\})
	\Big]^* .
\end{align}
where 
$
\gamma_0\gamma_\nu^\dagger\gamma_0=\gamma_\nu$ and $\gamma_0(\gamma_\nu\gamma_5)^\dagger\gamma_0
=\gamma_\nu\gamma_5$ are used.
\section{ One-Loop Order Corrections}\label{one-loop calculations}
Before giving the one-loop derivation, we collect the notation used in Apps.~D--F. We follow the conventions introduced in Sec.~\ref{Real-time formalism} and Tab.~\ref{tab:conventions}. In this table, the first block applies throughout, while the second block is used only for the two-loop computations discussed in the following appendices.

		\begin{table}[h]
		\caption{The self-energy and vertex conventions. The first block applies generally, while the second block is used only for the two-loop computations.}
		\label{tab:conventions}
		\centering
		\begin{tabular}{p{0.30\textwidth}p{0.7\textwidth}}
			\toprule
			Convention & Description \\
			\midrule
			Virtual corrections & $\Sigma_{X}$: self-energy; $\Gamma^\mu_X$: vertex; $X$ denotes topology type, gauge sectors, and statistical sectors. \\
			Dirac structures &  $\mathcal{S}$: self-energy; $\mathcal{V}$: vertex.\\
		Subscripts on $\mathcal{S}_{(x_s)}$ and $\mathcal{V}_{(x_v)}$& The subscripts denote that the shift momenta are applied to the corresponding Dirac numerator.\\
			Composite operations & $S_{(a_s g_s)}$: sequential application, first $a_s$ and then $g_s$, on $S$.\\
			Prime notation & $\mathcal{S}'_{(a_sc_s)}$: $P \leftrightarrows L$ transformation applied to $\mathcal{S}{(a_sc_s)}$.\\
			Compact momentum notations&  Plus signs in internal propagators are omitted; e.g., $\Delta_j(PLQ):=\Delta_j(P+L+Q)$ with $j=R,A,s$.\\
			\midrule
			Metric sector& $\Sigma_{{X}_{11}}$,\quad $\Gamma^\mu_{X_{11}}$.\\
			Metric--longitudinal  $(1-\xi)$& $\Sigma_{X_{12}}+\Sigma_{X_{21}}$,\quad $\Gamma^\mu_{X_{12}}+\Gamma^\mu_{X_{21}}$.\\
			Fully longitudinal $(1-\xi)^2$& $\Sigma_{X_{22}}$,\quad $\Gamma^\mu_{X_{22}}$ .\\
			Dirac structures in new basis & self-energy:  $\left\{{\mathcal{S}_{11},\mathcal{S}_{12}/\mathcal{S}_{21},\mathcal{S}_{22}}\right\}$,\quad vertex:$\left\{{\mathcal{V}_{11}, \mathcal{V}_{12}/\mathcal{V}_{21}, \mathcal{V}_{22}}\right\}$.	\vspace{0.15cm}\\
			Statistical sectors & $N_B(P)N_B(L)$, $N_B(P)N_F(L)$, $N_F(L)N_F(P)$ \\
			\bottomrule
		\end{tabular}
	\end{table}
	\begin{figure}
	\centering
	\includegraphics[width=\textwidth]{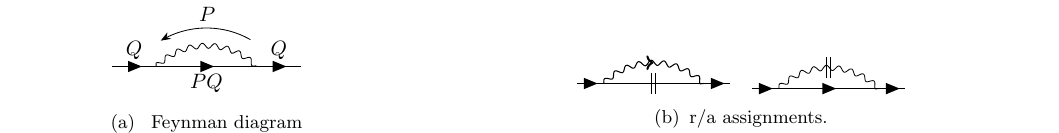}
	\caption{One-loop Feynman diagram and $r/a$ assignments.}\label{Fig:one-loop-self-energy}
	\phantomsubcaption
	\phantomsubcaption\label{Fig: r/a one-loop self-energy}
	\hrule
\end{figure}
	The $r/a$ assignments for the one-loop fermion self-energy, shown in Fig.~\ref{Fig:one-loop-self-energy}, are drawn using the propagator conventions of Fig.~\ref{Free propagators}. Together with Eq.~\eqref{5} and the ordinary QED Feynman rules, they give
	\begin{align}\label{eq:C1}
		-i\Sigma^{(1)}_{m}(Q)
		&= e^2\!\int_P \, \mathcal S_{m}
		\Big[\Delta_s(PQ)\Delta_A(P)+\tilde\Delta_s(P)\Delta_R(PQ)\Big],
		\nonumber\\
		-i\Sigma^{(1)}_{\xi}(Q)
		&= i e^2(1-\xi)\!\int_P \, \mathcal S_{\xi}
		\Big[\Delta_s(PQ)\Delta_A^{(2)}(P)+\tilde\Delta_s^{(2)}(P)\Delta_R(PQ)\Big],
	\end{align}
	with Dirac structures
	\begin{equation}\label{Dirac-one-loop}
		\mathcal S_{m}=\gamma_\mu(\slashed P+\slashed Q)\gamma^\mu,
		\qquad
		\mathcal S_{\xi}=\slashed P(\slashed P+\slashed Q)\slashed P.
	\end{equation}   
Defining the shift $a_s: P\to P-Q$,
	we shift the loop momentum in the first term so that the symmetric propagator depends only on \(P\). This allows us to impose the on-shell condition through $\Delta_d(P)$:
	\begin{align}\label{basic-self-energy}
		\Sigma^{(1)}_{m}(Q)
		&=i e^2\!\int_P
		\Big[\mathcal S_{m(as)}\,\Delta_s(P)\Delta_A(P-Q)+\mathcal S_{m}\,\tilde\Delta_s(P)\Delta_R(PQ)\Big],
		\nonumber\\
		\Sigma^{(1)}_{\xi}(Q)
		&=- e^2(1-\xi)\!\int_P
		\Big[\mathcal S_{\xi(as)}\,\Delta_s(P)\Delta_A^{(2)}(P-Q)+\mathcal S_{\xi}\,\tilde\Delta_s^{(2)}(P)\Delta_R(PQ)\Big].
	\end{align}
For the second term in the last line, we use Eq.~\eqref{a3}:
	\begin{align}\label{C4}
		\int_P \tilde{\Delta}_s^{(2)}(P) \mathcal{S}_\xi \Delta_R(PQ)=i\int_P \Delta_d(P)\frac{d}{dp^0}\Bigl[\frac{N_B(P)\mathcal{S}_\xi\Delta_R(PQ)}{2p^0}\Bigr].
	\end{align}
	In this case, the expression inside the brackets is differentiated with respect to $p^0$. The on-shell condition is then applied, and the HTL expansion is performed at the end.
	\paragraph{\bf LP Order}
		Following the expansion procedure explained in  Sec.~\ref{fermion self-energy}, one obtains the LP contribution as
	\begin{align}\label{A2}
		\Sigma_{m}^{0(1;LP)}=\frac{1}{4}\mathrm{Tr}\left[\gamma^0  \Sigma_{m}^{(1;LP)}(Q)\right]&=-\frac{ e^2\mathcal{D}}{4}\int_P \Delta_d(P)\left[N_B(P)-N_F(P)\right]\,\frac{{-2p^0}}{P.Q}\no\\
		&=\frac{e^2\mathcal{D}}{2}\mathcal{N}\int_{\vec{p}}\frac{1}{p}\left[N_B(P)-N_F(P)\right]\int_z \frac{1}{v.Q}\no\\
		&=\frac{e^2T^2}{8}L(Q)\equiv m^2(T)\, L(Q),
	\end{align} 
where the functions given in App.~\ref{DR} have been used. The spatial component is similarly given by
	\begin{align}\label{A3}
		\Sigma_{m}^{\,i (1;LP)}=-\frac{1}{4}\mathrm{Tr}\left[\gamma^i\Sigma^{(1;LP)}_{m}(Q)\right]
		&=-\frac{e^2}{2}\mathcal{D}\mathcal{N}(\mathcal{R}_2+\mathcal{R}_1)\mathcal{A}^i_1= m^2(T)\,\frac{\hat{q}^i}{q} \left[1-q^0 L(Q)\right].
	\end{align} 
At this power order, the longitudinal-sector integrand is odd under the on-shell momentum reversal and therefore vanishes.
	\paragraph{\bf NLP Order}
At NLP order, we first compute the temporal and spatial projections and then recombine them into Dirac form. 
The expanded results are
	\begin{subequations}\label{Eq:NLP self-energy}
		\begin{align}
			\Sigma^{(1;NLP)}_{\rm m}(Q)
			&=\frac{e^2}{4}\mathcal{D}\int_P \Delta_d(P)\biggl\{\left[N_B(P)-N_F(P)\right]\frac{\slashed{P}\,Q^4}{2 (P\!\cdot\! Q)^3}
			-N_B(P)\frac{\slashed{Q}\,Q^2 }{(P\!\cdot\! Q)^2}\biggr\},\label{NLP-ph-self}\\
			\Sigma^{(1;NLP)}_{\xi}(Q)
			&=\frac{e^2}{4}(1-\xi)\int_P \Delta_d(P)\biggl\{\left[N_B(P)+N_F(P)\right]\Bigl[\frac{2 Q^2 \slashed{Q}}{(P\!\cdot\! Q)^2}-\frac{Q^4 \slashed{P}}{(P\!\cdot\! Q)^3}\Bigr]\nonumber\\
			&\hspace{2.7cm}+N_B(P)\Bigl[\frac{Q^2 \slashed{P}}{({p^0})^2 (P\!\cdot\! Q)}+\frac{Q^2 q^0 \slashed{P}}{{p^0} (P\!\cdot\! Q)^2}-\frac{Q^2 \slashed{Q}}{(P\!\cdot\! Q)^2}-\frac{\gamma^0 Q^2}{{p^0} (P\!\cdot\! Q)}\Bigr]
			-\frac{Q^2 \slashed{P} }{{p^0} (P\!\cdot\! Q)} N^\prime_B(P)\biggr\}\label{NLP-xi-self},
		\end{align}
	\end{subequations}
	with $\mathcal{D}=D-2$.
The integrated results are shown in Eq.~\eqref{Eq: NLP final results}, where the radial and angular integrations in App.~\ref{DR} are used.
\section{Two-Loop Fermion Self-Energy: Intermediate Expressions}\label{two-loop self-energy computations}
We present the intermediate steps for obtaining pinch-free pre-HTL results for the two-loop self-energy, using the conventions defined below.
The loop-momentum transformations used in this appendix are
\begin{align}\label{self-energy transformation}
	&a_s: P\rightarrow P-Q,\qquad 
	b_s: L\rightarrow L-Q,\qquad 
	c_s: L\rightarrow L-P,\qquad
	d_s: P\rightarrow P-L,
	\nonumber\\
	&e_s: L\rightarrow L-P-Q,\qquad
	f_s: P\rightarrow P-L-Q,\qquad
	g_s: L\rightarrow L-P+Q .
\end{align}
When a label such as $(a_s)$ is attached to a Dirac numerator, it denotes the corresponding substitution in Eq.~\eqref{self-energy transformation} applied to that numerator.
Composite labels denote sequential substitutions, as summarized in Tab.~\ref{tab:conventions}.
A prime on a Dirac structure denotes the exchange $P\leftrightarrow L$ acting only on the Dirac structure.
We use a compact notation to treat the three gauge sectors discussed in Sec.~\ref {Intermediate self-energy} simultaneously.

Let the pair $(i,j)$, with $i\in\{1,2\}$ and $j\in \{1,2\}$, represent the gauge sectors as
\begin{align}\label{gauge sector map app}
	\Sigma_{X_{11}}\equiv \Sigma_{X_{m}},\qquad
	\Sigma_{X_{12}}+\Sigma_{X_{21}}\equiv \Sigma_{X_{\xi}},\qquad
	\Sigma_{X_{22}}\equiv \Sigma_{X_{\xi^2}},\quad X\in\{R,B,C\},
\end{align}
where $X$ labels the topology: $R$, $B$, and $C$ denote the rainbow, bubble, and cross-photon topologies, respectively.
The corresponding gauge factors are then
\begin{align}\label{gauge factors app}
	\hat{\xi}_{11}\equiv 1,\qquad
	\hat{\xi}_{12}=\hat{\xi}_{21}\equiv i(1-\xi),\qquad
	\hat{\xi}_{22}\equiv -(1-\xi)^2 .
\end{align}
To write the Dirac numerators compactly, we introduce
\begin{align}\label{compact Dirac}
	\mathcal{A}_1 \equiv \gamma^\alpha,\quad
	\mathcal{A}_2 \equiv \slashed{P},\qquad
	\mathcal{B}_1 \equiv \gamma^\nu,\quad
	\mathcal{B}_2 \equiv \slashed{L},\qquad
	\mathcal{C}_1 \equiv \gamma^\nu,\quad
	\mathcal{C}_2 \equiv \slashed{P},\qquad
	\mathcal{D}_1 \equiv \gamma^\alpha,\quad
	\mathcal{D}_2 \equiv \slashed{L},
\end{align}
with repeated Lorentz indices contracted. For example, the Dirac numerator corresponding to the rainbow self-energy is
\begin{align}\label{rainbow compact dirac app}
	\mathcal{S}_{R_{ij}}
	=
	\mathcal{A}_i(\slashed{P}+\slashed{Q})
	\mathcal{B}_j(\slashed{P}+\slashed{L}+\slashed{Q})
	\mathcal{B}_j(\slashed{P}+\slashed{Q})
	\mathcal{A}_i,
	\qquad i,j=1,2,
\end{align}
which gives the explicit structures in Eq.~\eqref{Dirac rainbow}.
The same compact organization is used throughout for the remaining two-loop self-energy and vertex topologies. These conventions are summarized in Tab.~\ref{tab:conventions}.

The different gauge contributions arise from the photon propagators. For the two internal photon lines, one assigns the labels $i$ and $j$ to indicate the gauge structure. The Dirac numerator $\mathcal{S}_{ij}$ must be matched to the corresponding explicit Dirac numerators; see, for instance, Eq.~\eqref{Dirac rainbow}. Afterward, the remaining steps proceed in the same way, and this notation retains all gauge sectors simultaneously.
\subsection{The Rainbow Self-Energy}
\begin{figure}
	\centering
	\includegraphics[width=\textwidth]{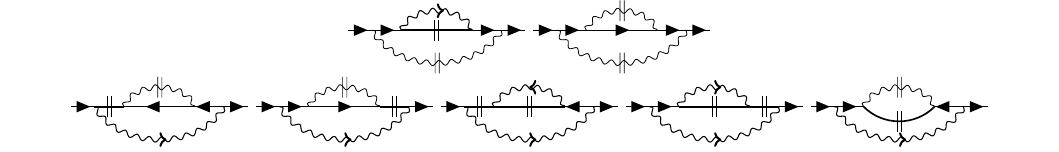}
	\caption{$r/a$ assignments for the rainbow self-energy. Reading the diagrams from left to right gives the terms in the direct sum of Eq.~\eqref{rainbow self-energy direct}.}\label{rainbow self-energy assignments}
	\phantomsubcaption
	\hrule
\end{figure} 
All allowed $r/a$ assignments are shown in Fig.~\ref{rainbow self-energy assignments}. 
Using the notation above, the sum of $r/a$ assignments gives
\begin{align}\label{rainbow self-energy direct}
	-i\Sigma^{(2)}_{R_{ij}}(Q)
	&=
	e^4 \int_P\int_L 
	\hat{\xi}_{ij}\,
	\mathcal{S}_{R_{ij}}
	\Bigl\{
	\tilde{\Delta}^{(i)}_s(P)\Delta_R^{(2)}(PQ)
	\Bigl[
	\Delta_s(PLQ)\Delta_A^{(j)}(L)
	+\tilde{\Delta}^{(j)}_s(L)\Delta_R(PLQ)
	\Bigr]
	\nonumber\\
	&\quad
	+\Delta_s(PQ)\tilde{\Delta}^{(j)}_s(L)\Delta_A^{(i)}(P)
	\Bigl[
	\Delta_A(PLQ)\Delta_A(PQ)
	+\Delta_R(PLQ)\Delta_R(PQ)
	\Bigr]
	\nonumber\\
	&\quad
	+\Delta_s(PQ)\Delta_s(PLQ)\Delta_A^{(i)}(P)
	\Bigl[
	\Delta_R^{(j)}(L)\Delta_A(PQ)
	+\Delta_R(PQ)\Delta_A^{(j)}(L)
	\Bigr]
	\nonumber\\
	&\quad
	+\Delta_s(PLQ)\tilde{\Delta}^{(j)}_s(L)
	\Delta_R(PQ)\Delta_A(PQ)\Delta_A^{(i)}(P)
	\Bigr\}.
\end{align}
The last line contains the pinch structure $\Delta_R(PQ)\Delta_A(PQ)$. 
To remove the pinch structure, we first sum the contributions from the second, third, and fourth lines. We then use the KMS relation, Eq.~\eqref{KMS relation}, to rearrange them into a pinch-free form. In the diagrammatic representation, these terms correspond to the five diagrams shown in the second block of Fig.~\ref{rainbow self-energy assignments}. Afterward, the pinch-free expression becomes
\begin{align}\label{rainbow self-energy pinchfree}
	-i\Sigma^{(2)}_{R_{ij}}(Q)
	&=
	e^4 \int_P\int_L 
	\hat{\xi}_{ij}\,
	\mathcal{S}_{R_{ij}}
	\Bigl\{
	\tilde{\Delta}^{(i)}_s(P)\Delta_R^{(2)}(PQ)
	\Bigl[
	\Delta_s(PLQ)\Delta_A^{(j)}(L)
	+\tilde{\Delta}^{(j)}_s(L)\Delta_R(PLQ)
	\Bigr]
	\nonumber\\
	&\quad
	+\tilde{\Delta}^{(j)}_s(L)\Delta_A^{(i)}(P)N_F(PQ)
	\Bigl[
	\Delta_R^{(2)}(PQ)\Delta_R(PLQ)
	-\Delta_A^{(2)}(PQ)\Delta_A(PLQ)
	\Bigr]
	\nonumber\\
	&\quad
	+\Delta_s(PLQ)\Delta_A^{(i)}(P)N_F(PQ)
	\Bigl[
	\Delta_R^{(2)}(PQ)\Delta_A^{(j)}(L)
	-\Delta_A^{(2)}(PQ)\Delta_R^{(j)}(L)
	\Bigr]
	\Bigr\}.
\end{align}
We next rewrite the integrand so that every spectral or difference function depends only on $P$ or only on $L$. 
The second and third lines of Eq.~\eqref{rainbow self-energy pinchfree} contain only one explicit spectral factor. 
The second on-shell structure is generated by adding and subtracting advanced propagators, equivalently by introducing difference propagators. 
For example, for the second line we can write
\begin{align}\label{rainbow difference example}
	X
	&=
	\tilde{\Delta}^{(j)}_s(L)\Delta_A^{(i)}(P)N_F(PQ)
	\Bigl[
	\Delta_R^{(2)}(PQ)\Delta_R(PLQ)
	-\Delta_A^{(2)}(PQ)\Delta_A(PLQ)
	\Bigr]
	\nonumber\\
	&=
	\tilde{\Delta}^{(j)}_s(L)\Delta_A^{(i)}(P)N_F(PQ)
	\Bigl[
	\Delta_d^{(2)}(PQ)\Delta_R(PLQ)
	+\Delta_A^{(2)}(PQ)\Delta_d(PLQ)
	\Bigr].
\end{align}
The same rearrangement applies to the third line of Eq.~\eqref{rainbow self-energy pinchfree}. After the momentum shifts in Eq.~\eqref{self-energy transformation} are performed, the rainbow contribution becomes
\begin{align}\label{rainbow compact}
	-i\Sigma^{(2)}_{R_{ij}}(Q)
	&=
	e^4 \int_P\int_L \hat{\xi}_{ij}\no\\
	&\times\Bigl\{
	\Delta_d^{(i)}(P)N_B(P)\Delta_R^{(2)}(PQ)
	\Bigl[
	\Delta_d(L)N_F(L)\Delta_A^{(j)}(L-P-Q)\,
	\mathcal{S}_{R_{ij}(e_s)}
	+\Delta_d^{(j)}(L)N_B(L)\Delta_R(PLQ)\,
	\mathcal{S}_{R_{ij}}
	\Bigr]
	\nonumber\\
	&\quad
	+\Delta_d^{(2)}(L)N_F(L)\Delta_A^{(i)}(L-Q)
	\Bigl[
	\Delta_d^{(j)}(P)N_B(P)\Delta_R(PL)\,
	\mathcal{S}'_{R_{ij}(a_s)}
	+\Delta_d(P)N_F(P)\Delta_A^{(j)}(P-L)\,
	\mathcal{S}'_{R_{ij}(a_sc_s)}
	\Bigr]
	\nonumber\\
	&\quad
	+\Delta_d(L)\Delta_d^{(j)}(P)N_F(L)N_B(P)
	\Delta_A^{(i)}(L-P-Q)\Delta_A^{(2)}(L-P)\,
	\mathcal{S}'_{R_{ij}(f_s)}
	\Bigr\}.
\end{align}
The on-shell condition $P^2=L^2=0$ is imposed for terms proportional to $\Delta_d(P)\Delta_d(L)$. Higher difference propagators, such as $\Delta_d^{(2)}(L)$ or products of the form $\Delta_d^{(n)}(P)\Delta_d^{(m)}(L)$, are first converted to spectral functions using Eq.~\eqref{a3}. For example, the metric component of the second line in the equation above becomes
\begin{align}\label{rainbow derivative example}
	i\int_P \Delta_d(P)\int_L \Delta_d(L)
	\frac{d}{dl^0}
	\Biggl\{
	\frac{\Delta_A(L-Q)N_F(L)}{2l^0}
	\Bigl[
	N_B(P)\Delta_R(PL)\mathcal{S}'_{R_{11}(a_s)}
	+N_F(P)\Delta_A(P-L)\mathcal{S}'_{R_{11}(a_sc_s)}
	\Bigr]
	\Biggr\}.
\end{align}
The derivative acts only on the $l^0$-dependent factors inside the curly bracket. Since no derivative acts on $p^0$, one may first impose $P^2=0$. The $l^0$ derivative is then taken, after which $L^2=0$ is imposed. Finally, the HTL expansion is carried out. Equivalently, one may take the $l^0$ derivative, and then impose $P^2=L^2=0$; the result is unchanged. The same prescription applies to the remaining sectors.
\subsection{The Cross-Photon and Bubble Topologies}\label{app: the remaining topologies of the self-energy}
\begin{figure}
	\centering
	\includegraphics[width=\textwidth]{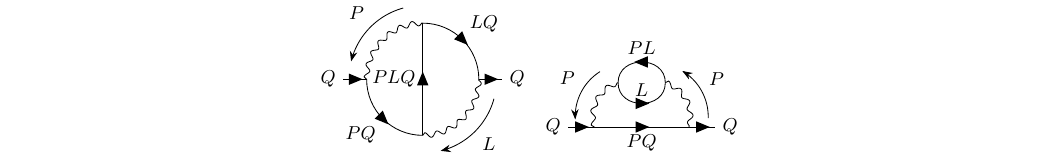}
	\caption{From left to right: cross-photon and bubble diagrams.}
	\label{topologies of fermion self-energy}
	\phantomsubcaption
	\hrule
\end{figure}
For the cross-photon and bubble topologies (see Fig. \ref{topologies of fermion self-energy}), we obtain the final results as
\begin{subequations}\label{self-energy compact}
\begin{align}
	&-i\Sigma^{(2)}_{C_{ij}}(Q)=e^4 \int_P\int_L\hat{\xi}_{ij}	\nonumber\\
	&\times\biggl\{	
	\Delta_d^{(j)}(L)\Delta_d^{(i)}(P)N_B(L)N_B(P)\Delta_{R}(PQ)\Delta_{R}(PLQ)\Delta_{R}(LQ)\mathcal{S}_{C_{ij}}\nonumber\\
	&\quad+\Delta_d(L)\Delta_d(P)N_F(L)N_F(P)\Bigl\{\Delta_A^{(i)}(P-Q)\Delta_A^{(j)}(L-Q)\Delta_A(P+L-Q)\mathcal{S}_{C_{ij}(a_sb_s)}\nonumber\\
	&
	\hspace{3cm}+\Delta_{R}(L-P+Q)\left[\Delta_A^{(i)}(P-Q)\Delta_{R}^{(j)}(L-P)\mathcal{S}_{C_{ij}(a_sc_s)}+\Delta_A^{(j)}(P-Q)\Delta_A^{(i)}(L-P)\mathcal{S}^\prime_{C_{ij}(b_sd_s)}\right]\Bigr\}\nonumber\\
	&\quad+\Delta_d(L)\Delta_d^{(i)}(P)N_F(L)N_B(P)\Delta_{R}(PQ)\Bigl\{\Delta_A^{(j)}(L-Q)\Delta_{R}(PL)\mathcal{S}_{C_{ij}(b_s)}+\Delta_A^{(j)}(L-P-Q)\Delta_A(L-P)\mathcal{S}_{C_{ij}(e_s)}\Bigr\}\no\\
	&\quad+\Delta_d(L)\Delta_d^{(j)}(P)N_F(L)N_B(P)\Delta_{R}(PQ)\Bigl\{\Delta_A^{(i)}(L-Q)\Delta_A(PL)\mathcal{S}^\prime_{C_{ij}(a_s)}+\Delta_A^{(i)}(L-P-Q)\Delta_{R}(L-P)\mathcal{S}^\prime_{C_{ij}(f_s)}\Bigr\}\biggr\},\\
	&-i\Sigma^{(2)}_{B_{ij}}(Q)=-e^4 \int_P\int_L\hat{\xi}_{ij}\Delta_d(L) \nonumber\\
	&\times\biggl\{	\Delta_d^{(i+j)}(P)N_F(L)N_B(P)\Delta_R(PQ)\left[\Delta_{R}(PL)\mathcal{S}_{B_{ij}}+\Delta_A(L-P)\mathcal{S}_{B_{ij}(c_s)}\right]\nonumber\\
	&\quad
	+\Delta_d(P)N_F(L)N_F(P)\Bigl[\Delta_{A}^{(i+j)}(P-Q)\left[\Delta_{R}(L-P+Q)\mathcal{S}_{B_{ij}(a_sg_s)}+\Delta_A(P+L-Q)\mathcal{S}_{B_{ij}(a_s)}\right]\nonumber\\
	&\hspace{3.7cm}+\Delta_{R}(P-L+Q)\Delta_{A}^{(i+j)}(P-L)\mathcal{S}_{B_{ij}(d_s)} \Bigr]\biggr\},\label{eq:bubble_xi2_SE}
\end{align}
\end{subequations}
where the Dirac structures are 
\begin{align}
	&\mathcal{S}_{C_{ij}}\equiv\mathcal{A}_i(\slashed{P}+\slashed{Q})\mathcal{B}_j(\slashed{P}+\slashed{L}+\slashed{Q})\mathcal{A}_i(\slashed{L}+\slashed{Q})\mathcal{B}_j,\qquad \mathcal{S}_{B_{ij}}\equiv\mathcal{A}_i(\slashed{P}+\slashed{Q})\mathcal{C}_j\,\mathrm{Tr}\left[\mathcal{A}_i\,\slashed{L}\,\mathcal{C}_j(\slashed{P}+\slashed{L})\right]
\end{align}
The bubble contribution also contains pinch structures at an intermediate stage. After using the KMS relation, the final bubble expressions written above are pinch-free.

Now, consider the purely longitudinal sector, $i=j=2$. In this case, the second line of Eq.~\eqref{eq:bubble_xi2_SE} can be rewritten in spectral form with derivatives as
\begin{align}\label{D15}
	\frac{(-i)^5}{3!}\,N_F(L)\Delta_d(L)\Delta_d(P)\Bigl\{2p^0\Bigl[\frac{-1}{2p^0}\frac{d}{dp^0}\Bigr]^3
	\Bigl[-\frac{N_B(P)\Delta_R(PQ)}{2p^0}\bigl(\Delta_{R}(PL)\mathcal{S}_{B_{ij}}+\Delta_A(L-P)\mathcal{S}_{B_{ij}(c_s)}\bigr)\Bigr]\Bigr\}.
\end{align}
Since $\Delta_d(L)$ enforces $L^2=0$, the contribution above vanishes after evaluation. Therefore, it is useful to impose the on-shell condition associated with the spectral function first and then take the derivatives with respect to the other momentum. The same prescription is used for the remaining sectors and for the vertex corrections.
\section{Two-Loop Order Pre-HTL Vertex}\label{intermediate vertex}
\begin{figure}[htbp]
	\includegraphics[width=\textwidth]{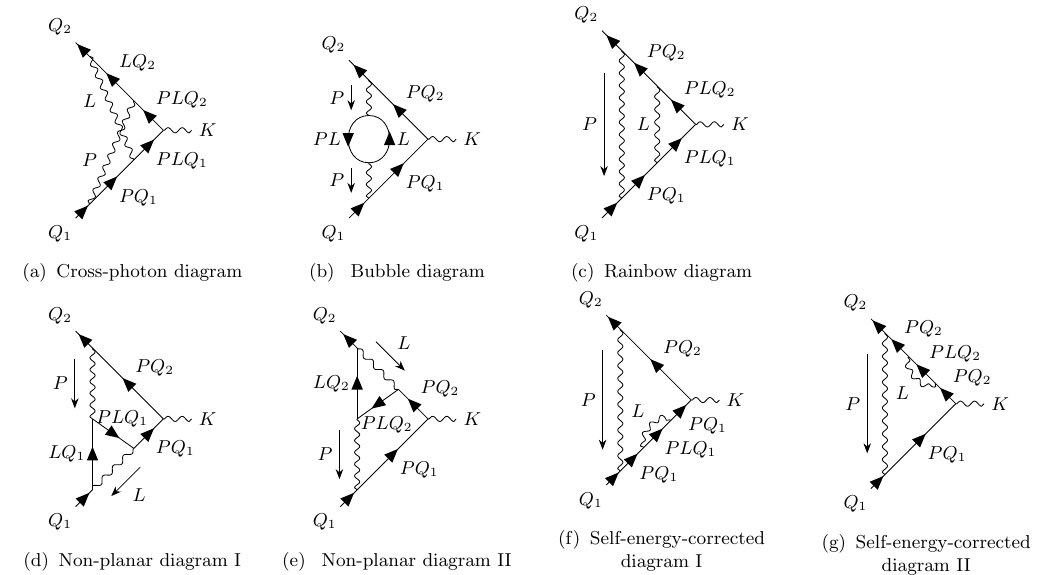}
	\caption{Distinct topologies contributing to the vertex diagram at two-loop order.}\label{vertex diagrams at two-loop order}
	\phantomsubcaption\label{cross-photon vertex diagram}
	\phantomsubcaption\label{bubble vertex diagram}
	\phantomsubcaption\label{rainbow vertex diagram}
	\phantomsubcaption\label{non-planar vertex diagram I}
	\phantomsubcaption\label{non-planar vertex diagram II}
	\phantomsubcaption\label{Self-energy-corrected diagram I}
	\phantomsubcaption
	\hrule
\end{figure}
This appendix collects the pinch-free pre-HTL expressions for the fermion--photon vertex. The seven two-loop topologies are shown in Fig.~\ref{vertex diagrams at two-loop order}. As discussed in Sec.~\ref{vertex two-loop section}, the type-II non-planar and self-energy-corrected diagrams are generated from the corresponding type-I diagrams by the exchange $Q_1\leftrightarrow Q_2$. The bubble and self-energy-corrected contributions contain pinch structures at intermediate stages, which are removed using the KMS relation. We use the following momentum transformations:
\begin{align}\label{vertex transformations}
	&a_v: P\rightarrow P-Q_1,\qquad\quad b_v:L\rightarrow L-Q_2,\quad\qquad c_v:L\rightarrow L-P,\qquad\qquad\qquad\quad
	d_v:P\rightarrow P-L+Q_2-Q_1,\no\\
	&e_v:L\rightarrow L-P-Q_1,\quad
	f_v:P\rightarrow P-L-Q_1,\quad
	g_v:L\rightarrow L-P-Q_2+Q_1,\qquad h_v:P\rightarrow P-L,\no\\
	&j_v: L\rightarrow L-P-Q_2,\quad
	k_v:P\rightarrow P-L-Q_2,\quad m_v: P\rightarrow P-Q_2,\quad\qquad\qquad\quad n_v:L\rightarrow L-P+Q_2-Q_1,\no\\
	&r_v:L\rightarrow L-P+Q_1,\quad s_v:L\rightarrow L-P+Q_2,\quad\hspace{0.2cm}	t_v: L\rightarrow L-Q_1.
\end{align}
When one of these labels is attached to a Dirac numerator, the corresponding substitution is applied to that numerator. The cross-photon, rainbow, bubble, type-I non-planar, and type-I self-energy-corrected contributions are given below,
\begin{subequations}\label{vertex compact}
\begin{align}
	&-ie\Gamma_{C_{ij}}^{\mu(2)}(Q_2,Q_1)=-ie^5\int_P\int_L	\hat{\xi}_{ij}\, \nonumber\\
	&\times\Bigl\{\Delta^{(j)}_d(L)\Delta^{(i)}_d(P)N_B(L)N_B(P)\Delta_{R}(PQ_1)\Delta_{R}(PLQ_1)\Delta_{R}(PLQ_2)\Delta_{R}(LQ_2)\mathcal{V}_{C_{ij}}^\mu\nonumber\\
	&\quad+\Delta_d(L)\Delta_d(P)N_F(L)N_F(P)\Bigl[
	\Delta_A^{(i)}(P-Q_1)\Delta_A(P+L-Q_2)\Delta_A(P+L-Q_1)\Delta_A^{(j)}(L-Q_2)\mathcal{V}_{C_{ij}(a_vb_v)}^\mu\no\\
	&\hspace{4cm}+\Delta_A^{(i)}(P-Q_1)\Delta_{R}^{(j)}(L-P)\Delta_A(L+Q_2-Q_1)\Delta_{R}(L-P+Q_2)\mathcal{V}_{C_{ij}(a_vc_v)}^\mu\nonumber\\
	&\hspace{4cm}+\Delta_{R}(P-L+Q_2)\Delta_A^{(i)}(P-L+Q_2-Q_1)\Delta_A(P+Q_2-Q_1)\Delta_A^{(j)}(L-Q_2)\mathcal{V}_{C_{ij}(b_vd_v)}^\mu\nonumber\\
	&\hspace{4cm}+\Delta_A^{(i)}(P-Q_1)\Delta_{R}^{(j)}(L-P-Q_2+Q_1)\Delta_{R}(L-Q_2+Q_1)\Delta_{R}(L-P+Q_1)\mathcal{V}_{C_{ij}(a_vg_v)}^\mu\nonumber\\
	&\hspace{4cm}+\Delta_{R}(P-L+Q_1)\Delta_A^{(i)}(P-L)\Delta_{R}(P-Q_2+Q_1)\Delta_A^{(j)}(L-Q_2)\mathcal{V}_{C_{ij}(b_vh_v)}^\mu\Bigr]\nonumber\\
	&\quad+\Delta_d(L)\Delta^{(j)}_d(P)N_F(L)N_B(P)\Bigl[\Delta_A^{(i)}(L-Q_1)\Delta_A(PL)\Delta_A(P+L+Q_2-Q_1)\Delta_{R}(PQ_2){\mathcal{V}^\prime}_{C_{ij}(a_v)}^\mu\nonumber\\
	&\hspace{4cm}+\Delta_A^{(i)}(L-P-Q_1)\Delta_{R}(L-P)\Delta_A(L+Q_2-Q_1)\Delta_{R}(PQ_2){\mathcal{V}^\prime}_{C_{ij}(f_v)}^\mu\nonumber\\
	&\hspace{4cm}+\Delta_A^{(i)}(L-P-Q_2)\Delta_{R}(L-P-Q_2+Q_1)\Delta_{R}(L-Q_2+Q_1)\Delta_{R}(PQ_2){\mathcal{V}^\prime}_{C_{ij}(k_v)}^\mu\Bigr]\no\\
	&\quad+\Delta_d(L)\Delta^{(i)}_d(P)N_F(L)N_B(P)\Bigl[
	\Delta_{R}(PQ_1)\Delta_A^{(j)}(L-Q_2)\Delta_{R}(P+L-Q_2+Q_1)\Delta_{R}(PL)\mathcal{V}_{C_{ij}(b_v)}^\mu\nonumber\\
	&\hspace{4cm}+\Delta_{R}(PQ_1)\Delta_A^{(j)}(L-P-Q_1)\Delta_A(L+Q_2-Q_1)\Delta_A(L-P+Q_2-Q_1)\mathcal{V}_{C_{ij}(e_v)}^\mu\nonumber\\
	&\hspace{4cm}+\Delta_{R}(PQ_1)\Delta_{R}(L-Q_2+Q_1)\Delta_A^{(j)}(L-P-Q_2)\Delta_A(L-P)\mathcal{V}_{C_{ij}(j_v)}^\mu\Bigr]
	\Bigr\},\\
			&-ie\Gamma_{R_{ij}}^{\mu(2)}(Q_2,Q_1)=-ie^5\int_P\int_L\hat{\xi}_{ij} \nonumber\\
	&\times\Bigl\{\Delta_d^{(j)}(L)\Delta_d^{(i)}(P)N_B(L)N_B(P)\Delta_{R}(P	Q_1)\Delta_{R}(PLQ_1)\Delta_{R}(PLQ_2)\Delta_{R}(PQ_2)\mathcal{V}_{R_{ij}}^\mu\nonumber\\
	&\quad+\Delta_d(L)\Delta_d(P)N_F(L)N_F(P)\Bigl[\Delta_A^{(i)}(P-Q_1)\Delta_{R}^{(j)}(L-P)\Delta_A(L+Q_2-Q_1)\Delta_A(P+Q_2-Q_1)\mathcal{V}_{R_{ij}(a_vc_v)}^\mu\nonumber\\
	&\hspace{4cm}+\Delta_A^{(i)}(P-Q_2)\Delta_{R}(P-Q_2+Q_1)\Delta_A^{(j)}(L-P+Q_2-Q_1)\Delta_A(L+Q_2-Q_1)\mathcal{V}^\mu_{R_{ij}(m_vn_v)}\nonumber\\
	&\hspace{4cm}+\Delta_A^{(i)}(P-Q_1)\Delta_{R}^{(j)}(L-P-Q_2+Q_1)\Delta_{R}(L-Q_2+Q_1)\Delta_A(P+Q_2-Q_1)\mathcal{V}^\mu_{R_{ij}(a_vg_v)}\nonumber\\
	&\hspace{4cm}+\Delta_{R}(P+Q_1-Q_2)\Delta_A^{(i)}(P-Q_2)\Delta_{R}(L+Q_1-Q_2)\Delta_A^{(j)}(L-P)\mathcal{V}^\mu_{R_{ij}(m_vc_v)}\Bigr]\no\\
	&\quad+\Delta_d(L)\Delta_d^{(j)}(P)N_F(L)N_B(P)\Bigl[\Delta_A^{(i)}(L-Q_1)\Delta_A(PL)\Delta_A(P+L+Q_2-Q_1)\Delta_A(L+Q_2-Q_1){\mathcal{V}^\prime}_{R_{ij}(a_v)}^\mu\nonumber\\
	&\hspace{4cm}+\Delta_{R}(L-Q_2+Q_1)\Delta_{R}(P+L-Q_2+Q_1)\Delta_{R}(PL)\Delta_A^{(i)}(L-Q_2){\mathcal{V}^\prime}^\mu_{R_{ij}(m_v)}\nonumber\\
	&\hspace{4cm}+\Delta_A^{(i)}(L-P-Q_1)\Delta_{R}(L-P)\Delta_A(L+Q_2-Q_1)\Delta_A(L-P+Q_2-Q_1){\mathcal{V}^\prime}^\mu_{R_{ij}(f_v)}\nonumber\\
	&\hspace{4cm}+\Delta_A^{(i)}(L-P-Q_2)\Delta_{R}(L-P-Q_2+Q_1)\Delta_{R}(L-Q_2+Q_1)\Delta_A(L-P){\mathcal{V}^\prime}^\mu_{R_{ij}(k_v)}\Bigr]\no\\
	&\quad+\Delta_d(L)\Delta_d^{(i)}(P)N_F(L)N_B(P)\Bigl[\Delta_{R}(PQ_1)\Delta_A^{(j)}(L-P-Q_1)\Delta_A(L+Q_2-Q_1)\Delta_{R}(PQ_2)\mathcal{V}^\mu_{R_{ij}(e_v)}\nonumber\\
	&\hspace{4cm}+\Delta_{R}(PQ_1)\Delta_{R}(L-Q_2+Q_1)\Delta_A^{(j)}(L-P-Q_2)\Delta_{R}(PQ_2)\mathcal{V}^\mu_{R_{ij}(j_v)}\Bigr]\Bigr\},\\
		&-ie\Gamma_{B_{ij}}^{\mu(2)}(Q_2,Q_1)=+ie^5\int_P\int_L\hat{\xi}_{ij}\Delta_d(L)\nonumber\\
	&\times\Bigl\{\Delta^{(i+j)}_d(P)N_F(L)N_B(P)\Delta_{R}(P	Q_1)\Delta_{R}(PQ_2)\left[\Delta_{R}(PL)\mathcal{V}^\mu_{B_{ij}}+\Delta_A(L-P)\mathcal{V}^\mu_{B_{ij}(c_v)}\right]\nonumber\\
	&+\Delta_d(P) N_F(L)N_F(P)\Big[\Delta_{R}(P-L+Q_1)\Delta_{R}(P-L+Q_2)\Delta_{A}^{(i+j)}(P-L)\mathcal{V}^\mu_{B_{ij}(h_v)}\no\\
	&\quad\qquad\qquad\qquad+
	\Delta_A(P+Q_2-Q_1)\Delta_{A}^{(i+j)}(P-Q_1)\Big(\Delta_{R}(L-P+Q_1)\mathcal{V}^\mu_{B_{ij}(a_vr_v)}+\Delta_A(P+L-Q_1)\mathcal{V}^\mu_{B_{ij}(a_v)}\Big)\nonumber\\
	&\quad\qquad\qquad\qquad+\Delta_{R}(P-Q_2+Q_1)\Delta_{A}^{(i+j)}(P-Q_2)\Big(\Delta_{R}(L-P+Q_2)\mathcal{V}^\mu_{B_{ij}(m_vs_v)}+\Delta_A(P+L-Q_2)\mathcal{V}^\mu_{B_{ij}(m_v)}\Big)\Big]\Bigr\},\\
	&-ie\Gamma_{N_{I_{ij}}}^{\mu(2)}(Q_2,Q_1)=-ie^5\int_P\int_L\hat{\xi}_{ij} \nonumber\\
&\times\Bigl\{	\Delta_d^{(i)}(L)\Delta_d^{(j)}(P)N_B(L)N_B(P)\Delta_{R}(LQ_1)\Delta_{R}(PLQ_1)\Delta_{R}(PQ_1)\Delta_{R}(PQ_2)\mathcal{V}^\mu_{N_{I_{ij}}}\nonumber\\
&\quad+\Delta_d(L)\Delta_d(P)N_F(L)N_F(P)\Bigl[\Delta_{R}^{(j)}(P-L)\Delta_A^{(i)}(L-Q_1)\Delta_{R}(P-L+Q_1)\Delta_{R}(P-L+Q_2)\mathcal{V}^\mu_{N_{I_{ij}}(t_vh_v)}\no\\	&\hspace{3.5cm}+\Delta_A^{(i)}(L-Q_1)\Delta_A(P+L-Q_1)\Delta_A^{(j)}(P-Q_1)\Delta_A(P+Q_2-Q_1)\mathcal{V}^\mu_{N_{I_{ij}}(t_va_v)}\nonumber\\
&\hspace{3.5cm}+\Delta_A^{(i)}(L-Q_1)\Delta_A(P+L-Q_2)\Delta_A^{(j)}(P-Q_2)\Delta_{R}(P-Q_2+Q_1)\mathcal{V}^\mu_{N_{I_{ij}}(t_vm_v)}\nonumber\\
&\hspace{3.5cm}+\Delta_A^{(i)}(L-P)\Delta_{R}(L-P+Q_1)\Delta_A^{(j)}(P-Q_1)\Delta_A(P+Q_2-Q_1)\mathcal{V}^\mu_{N_{I_{ij}}(a_vc_v)}\nonumber\\
&\hspace{3.5cm}+\Delta_{R}(L-P+Q_2)\Delta_A^{(i)}(L-P+Q_2-Q_1)\Delta_{R}(P-Q_2+Q_1)\Delta_A^{(j)}(P-Q_2)\mathcal{V}^\mu_{N_{I_{ij}}(m_vn_v)}\Bigr]\no\\
&\quad+\Delta_d(L)\Delta_d^{(i)}(P)N_F(L)N_B(P)\Bigl[
\Delta_{R}(PQ_1)\Delta_A^{(j)}(L-P-Q_1)\Delta_A(L-P+Q_2-Q_1)\Delta_A(L-P){\mathcal{V}^\prime}^\mu_{N_{I_{ij}}(f_v)}\nonumber\\
&\hspace{3.5cm}+\Delta_{R}(PQ_1)\Delta_{R}(PL)\Delta_A^{(j)}(L-Q_1)\Delta_A(L+Q_2-Q_1){\mathcal{V}^\prime}^\mu_{N_{I_{ij}}(a_v)}\nonumber\\
&\hspace{3.5cm}+\Delta_{R}(PQ_1)\Delta_{R}(P+L-Q_2+Q_1)\Delta_A^{(j)}(L-Q_2)\Delta_{R}(L-Q_2+Q_1){\mathcal{V}^\prime}^\mu_{N_{I_{ij}}(m_v)}\nonumber\Bigr]\no\\
&\quad+\Delta_d(L)\Delta_d^{(j)}(P)N_F(L)N_B(P)\Bigl[\Delta_A^{(i)}(L-Q_1)\Delta_{R}(PQ_1)\Delta_{R}(PQ_2)\Delta_A(PL)\mathcal{V}^\mu_{N_{I_{ij}}(t_v)}\nonumber\\
&\hspace{3.5cm}+\Delta_{R}(L-P)\Delta_A^{(i)}(L-P-Q_1)\Delta_{R}(PQ_1)\Delta_{R}(PQ_2){\mathcal{V}}^\mu_{N_{I_{ij}}(e_v)}\Bigr]
\Bigr\},\\
	&-ie\Gamma_{S_{I_{ij}}}^{\mu(2)}(Q_2,Q_1)=-ie^5\int_P\int_L \hat{\xi}_{ij}\nonumber\\
	&\times\Bigl\{\Delta^{(j)}_d(L)\Delta^{(i)}_d(P)N_B(L)N_B(P)\Delta_{R}^{(2)}(PQ_1)\Delta_{R}(PLQ_1)\Delta_{R}(PQ_2){\mathcal{V}}^\mu_{S_{I_{ij}}}\no\\
	&\quad+\Delta_d(L)\Delta_d(P)N_F(L)N_F(P)\Delta_A^{(i)}(P-Q_2)\Delta_{R}^{(2)}(P-Q_2+Q_1)\Delta_A^{(j)}(L-P+Q_2-Q_1){\mathcal{V}}^\mu_{S_{I_{ij}}(m_vn_v)}\nonumber\\
	&\quad+\Delta_d(L)\Delta^{(2)}_d(P) N_F(L)N_F(P)\Delta_A(P+Q_2-Q_1)\Delta_A^{(i)}(P-Q_1)\Delta_A^{(j)}(L-P)\mathcal{V}^\mu_{S_{I_{ij}}(a_vc_v)}\nonumber\\
	&\quad+\Delta_d(L)\Delta^{(j)}_d(P)N_F(L)N_B(P)\Bigl[\Delta_A^{(i)}(L-P-Q_1)\Delta_A(L-P+Q_2-Q_1)\Delta_{A}^{(2)}(L-P){\mathcal{V}^\prime}^\mu_{S_{I_{ij}}(f_v)}\nonumber\\
	&\quad\hspace{4.5cm}+\Delta_A^{(i)}(L-Q_2)\Delta_{R}^{(2)}(L-Q_2+Q_1)\Delta_{R}(P+L-Q_2+Q_1){\mathcal{V}^\prime}^\mu_{S_{I_{ij}}(m_v)}\Bigr]\no\\
	&\quad+\Delta_d(L)\Delta^{(i)}_d(P)N_F(L)N_B(P)\Delta_{R}^{(2)}(PQ_1)\Delta_A^{(j)}(L-P-Q_1)\Delta_{R}(PQ_2){\mathcal{V}}^\mu_{S_{I_{ij}}(e_v)}\no\\
	&\quad+\Delta^{(2)}_d(L)\Delta^{(j)}_d(P)N_F(L)N_B(P)\Delta_A^{(i)}(L-Q_1)\Delta_A(L+Q_2-Q_1)\Delta_{R}(PL){\mathcal{V}^\prime}^\mu_{S_{I_{ij}}(a_v)}
	\Bigr\},
\end{align}
\end{subequations}
where the Dirac structures are
\begin{subequations}
\begin{align}\label{vertex cross-photon Dirac}
	&\mathcal{V}_{C_{ij}}^\mu\equiv\mathcal{A}_i(\slashed{P}+\slashed{Q_1})\mathcal{B}_j(\slashed{P}+\slashed{L}+\slashed{Q_1})\gamma^\mu(\slashed{P}+\slashed{L}+\slashed{Q_2})\mathcal{A}_i(\slashed{L}+\slashed{Q_2})\mathcal{B}_j,\\
		&\mathcal{V}_{R_{ij}}^\mu\equiv	\mathcal{A}_i(\slashed{P}+\slashed{Q_1})\mathcal{B}_j(\slashed{P}+\slashed{L}+\slashed{Q_1})\gamma^\mu(\slashed{P}+\slashed{L}+\slashed{Q_2})\mathcal{B}_j(\slashed{P}+\slashed{Q_2})\mathcal{A}_i,\\
		&\mathcal{V}_{B_{ij}}^\mu\equiv\mathcal{A}_i(\slashed{P}+\slashed{Q_1})\gamma^\mu(\slashed{P}+\slashed{Q_2})\mathcal{C}_j\,\mathrm{Tr}\left[\mathcal{A}_i\,\slashed{L}\,\mathcal{C}_j(\slashed{P}+\slashed{L})\right],\\
		&\mathcal{V}^\mu_{N_{I_{ij}}}=\mathcal{D}_i(\slashed{L}+\slashed{Q_1})\mathcal{C}_j(\slashed{P}+\slashed{L}+\slashed{Q_1})\mathcal{D}_i(\slashed{P}+\slashed{Q_1})\gamma^\mu(\slashed{P}+\slashed{Q_2})\mathcal{C}_j,\\
		&\mathcal{V}^\mu_{S_{I_{ij}}}=\mathcal{A}_i(\slashed{P}+\slashed{Q_1})\mathcal{B}_j(\slashed{P}+\slashed{L}+\slashed{Q_1})\mathcal{B}_j(\slashed{P}+\slashed{Q_1})\gamma^\mu(\slashed{P}+\slashed{Q_2})\mathcal{A}_i.
\end{align}
\end{subequations}
Higher difference functions are converted to spectral form using Eq.~\eqref{a3}. When products of spectral and difference functions occur, the derivative-free on-shell constraint is imposed first, as in Eq.~\eqref{D15}.

As an explicit example of the soft-expansion bookkeeping, consider the rainbow contribution with $i=j=2$. Terms whose denominators depend on a single soft momentum, either $Q_1$ or $Q_2$, are left unchanged. Terms containing linear combinations of $Q_1$ and $Q_2$ are rewritten in terms of the transfer momentum $K\equiv Q_2-Q_1$. For the second and third terms, this gives
\begin{align}
	\Delta_d(P)\Delta_d(L)N_F(L)N_F(P)&\Bigl[\Delta^2_A(P-Q_1)\Delta_R^2(L-P)\Delta_A(L+K)\Delta_A(P+K)\mathcal{V}_{R_{22}(a_vc_v)}\no\\
	&+\Delta_A^2(P-Q_2)\Delta_A(P-K)\Delta^2_A(L-P+K)\Delta_A(L+K)\mathcal{V}_{R_{22}(m_vn_v)}\Bigr],
\end{align}
where the numerators are rewritten as
\begin{align}
	&\mathcal{V}_{R_{22}(a_vc_v)}=(\slashed{P}-\slashed{Q}_1)\slashed{P}(\slashed{L}-\slashed{P})\slashed{L}\gamma^\mu (\slashed{L}+\slashed{K})(\slashed{L}-\slashed{P})(\slashed{P}+\slashed{K})(\slashed{P}-\slashed{Q}_1),\no\\
	&\mathcal{V}_{R_{22}(m_vn_v)}=(\slashed{P}-\slashed{Q}_2)(\slashed{P}-\slashed{K})(\slashed{L}-\slashed{P}+\slashed{K})\slashed{L}\gamma^\mu (\slashed{L}+\slashed{K})(\slashed{L}-\slashed{P}+\slashed{K})\slashed{P}(\slashed{P}-\slashed{Q}_2).
\end{align}
All other vertex expressions follow the same prescription.
\section{Zero-temperature Counterterm Insertions}\label{APP-ct}
\begin{figure}[t]
	\centering
	\includegraphics[width=\textwidth]{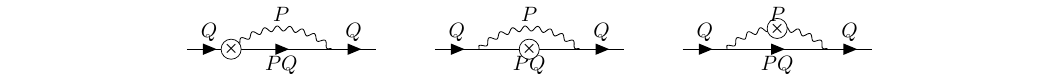}
	\caption{Counterterm insertions, from left to right: vertex insertion, fermion-line insertion, and photon-line insertion. The vertex-counterterm can be inserted at either the left or right vertex.}
	\label{fig:selfenergy-ct}
\end{figure}
\subsection{Self-Energy}\label{App-self-energy-CT}
The corresponding one-loop diagrams, containing the vertex counterterm and the counterterm insertions on the internal fermion and photon lines, are shown in Fig.~\ref{fig:selfenergy-ct}. Using the \(r/a\) representations of the counterterm lines in Eq.~\eqref{ra-ct}, together with the free \(r/a\) propagators in Eq.~\eqref{5}, one obtains
\begin{align}\label{mainCT-Self}
	&-i\Sigma^{(1)}_{V_{m},{\rm CT}}
	=
	2e^2\delta^{(1)}_2\int_P
	\Bigl[
	\Delta_s(P)\Delta_A(P-Q)\mathcal{S}_{m{(a_s)}}
	+
	\tilde{\Delta}_s(P)\Delta_R(PQ)\mathcal{S}_{m}
	\Bigr]=-2i\delta^{(1)}_2 \Sigma^{(1)}_{m}(Q),
	\no\\
	&-i\Sigma^{(1)}_{V_{\xi},{\rm CT}}
	=
	2ie^2\delta^{(1)}_2(1-\xi)\int_P
	\Bigl[
	\Delta_s(P)\Delta_A^{(2)}(P-Q)\mathcal{S}_{\xi{(a_s)}}
	+
	\tilde{\Delta}_s^{(2)}(P)\Delta_R(PQ)\mathcal{S}_{\xi}
	\Bigr]=-2i\delta^{(1)}_2\Sigma_\xi^{(1)}(Q),
	\no\\
	&-i\Sigma^{(1)}_{f_{m},{\rm CT}}
	=
	ie^2\delta^{(1)}_2\int_P
	\Bigl[
	\mathcal{S}_{m{(a_s)}}P^2\Delta_s^{(2)}(P)\Delta_A(P-Q)
	+
	\mathcal{S}_{m}(P+Q)^2\tilde{\Delta}_s(P)\Delta_R^{(2)}(PQ)
	\Bigr],
	\no\\
	&-i\Sigma^{(1)}_{f_{\xi},{\rm CT}}
	=
	-e^2\delta^{(1)}_2(1-\xi)\int_P
	\Bigl[
	\mathcal{S}_{\xi(a_s)}P^2\Delta_s^{(2)}(P)\Delta_A^{(2)}(P-Q)
	+
	\mathcal{S}_{\xi}(P+Q)^2\tilde{\Delta}_s^{(2)}(P)\Delta_R^{(2)}(PQ)
	\Bigr],
	\no\\
	&-i\Sigma^{(1)}_{\gamma_{m},{\rm CT}}
	=
+ie^2\delta^{(1)}_3\int_P
	\Bigl\{
	\Delta_s(P)\Delta_A^{(2)}(P-Q)
	\Bigl[
	(P-Q)^2\mathcal{S}_{m(a_s)}-\mathcal{S}_{\xi(a_s)}
	\Bigr]
	+
	\tilde{\Delta}_s^{(2)}(P)\Delta_R(PQ)
	\Bigl[
	P^2\mathcal{S}_{m}-\mathcal{S}_{\xi}
	\Bigr]
	\Bigr\}.
\end{align}
Here, the subscripts $V$, $f$, and $\gamma$ label the counterterm insertions at the vertex, on the fermion line, and on the photon line, respectively. The factor of two in the vertex-counterterm contribution accounts for its two possible placements, at the left and right vertices. The Dirac structures $\mathcal{S}_{m}$ and $\mathcal{S}_{\xi}$ are defined in Eq.~\eqref{Dirac-one-loop}, and the label $a_s$ denotes the shift $P\to P-Q$. 
Since the photon counterterm tensor is transverse and gauge-parameter independent,  there is no separate contribution proportional to \((1-\xi)\).
The LP contributions are
\begin{align}\label{LP-CT-Self}
	\Sigma^{(1;\rm LP)}_{{V}_{m},CT}=2\delta^{(1)}_2 \mathcal{H}(T),\quad
	\Sigma^{(1;\rm LP)}_{f_{m},CT}=-\delta^{(1)}_2 \mathcal{H}(T),\quad
	\Sigma^{(1;\rm LP)}_{f_\xi,CT}=\Sigma^{(1;LP)}_{{V}_{\xi},CT}=0,\quad
	\Sigma^{(1;\rm LP)}_{\gamma_{m},{\rm CT}}=-\delta^{(1)}_3\mathcal{H}(T),
\end{align}
where $\mathcal{H}(T)$ is the bare one-loop LP contribution, Eq.~\eqref{14}.
The nonvanishing terms are collected in Eq.~\eqref{self-summing-cts}. Note that here only the part of the counterterm insertions relevant
to the $1/\epsilon$ divergence is displayed. The $O(\epsilon)$ terms
of the $D$-dimensional integrals in Eq.~\eqref{mainCT-Self} can combine with the
$1/\epsilon$ counterterms only to give finite contributions and therefore
do not affect the divergence in Eq.~\eqref{divergences in hard sector}.
\subsection{Vertex}\label{App-vertex-CT}
\begin{figure}[t]
	\centering
	\includegraphics[width=\textwidth]{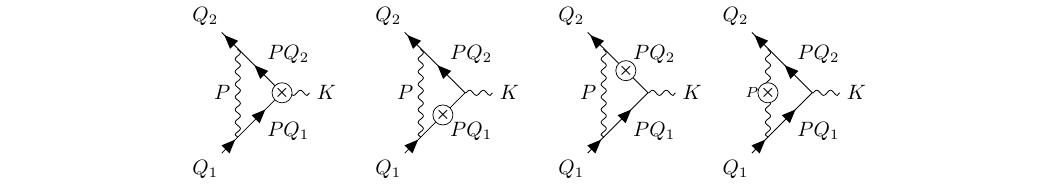}
	\caption{Counterterm insertions in the one-loop fermion--photon vertex: a representative vertex insertion, the two internal-fermion-line insertions, and the internal-photon-line insertion.}
	\label{fig:vertex-ct}
\end{figure}
The one-loop diagrams containing the counterterm insertions are shown in Fig.~\ref{fig:vertex-ct}, yielding the contributions
	\begin{align}\label{Vertex-CT-one-loop}
		&-i e\Gamma_{V_{m},CT}^{\mu(1)}(Q_2,Q_1)=-3ie\,\delta^{(1)}_2\Gamma^{\mu(1)}_{m}(Q_2,Q_1),\quad-ie\Gamma_{V_{\xi},CT}^{\mu}(Q_2,Q_1)=-3ie\,\delta^{(1)}_2\Gamma^{\mu(1)}_{\xi}(Q_2,Q_1),\no\\
		&-ie\Gamma^{\mu(1)}_{f_{I_{m}},CT}(Q_2,Q_1)=e^3\delta^{(1)}_2\int_P\Bigl\{(P+Q_1)^2 \Delta^{(2)}_R(PQ_1)\Delta_R(PQ_2)\tilde{\Delta}_s(P)\mathcal{V}_{m}\no\\
		&\hspace{5cm}+P^2\Delta^{(2)}_s(P)\Delta_A(P+Q_2-Q_1)\Delta_A(P-Q_1)\mathcal{V}_{m(a_v)}\no\\
	&\hspace{5cm}+(P+Q_1-Q_2)^2\Delta_s(P)\Delta^{(2)}_R(P+Q_1-Q_2)\Delta_A(P-Q_2)\mathcal{V}_{m(m_v)}\Bigr\},\no\\
		&-ie\Gamma^{\mu(1)}_{f_{I_{\xi}},CT}(Q_2,Q_1)=ie^3\delta^{(1)}_2\int_P(1-\xi)\Bigl\{(P+Q_1)^2
		 \Delta^{(2)}_R(PQ_1)\Delta_R(PQ_2)\tilde{\Delta}^{(2)}_s(P)\mathcal{V}_{\xi}\no\\
	&\hspace{5cm}+P^2\Delta^{(2)}_s(P)\Delta_A(P+Q_2-Q_1)\Delta^{(2)}_A(P-Q_1)\mathcal{V}_{\xi(a_v)}\no\\
	&\hspace{5cm}+(P+Q_1-Q_2)^2\Delta_s(P)\Delta^{(2)}_R(P+Q_1-Q_2)\Delta^{(2)}_A(P-Q_2)\mathcal{V}_{\xi(m_v)}\Bigr\},\no\\
		&-i e\Gamma_{{\gamma}_m,CT}^{\mu(1)}(Q_2,Q_1)=+e^3\delta^{(1)}_3\int_P\Bigl\{\Bigl[P^2 \mathcal{V}_{m}-\mathcal{V}_\xi\Bigr]\tilde{\Delta}_s^{(2)}(P)\Delta_R(PQ_1)\Delta_R(PQ_2)\no\\
		&\hspace{5cm}+\Bigl[(P-Q_1)^2\mathcal{V}_{m(a_v)}-\mathcal{V}_{\xi(a_v)}\Bigr]\Delta_s(P)\Delta_A(P+Q_2-Q_1)\Delta^{(2)}_A(P-Q_1)\no\\
		&\hspace{5cm}+\Bigl[(P-Q_2)^2\mathcal{V}_{m(m_v)}-\mathcal{V}_{\xi(m_v)}\Bigr]\Delta_s(P)\Delta_R(P+Q_1-Q_2)\Delta^{(2)}_A(P-Q_2)\Bigr\}.
	\end{align}
The vertex counterterm can be inserted at any of the three vertices, and the factor of three accounts for these possible placements. The counterterm insertion on the fermion line, for the second diagram in Fig.~\ref{fig:vertex-ct}, is given above. The remaining diagram is obtained by the exchange $Q_1\leftrightarrow Q_2$, with the corresponding vector and axial projections following from the exchange relations in Eq.~\eqref{eq:exchangeSummaryGeneral}. The LP corrections then become
\begin{align}\label{Vertex-CT-one-loop-LP}
&\Gamma_{V_{m},CT}^{\mu(1;\rm LP)}(Q_2,Q_1)=3\,\delta^{(1)}_2\Gamma^{\mu(1;\rm LP)}(Q_2,Q_1),\quad \Gamma^{\mu(1;\rm LP)}_{f_{{\rm I}_{m}},CT}(Q_2,Q_1)+\Gamma^{\mu(1;\rm LP)}_{f_{{\rm{II}}_{m}},CT}(Q_2,Q_1)=-2\delta_2^{(1)}\Gamma^{\mu(1;\rm LP)}(Q_2,Q_1),\no\\
&\Gamma_{\gamma_{m},CT}^{\mu(1;\rm LP)}(Q_2,Q_1)=-\delta_3^{(1)}\Gamma^{\mu(1;\rm LP)}(Q_2,Q_1),\quad\Gamma^{\mu(1;\rm LP)}_{f_{{\rm I}_{\xi}},CT}(Q_2,Q_1)=\Gamma^{\mu(1;\rm LP)}_{f_{{\rm II}_{\xi}},CT}(Q_2,Q_1)=\Gamma_{V_{\xi},CT}^{\mu(1;\rm LP)}(Q_2,Q_1)=0,
\end{align}
where $\Gamma^{\mu(1;LP)}(Q_2,Q_1)$ is given in Eq.~\eqref{LP-Vertex}.
The nonvanishing terms are collected in Eq.~\eqref{CT-vertex-result}.

		\section{Two-Loop Self-Energy Results}\label{self-energy computations two-loop}
	After the soft expansion, the self-energy contributions become
		\begin{subequations}\label{HTL-self-energy-results}
			\begin{align}
				&-i\Sigma^{(2;\rm LP)}_{m}=-\frac{ie^4}{4} \int_P\int_L \Delta_d(L)\Delta_d(P)\no\\
				&\times\Bigl\{N_B(L) N_B(P) \Bigl[\mathcal{D}^2 \Bigl(-\frac{\slashed{L}}{L.P P.Q}-\frac{Q^2 \slashed{P}}{(P.Q)^3}+\frac{\slashed{Q}}{(P.Q)^2}\Bigr)+\frac{2 \mathcal{D} \, Q^2 \slashed{L}}{L.Q (P.Q)^2}\Bigr]+N_F(L)N_F^\prime(P)\Bigl[\frac{\mathcal{D}^2 \slashed{P}}{{p^0} P.Q}\Bigr]\no\\
				&\quad+ N_F(L) N_F(P)\Bigl[\mathcal{D}^2 \Bigl(-\frac{\slashed{L}}{{l^0}^2 L.Q}+\frac{\gamma_0}{{l^0} L.Q}-\frac{{q^0} \slashed{L}}{{l^0} (L.Q)^2}-\frac{\slashed{P}}{L.P L.Q}+\frac{Q^2 \slashed{L}}{(L.Q)^3}\Bigr)\no\\
				&\hspace{2.8cm}+\mathcal{D} \Bigl(-\frac{2 Q^2 \slashed{L}}{L.Q (P.Q)^2}-\frac{4 Q^2 \slashed{P}}{(P.Q)^3}\Bigr)-\frac{8 \slashed{L}}{L.P P.Q}-\frac{8 \slashed{P} L.Q}{L.P (P.Q)^2}+\frac{8 \slashed{Q}}{(P.Q)^2}\Bigr]\no\\
				&\quad+ N_F(L)N_B(P)\Bigl[\mathcal{D}^2\Bigl( \frac{\slashed{L}}{{l^0}^2 L.Q}+\frac{\slashed{L}{q^0}}{{l^0} (L.Q)^2}+\frac{\slashed{L}}{L.P P.Q}-\frac{\slashed{L}Q^2}{(L.Q)^3}-\frac{\gamma_0}{{l^0} L.Q}+ \frac{\slashed{P}}{L.P L.Q}+\frac{\slashed{P}Q^2}{(P.Q)^3}-\frac{\slashed{Q}}{(P.Q)^2}\Bigr)\no\\
				&\hspace{2.8cm}+\mathcal{D}\Bigl(\frac{2\slashed{P}  Q^2}{(L.Q)^2 P.Q}+\frac{4\slashed{P} }{{p^0}^2 P.Q}+\frac{4\slashed{P}  {q^0}}{{p^0} (P.Q)^2}-\frac{4\slashed{P}  Q^2}{(P.Q)^3}-\frac{2 Q^2 \slashed{L}}{L.Q (P.Q)^2}-\frac{4 \gamma_0}{{p^0} P.Q}+\frac{4 \slashed{Q}}{(P.Q)^2}\Bigr)\no\\
				&\hspace{2.8cm}+\frac{8\slashed{L}}{L.P P.Q}+\frac{8\slashed{P} L.Q}{L.P (P.Q)^2}-\frac{8\slashed{Q}}{(P.Q)^2}\Bigr]\no\\
				&\quad-N_B(P)N_F^\prime(L)\Bigl(\frac{\mathcal{D}^2 \slashed{L}}{{l^0} L.Q}\Bigr)-N_F(L)N_B^\prime(P)\Big(\frac{4\mathcal{D} \slashed{P}}{{p^0} P.Q}\Big)\Bigr\},\label{self-energy results}\\
				&-i\Sigma^{(2;\rm LP)}_{\xi}=\frac{ie^4\mathcal{D}}{4} \int_P\int_L(1-\xi) \Delta_d(L)\Delta_d(P)\no\\
				&\times\Bigl\{N_B(L) N_B(P)\Bigl[\slashed{P}\Bigl(\frac{1}{{p^0}^2 P.Q}+\frac{{q^0}}{{p^0} (P.Q)^2}-\frac{2 Q^2}{(P.Q)^3}\Bigr)-\frac{\gamma^0}{{p^0} P.Q}\no\\
				&\hspace{2.5cm}+\slashed{Q}\Bigl(\frac{{l^0}}{{p^0} L.Q P.Q}-\frac{L.P}{{p^0}^2 L.Q P.Q}-\frac{{q^0} L.P}{{p^0} L.Q (P.Q)^2}+\frac{Q^2 L.P}{L.Q (P.Q)^3}+\frac{2}{(P.Q)^2}\Bigr)\Bigr]\no\\
				&\quad+N_B(L)N_B^\prime(P)\Bigl[\frac{\slashed{Q} L.P}{{p^0} L.Q P.Q}-\frac{\slashed{P}}{{p^0} P.Q}\Bigr]+N_F(L)N_F(P)\Bigl[-\frac{Q^2 \slashed{L}}{L.Q (P.Q)^2}-\frac{Q^2 \slashed{Q} L.P}{(L.Q)^3 P.Q}+\frac{2 \slashed{Q}}{(P.Q)^2}\Bigr]\no\\
				&\quad+ N_F(L) N_B(P)\Bigl[\slashed{P}\Bigl(\frac{Q^2}{(L.Q)^2 P.Q}-\frac{1}{{p^0}^2 P.Q}-\frac{{q^0}}{{p^0} (P.Q)^2}+\frac{2 Q^2}{(P.Q)^3}\Bigr)+\frac{\gamma^0}{{p^0} P.Q}\no\\
				&\hspace{2.5cm}+\slashed{Q}\Bigl(-\frac{{l^0}}{{p^0} L.Q P.Q}+\frac{L.P}{{p^0}^2 L.Q P.Q}+\frac{{q^0} L.P}{{p^0} L.Q (P.Q)^2}+\frac{Q^2 L.P}{(L.Q)^3 P.Q}-\frac{Q^2 L.P}{L.Q (P.Q)^3}-\frac{2}{(L.Q)^2}-\frac{2}{(P.Q)^2}\Bigr)\Bigr]\no\\
				&\quad+N_F(L)N_B^\prime(P)\Bigl[-\frac{\slashed{Q} L.P}{{p^0} L.Q P.Q}+\frac{\slashed{P}}{{p^0} P.Q}\Bigr]\Bigr\}.\label{self-energy results xi}
			\end{align}
		\end{subequations}
	To compute the collinear singularities, $\frac{1}{P.L}$, we collect the
		required collinear moments in
		Eqs.~\eqref{collinear integrals1}--\eqref{collinear integrals3}. Using these
		identities, the coupled angular dependence between the two hard directions
		reduces to a combination of ordinary HTL angular moments and the collinear
		moment. For example,
		\begin{equation}
			\int_{v}\int_{v'}\frac{\slashed v\,(v^{\prime}\!\cdot Q)}
			{(v\!\cdot v^{\prime})\,(v\!\cdot Q)^2}
			=
			q^0\,\mathcal A_0\,\slashed{\mathcal A}_2
			+(\mathcal C_1-\mathcal A_0)\,\slashed{\mathcal A}_1,
			\label{eq:example-collinear-reduction}
		\end{equation}
		where Eqs.~\eqref{Angular Integrals} and \eqref{collinear integrals3} have
		been used.
		 Here,
		\(\slashed{\mathcal A}_\alpha\equiv
		\mathcal A_\alpha\gamma^0-\mathcal A_\alpha^{\,i}\gamma_i\), and
		\(\mathcal C_1\) denotes the collinear moment.
		
		We use
		Eq.~\eqref{Rdial integral} for the radial integrals,
		Eqs.~\eqref{Angular Integrals} and \eqref{Angular integrals-spatial} for the
		non-collinear angular integrals, and
		Eqs.~\eqref{collinear integrals1}--\eqref{collinear integrals3} for the
		collinear moments. This yields
		\begin{subequations}\label{28}
			\begin{align}
				&\Sigma^{(2;\rm LP)}_{m}(Q)=\frac{e^4}{4}\mathcal{N}^2\no\\
				&\times\biggl\{\mathcal{A}_0\Bigl\{\mathcal{D}^2 (\mathcal{R}_1+\mathcal{R}_2) \Bigl[\mathcal{R}_3 \left(-\mathcal{A}_1\gamma_0+\mathcal{A}_2\slashed{Q}-Q^2 \slashed{\mathcal{A}}_3+\slashed{\mathcal{A}}_1\right)-Q^2 \mathcal{R}_4 \slashed{\mathcal{A}}_3+q^0 \mathcal{R}_4 \slashed{\mathcal{A}}_2-\mathcal{R}_6 \slashed{\mathcal{A}}_1\Bigr]\no\\
				&\hspace{1.2cm}+4 \mathcal{D} \mathcal{R}_2\Bigl[-\mathcal{R}_3 \bigl[-\mathcal{A}_1\gamma_0+\mathcal{A}_2\slashed{Q}-Q^2 \slashed{\mathcal{A}}_3+q^0 \slashed{\mathcal{A}}_2+\slashed{\mathcal{A}}_1\bigr]+Q^2 \mathcal{R}_4 \slashed{\mathcal{A}}_3+\mathcal{R}_5 \slashed{\mathcal{A}}_1\Bigr]\no\\
				&\hspace{1.2cm}+8 \mathcal{R}_2 (\mathcal{R}_3-\mathcal{R}_4) \bigl[-\mathcal{A}_1\gamma_0+\mathcal{A}_2\slashed{Q}-q^0 \slashed{\mathcal{A}}_2+2 \slashed{\mathcal{A}}_1\bigr]\Bigr\}+2\mathcal{D} Q^2 \mathcal{A}_2 \slashed{\mathcal{A}}_1 (\mathcal{R}_1+\mathcal{R}_2) (\mathcal{R}_3+\mathcal{R}_4)\no\\
				&\quad+\mathcal{C}_1 \slashed{\mathcal{A}}_1 (\mathcal{R}_4-\mathcal{R}_3) \bigl[\mathcal{D}^2 (\mathcal{R}_1+\mathcal{R}_2)+16 \mathcal{R}_2\bigr]\biggr\},\label{physical sector result}\\
				&\Sigma^{(2;\rm LP)}_{\xi}(Q)=-\frac{e^4}{4}(1-\xi)\mathcal{D}\mathcal{N}^2(\mathcal{R}_1 +\mathcal{R}_2 )\no\\
				&\times\Bigl\{\mathcal{A}_0 \Bigl[-2 \mathcal{A}_2 \mathcal{R}_4  \slashed{Q}-\mathcal{R}_5  \slashed{\mathcal{A}}_1+\mathcal{R}_3 \bigl[-\mathcal{A}_1 \gamma_0+2 \mathcal{A}_2 \slashed{Q}-2 Q^2 \slashed{\mathcal{A}}_3+q^0 \slashed{\mathcal{A}}_2+\slashed{\mathcal{A}}_1\bigr]\Bigr]+\mathcal{A}_2 Q^2 \mathcal{R}_4  \slashed{\mathcal{A}}_1\no\\
				&\quad+\slashed{Q}\Bigl[\mathcal{A}^j_1\mathcal{A}^j_1 (\mathcal{R}_3 -\mathcal{R}_5 )+\mathcal{A}_1^j\mathcal{A}_2^j q^0 \mathcal{R}_3 -\mathcal{A}^j_1\mathcal{A}^j_3 Q^2 (\mathcal{R}_3 +\mathcal{R}_4 )+\mathcal{A}_1 \big[\mathcal{A}_1 \mathcal{R}_5 -\mathcal{A}_2 q^0 \mathcal{R}_3 +\mathcal{A}_3 Q^2 (\mathcal{R}_3 +\mathcal{R}_4 )\big]\Bigr]\Bigr\}.\label{xi sector result}
			\end{align}
		\end{subequations}
		where $\mathcal{N}$ is given in Eq.~\eqref{Eq:N-expansion}.
		Radial singularities arise only through \(\mathcal R_3\) and
		\(\mathcal R_4\), whose singular parts are identical.
		Hence, the difference \(\mathcal R_4-\mathcal R_3\) is free of radial
		singularities and leaves a finite remainder
		proportional to \(\ln(2)\), with further \(O(\epsilon)\) contributions.
		In the last term of Eq.~\eqref{physical sector result}, this radial difference
		multiplies the collinear moment \(\mathcal C_1\). Therefore, the resulting contribution contains a collinear pole whose coefficient is proportional to
		\(\ln(2)\). This makes the collinear singularity distinguishable from the purely radial singularities; in particular, this term does not generate a radial--collinear double pole.
		
		We expand both the angular and radial integrals through \(O(\epsilon)\). This
		is necessary for two reasons. First, in the collinear contributions, the
		combination \(\mathcal R_3-\mathcal R_4\) cancels the \(\ln(2\pi T)\) term at
		\(O(\epsilon^0)\); see App.~\ref{DR}. 
		Since the scale dependence must enter through the dimensionless ratio of the renormalization scale \(\bar{\Lambda}\)
		to the thermal scale \(2\pi T\), the radial expansions must be kept through
		\(O(\epsilon)\). Only then do they reproduce the correct scale-dependent
		logarithm,
		\(\ln(e^{\gamma_E}\bar{\Lambda}/2\pi T)\).
		Second, products of radial and angular
		integrals can generate finite contributions through
		\(O(\epsilon)\times 1/\epsilon\). In particular, the poles in
		\(\mathcal R_{3,4}\) and in $\mathcal{C}_1$ can multiply the \(O(\epsilon)\) parts of the remaining
		radial integrals or of the angular moments, producing finite terms. Thus, both
		the radial and angular integrals must be kept through \(O(\epsilon)\) to
		obtain the complete finite part.
		We also keep the calculation explicitly \(D\)-dimensional throughout, because
		\(D\)-dependent prefactors, such as \(\mathcal D=D-2=2-2\epsilon\), carry
		\(O(\epsilon)\) corrections that multiply \(1/\epsilon\) poles and generate
		finite contributions.
			\section{Fermion--Photon Vertex Results}\label{final results}
\subsection{One-Loop NLP Vertex}\label{power corections}
	\begin{small}
		\begin{subequations}\label{metric-NLP}
\begin{align}\label{NLP vector vertex}
	&-ie\Bigl[V_{m}^{\mu\rho(1;\rm NLP)}\Bigr]\gamma_\rho=\frac{ie^3\mathcal{D}}{4}\int_P\Delta_d(P)\no\\
	&\times\biggl\{N_F(P) \Bigl\{\slashed{P}\Bigl[\frac{K^4 {P^\mu}}{2 (K.P)^2 P.Q_1 P.Q_2}-\frac{K^2 {K^\mu}}{2 (K.P)^2 P.Q_1}+\frac{K^2 {K^\mu}}{2 (K.P)^2 P.Q_2}-\frac{K^2 {P^\mu} Q_1^2}{2 (K.P)^2 (P.Q_1)^2}-\frac{K^2 {P^\mu} Q_2^2}{2 (K.P)^2 (P.Q_2)^2}+\frac{{K^\mu} Q_1^2}{2 K.P (P.Q_1)^2}\no\\
	&\hspace{2cm}+\frac{{K^\mu} Q_2^2}{2 K.P (P.Q_2)^2}+\frac{{P^\mu} Q_1^4}{2 K.P (P.Q_1)^3}-\frac{{P^\mu} Q_2^4}{2 K.P (P.Q_2)^3}\Bigr]+{\gamma}^\mu\Bigl[\frac{K^2}{2 K.P P.Q_1}-\frac{K^2}{2 K.P P.Q_2}-\frac{Q_1^2}{2 (P.Q_1)^2}-\frac{Q_2^2}{2 (P.Q_2)^2}\Bigr]\no\\
	&\hspace{2cm}+\slashed{K}\Bigl[-\frac{K^2 {P^\mu}}{2 (K.P)^2 P.Q_1}+\frac{K^2 {P^\mu}}{2 (K.P)^2 P.Q_2}+\frac{{P^\mu} Q_1^2}{2 K.P (P.Q_1)^2}+\frac{{P^\mu} Q_2^2}{2 K.P (P.Q_2)^2}\Bigr]\Bigr\}\no\\
	&\quad+N_B(P)\Bigl\{\slashed{P}\Bigl[-\frac{{P^\mu} Q_1^4}{2 (P.Q_1)^3 P.Q_2}-\frac{{P^\mu} Q_1^2 Q_2^2}{2 (P.Q_1)^2 (P.Q_2)^2}-\frac{{P^\mu} Q_2^4}{2 P.Q_1 (P.Q_2)^3}+\frac{Q_1^2 {Q_1^\mu}}{2 (P.Q_1)^2 P.Q_2}+\frac{{Q_1^\mu} Q_2^2}{2 P.Q_1 (P.Q_2)^2}+\frac{Q_2^2 {Q_2^\mu}}{2 P.Q_1 (P.Q_2)^2}\no\\
	&\hspace{2cm}+\frac{Q_1^2 {Q_2^\mu}}{2 (P.Q_1)^2 P.Q_2}\Bigr]+{\gamma}^\mu\Bigl[-\frac{Q_1^2}{2 P.Q_1 P.Q_2}-\frac{Q_1^2}{2 (P.Q_1)^2}-\frac{Q_2^2}{2 P.Q_1 P.Q_2}+\frac{Q_1.Q_2}{P.Q_1 P.Q_2}-\frac{Q_2^2}{2 (P.Q_2)^2}\Bigr]\no\\
	&\hspace{2cm}+\slashed{Q_2}\Bigl[\frac{{P^\mu} Q_1^2}{2 (P.Q_1)^2 P.Q_2}+\frac{{P^\mu} Q_2^2}{2 P.Q_1 (P.Q_2)^2}-\frac{{Q_1^\mu}}{P.Q_1 P.Q_2}\Bigr]+\slashed{Q_1}\Bigl[\frac{{P^\mu} Q_1^2}{2 (P.Q_1)^2 P.Q_2}+\frac{{P^\mu} Q_2^2}{2 P.Q_1 (P.Q_2)^2}-\frac{{Q_2^\mu}}{P.Q_1 P.Q_2}\Bigr]\Bigr\}\biggr\}.
\end{align}
	\begin{align}\label{NLP axial vertex}
		&-ie\Bigl[A_{m}^{\mu\rho(1;\rm NLP)}\Bigr]\gamma_\rho\gamma_5=\frac{ie^3}{4}\int_P\Delta_d(P)(D-6)\epsilon ^{\mu \alpha \beta \rho } \gamma_\rho\gamma_5\no\\
		&\times\biggl\{-iN_F(P)\Bigl[\frac{K^2 K_\alpha P_\beta}{2 (K.P)^2 P.Q_1}+\frac{K^2 K_\alpha P_\beta}{2 (K.P)^2 P.Q_2}-\frac{K_\alpha P_\beta Q_1^2}{2 K.P (P.Q_1)^2}+\frac{K_\alpha P_\beta Q_2^2}{2 K.P (P.Q_2)^2}\Bigr]\no\\
		&\quad-i N_B(P) \Bigl[\frac{P_\alpha Q_1^2 K_\beta}{2 (P.Q_1)^2 P.Q_2}+\frac{P_\alpha Q_2^2 K_\beta}{2 P.Q_1 (P.Q_2)^2}-\frac{{Q_1}_\alpha {Q_2}_\beta}{P.Q_1 P.Q_2}\Bigr]\biggr\}.
	\end{align}
		\end{subequations}
	\begin{subequations}\label{longitudinal-NLP}
	\begin{align}\label{pow xi vector vertex}
		&-ie\Bigl[V_{\xi}^{\mu\rho(1;\rm NLP)}\Bigr]\gamma_\rho=-\frac{ie^3}{4}\int_P(1-\xi)\Delta_d(P)\no\\
		&\times\biggl\{N_F(P)\Bigl\{\slashed{P}\Bigl[\frac{K^2 {P^\mu} {Q_1}^2}{2 (K.P)^2 ({P.Q_1})^2}+\frac{K^2 {P^\mu} K.{Q_1}}{2 (K.P)^2 ({P.Q_1})^2}+\frac{K^2 {P^\mu} {Q_2}^2}{2 (K.P)^2 ({P.Q_2})^2}-\frac{K^2 {P^\mu} K.{Q_2}}{2 (K.P)^2 ({P.Q_2})^2}-\frac{K^2 {Q_1^\mu}}{2 K.P ({P.Q_1})^2}+\frac{K^2 {Q_2^\mu}}{2 K.P ({P.Q_2})^2}\no\\
		&\hspace{2cm}-\frac{K^\mu {Q_1}^2}{2 K.P ({P.Q_1})^2}-\frac{K^\mu {Q_2}^2}{2 K.P ({P.Q_2})^2}-\frac{{P^\mu} {Q_1}^4}{K.P ({P.Q_1})^3}-\frac{{P^\mu} {Q_1}^2 K.{Q_1}}{K.P ({P.Q_1})^3}+\frac{{P^\mu} {Q_2}^4}{K.P ({P.Q_2})^3}-\frac{{P^\mu} {Q_2}^2 K.{Q_2}}{K.P ({P.Q_2})^3}+\frac{{Q_1}^2 {Q_1^\mu}}{({P.Q_1})^3}\no\\
		&\hspace{2cm}+\frac{{Q_2}^2 {Q_2^\mu}}{({P.Q_2})^3}\Bigr]+{\gamma}^\mu\Bigl[\frac{K^2}{2 K.P {P.Q_1}}-\frac{K^2}{2 K.P {P.Q_2}}-\frac{{Q_1}^2}{2 ({P.Q_1})^2}-\frac{{Q_2}^2}{2 ({P.Q_2})^2}\Bigr]\no\\
		&\hspace{1.2cm}+\slashed{Q_2}\Bigl[-\frac{K^2 {P^\mu}}{2 (K.P)^2 {P.Q_1}}-\frac{K^2 {P^\mu}}{2 (K.P)^2 {P.Q_2}}+\frac{K^\mu}{K.P {P.Q_2}}+\frac{{P^\mu} {Q_1}^2}{2 K.P ({P.Q_1})^2}-\frac{3 {P^\mu} {Q_2}^2}{2 K.P ({P.Q_2})^2}+\frac{{P^\mu} K.{Q_2}}{K.P ({P.Q_2})^2}-\frac{{Q_2^\mu}}{({P.Q_2})^2}\Bigr]\no\\
		&\hspace{1.2cm}+\slashed{Q_1}\Bigl[-\frac{K^2 {P^\mu}}{2 (K.P)^2 {P.Q_1}}-\frac{K^2 {P^\mu}}{2 (K.P)^2 {P.Q_2}}+\frac{K^\mu}{K.P {P.Q_1}}+\frac{3 {P^\mu} {Q_1}^2}{2 K.P ({P.Q_1})^2}+\frac{{P^\mu} K.{Q_1}}{K.P ({P.Q_1})^2}-\frac{{P^\mu} {Q_2}^2}{2 K.P ({P.Q_2})^2}-\frac{{Q_1^\mu}}{({P.Q_1})^2}\Bigr]\Bigr\}\no\\
		&\quad +N_B(P)\Bigl\{\slashed{P}\Bigl[-\frac{{Q_1^\mu}}{{p^0}^2 P.Q_1}-\frac{{Q_2^\mu}}{{p^0}^2 P.Q_2}-\frac{{q_1^0} {Q_1^\mu}}{{p^0} (P.Q_1)^2}-\frac{{q_2^0} {Q_2^\mu}}{{p^0} (P.Q_2)^2}+\frac{Q_1^2 {Q_1^\mu}}{(P.Q_1)^3}+\frac{Q_1^2 {Q_2^\mu}}{2 P.Q_1 (P.Q_2)^2}+\frac{{Q_1^\mu} Q_2^2}{2 P.Q_1 (P.Q_2)^2}+\frac{Q_2^2 {Q_2^\mu}}{(P.Q_2)^3}\no\\
		&\hspace{2cm}+\frac{{P^\mu} {q_2^0} Q_1.Q_2}{{p^0} P.Q_1 (P.Q_2)^2}-\frac{{P^\mu} Q_1^2 Q_1.Q_2}{2 (P.Q_1)^2 (P.Q_2)^2}-\frac{{P^\mu} Q_2^2 Q_1.Q_2}{2 (P.Q_1)^2 (P.Q_2)^2}-\frac{{P^\mu} Q_2^2 Q_1.Q_2}{P.Q_1 (P.Q_2)^3}+\frac{Q_1^2 {Q_2^\mu}}{2 (P.Q_1)^2 P.Q_2}+\frac{{Q_1^\mu} Q_2^2}{2 (P.Q_1)^2 P.Q_2}\no\\
		&\hspace{2cm}-\frac{g^{0\mu} Q_1.Q_2}{{p^0} P.Q_1 P.Q_2}+\frac{{P^\mu} Q_1.Q_2}{{p^0}^2 P.Q_1 P.Q_2}+\frac{{P^\mu} {q_1^0} Q_1.Q_2}{{p^0} (P.Q_1)^2 P.Q_2}-\frac{{P^\mu} Q_1^2 Q_1.Q_2}{(P.Q_1)^3 P.Q_2}\Bigr]\no\\
		&\hspace{1.2cm}+\gamma^0\Bigl[-\frac{{P^\mu} Q_1.Q_2}{{p^0} P.Q_1 P.Q_2}+\frac{{Q_1^\mu}}{{p^0} P.Q_1}+\frac{{Q_2^\mu}}{{p^0} P.Q_2}\Bigr]+{\gamma}^\mu\Bigl[-\frac{Q_1^2}{2 P.Q_1 P.Q_2}-\frac{Q_1^2}{2 (P.Q_1)^2}-\frac{Q_2^2}{2 P.Q_1 P.Q_2}+\frac{Q_1.Q_2}{P.Q_1 P.Q_2}-\frac{Q_2^2}{2 (P.Q_2)^2}\Bigr]\no\\
		&\hspace{1.2cm}+\slashed{Q_2}\Bigl[\frac{{P^\mu} Q_1^2}{2 (P.Q_1)^2 P.Q_2}+\frac{{P^\mu} Q_2^2}{2 P.Q_1 (P.Q_2)^2}-\frac{{Q_1^\mu}}{P.Q_1 P.Q_2}\Bigr]+\slashed{Q_1}\Bigl[\frac{{P^\mu} Q_1^2}{2 (P.Q_1)^2 P.Q_2}+\frac{{P^\mu} Q_2^2}{2 P.Q_1 (P.Q_2)^2}-\frac{{Q_2^\mu}}{P.Q_1 P.Q_2}\Bigr]\Bigr\}\no\\
		&\quad+N_B^\prime(P)\slashed{P}\Bigl[-\frac{{P^\mu} Q_1.Q_2}{{p^0} P.Q_1 P.Q_2}+\frac{{Q_1^\mu}}{{p^0} P.Q_1}+\frac{{Q_2^\mu}}{{p^0} P.Q_2}\Bigr]\biggr\}.
	\end{align}
		\begin{align}\label{pow xi axial vertex}
		&-ie\Bigl[A_{\xi}^{\mu\rho(1;\rm NLP)}\Bigr]\gamma_\rho\gamma_5=-\frac{ie^3}{4}\int_P(1-\xi)\Delta_d(P)(D-6)\gamma_\rho\gamma_5\no\\
		&\times\biggl\{-i N_F(P)\Bigl\{\epsilon ^{\mu \alpha \beta \rho } \Bigl[\frac{K^2 P^\alpha Q_1^\beta}{2 K.P (P.Q_1)^2}+\frac{K^2 P^\alpha Q_2^\beta}{2 K.P (P.Q_2)^2}+\frac{K^\alpha P^\beta Q_1^2}{2 K.P (P.Q_1)^2}-\frac{K^\alpha P^\beta Q_2^2}{2 K.P (P.Q_2)^2}-\frac{K^\alpha Q_1^\beta}{K.P P.Q_1}+\frac{K^\alpha Q_2^\beta}{K.P P.Q_2}+\frac{P^\alpha Q_1^\beta K.Q_1}{K.P (P.Q_1)^2}\no\\
		&\hspace{3cm}-\frac{P^\alpha Q_2^\beta K.Q_2}{K.P (P.Q_2)^2}-\frac{P^\alpha Q_1^2 Q_1^\beta}{(P.Q_1)^3}+\frac{P^\alpha Q_2^2 Q_2^\beta}{(P.Q_2)^3}\Bigr]\no\\
		&\hspace{1.3cm}+K^\alpha P^\beta \epsilon ^{\alpha \beta \nu \rho }\Bigl[-\frac{K^2 P^\mu Q_1^\nu}{2 (K.P)^2 (P.Q_1)^2}-\frac{K^2 P^\mu Q_2^\nu}{2 (K.P)^2 (P.Q_2)^2}+\frac{P^\mu Q_1^2 Q_1^\nu}{K.P (P.Q_1)^3}-\frac{P^\mu Q_2^2 Q_2^\nu}{K.P (P.Q_2)^3}-\frac{Q_1^\mu Q_1^\nu}{K.P (P.Q_1)^2}+\frac{Q_2^\mu Q_2^\nu}{K.P (P.Q_2)^2}\Bigr]\Bigr\}\no\\
		&-i N_B(P)\Bigl[\epsilon ^{\alpha \beta \nu \rho } \Bigl(\frac{g^{0\alpha}P^\mu Q_1^\beta Q_2^\nu}{{p^0}P.Q_1 P.Q_2}+\frac{g^{0\mu} P^\alpha Q_1^\beta Q_2^\nu}{{p^0}P.Q_1 P.Q_2}-\frac{P^\alpha P^\mu Q_1^\beta Q_2^\nu}{(p^0)^2 P.Q_1 P.Q_2}-\frac{P^\alpha P^\mu q_1^0Q_1^\beta Q_2^\nu}{{p^0}(P.Q_1)^2 P.Q_2}-\frac{P^\alpha P^\mu Q_1^\beta q_2^0Q_2^\nu}{{p^0}P.Q_1 (P.Q_2)^2}\Bigr)\no\\
		&\hspace{1.5cm}+\epsilon ^{\mu \alpha \beta \rho } \Bigl(-\frac{g^{0\alpha}Q_1^\beta}{{p^0}P.Q_1}+\frac{g^{0\alpha}Q_2^\beta}{{p^0}P.Q_2}+\frac{P^\alpha Q_1^\beta}{({p^0})^2 P.Q_1}-\frac{P^\alpha Q_2^\beta}{({p^0})^2 P.Q_2}+\frac{P^\alpha q_1^0Q_1^\beta}{{p^0}(P.Q_1)^2}-\frac{P^\alpha q_2^0Q_2^\beta}{{p^0}(P.Q_2)^2}\Bigr)\Bigr]\no\\
		&-iN_B'(P)\Bigl[\frac{\gamma^\rho  P^\alpha P^\mu Q_1^\beta Q_2^\nu \epsilon ^{\alpha \beta \nu \rho }}{{p^0}P.Q_1 P.Q_2}+\epsilon ^{\mu \alpha \beta \rho } \Bigl(\frac{P^\alpha Q_2^\beta}{{p^0}P.Q_2}-\frac{P^\alpha Q_1^\beta}{{p^0}P.Q_1}\Bigr)\Bigr]\biggr\}.
	\end{align}
		\end{subequations}
\subsection{Two-Loop LP Vertex}
\begin{align}\label{NBNB-metricsector-vector-vertex}
	&-ie\Bigl[V^{\mu\rho(2;\rm LP)}_{m,N_BN_B}\Bigr]\gamma_\rho=-\frac{ie^5}{4}\int_P\int_L\Delta_d(P)\Delta_d(L)N_B(L)N_B(P)\no\\
	&\times\biggl\{\mathcal{D}^2\biggl[\slashed{P}\Bigl[-\frac{P^\mu \text{Q1}^2}{(P.Q_1)^3 P.Q_2}-\frac{P^\mu \text{Q1}^2}{2 (P.Q_1)^2 (P.Q_2)^2}-\frac{P^\mu \text{Q2}^2}{2 (P.Q_1)^2 (P.Q_2)^2}-\frac{P^\mu \text{Q2}^2}{P.Q_1 (P.Q_2)^3}+\frac{Q_1^\mu}{2 (P.Q_1)^2 P.Q_2}+\frac{Q_1^\mu}{2 P.Q_1 (P.Q_2)^2}\no\\
	&\hspace{1.5cm}+\frac{Q_2^\mu}{2 (P.Q_1)^2 P.Q_2}+\frac{Q_2^\mu}{2 P.Q_1 (P.Q_2)^2}\Bigr]-\frac{P^\mu \slashed{L}}{L.P P.Q_1 P.Q_2}\no\\
	&\qquad\quad+\frac{P^\mu \slashed{Q}_1}{2 (P.Q_1)^2 P.Q_2}+\frac{P^\mu \slashed{Q}_1}{2 P.Q_1 (P.Q_2)^2}+\frac{P^\mu \slashed{Q}_2}{2 (P.Q_1)^2 P.Q_2}+\frac{P^\mu \slashed{Q}_2}{2 P.Q_1 (P.Q_2)^2}+\gamma^\mu \Bigl[-\frac{1}{2 (P.Q_1)^2}-\frac{1}{2 (P.Q_2)^2}\Bigr]\biggr]\no\\
    &\quad+\mathcal{D}\biggl[\slashed{L} \Bigl[-\frac{L^\mu}{L.P L.Q_2 P.Q_1}+\frac{L^\mu}{L.P L.Q_1 P.Q_2}+\frac{L^\mu}{L.P L.Q_1 P.Q_1}+\frac{L^\mu}{L.P L.Q_2 P.Q_2}+\frac{P^\mu \text{Q1}^2}{L.Q_1 (P.Q_1)^2 P.Q_2}+\frac{P^\mu \text{Q1}^2}{L.Q_2 (P.Q_1)^2 P.Q_2}\no\\
	&\qquad\qquad+\frac{P^\mu \text{Q2}^2}{L.Q_1 P.Q_1 (P.Q_2)^2}+\frac{P^\mu \text{Q2}^2}{L.Q_2 P.Q_1 (P.Q_2)^2}+\frac{2 P^\mu}{L.P P.Q_1 P.Q_2}-\frac{P^\mu L.Q_2}{L.P L.Q_1 P.Q_1 P.Q_2}-\frac{P^\mu L.Q_1}{L.P L.Q_2 P.Q_1 P.Q_2}\no\\
	&\qquad\qquad-\frac{Q_1^\mu}{L.Q_1 P.Q_1 P.Q_2}-\frac{Q_1^\mu}{L.Q_2 P.Q_1 P.Q_2}-\frac{Q_2^\mu}{L.Q_1 P.Q_1 P.Q_2}-\frac{Q_2^\mu}{L.Q_2 P.Q_1 P.Q_2}\Bigr]\no\\
	&\qquad\quad+\slashed{P}\Bigl[\frac{P^\mu \text{Q1}^2}{(L.Q_1)^2 P.Q_1 P.Q_2}+\frac{P^\mu \text{Q2}^2}{(L.Q_2)^2 P.Q_1 P.Q_2}-\frac{2 P^\mu}{L.P L.Q_2 P.Q_1}\Bigr]\biggr]\biggr\}.
\end{align} 
\begin{align}\label{NBNB-metricsector-axial-vertex}
	&-ie\Bigl[A^{\mu\rho(2;\rm LP)}_{m,N_BN_B}\Bigr]\gamma_\rho\gamma_5=-\frac{ie^5}{4}\int_P\int_L\Delta_d(P)\Delta_d(L)(-i)N_B(L)N_B(P)\gamma_\rho\gamma_5\no\\
	&\times\Bigl\{\mathcal{D}^2 \epsilon ^{\mu \rho \alpha \beta } \Bigl[\frac{K_\beta P_\alpha}{2 (P.Q_1)^2 P.Q_2}+\frac{K_\beta P_\alpha}{2 P.Q_1 (P.Q_2)^2}\Bigr]+\mathcal{D}\,\epsilon ^{\rho \alpha \beta \nu }\Bigl[\frac{K_\nu L_\alpha L^\mu P_\beta}{L.P L.Q_1 P.Q_1 P.Q_2}+\frac{K_\nu L_\alpha L^\mu P_\beta}{L.P L.Q_2 P.Q_1 P.Q_2}\Bigr]\no\\
	&+\mathcal{D}\,\epsilon ^{\mu \rho \alpha \beta } \Bigl[-\frac{K_\beta L_\alpha}{L.Q_1 P.Q_1 P.Q_2}-\frac{K_\beta L_\alpha}{L.Q_2 P.Q_1 P.Q_2}-\frac{2 K_\beta P_\alpha}{(P.Q_1)^2 P.Q_2}-\frac{2 K_\beta P_\alpha}{P.Q_1 (P.Q_2)^2}+\frac{L_\alpha P_\beta L.Q_2}{L.P L.Q_1 P.Q_1 P.Q_2}-\frac{L_\alpha P_\beta L.Q_1}{L.P L.Q_2 P.Q_1 P.Q_2}\Bigr]\Bigr\}.
\end{align}
	\end{small}
\end{appendix}
\end{appendices}
\bibliographystyle{apsrev4-2}
\bibliography{Refs}
\end{document}